\documentclass[printer]{aa}  

\usepackage{lscape}
\usepackage{natbib}
\usepackage{color}
\usepackage{xcolor}
\usepackage{natbib}
\usepackage{latexsym} 
\usepackage{times} 
\usepackage{hyperref}
\usepackage{mathrsfs}
\usepackage{tikz}
\usepackage{booktabs}
\usepackage{array}
\usepackage{multirow}
\usepackage{mathtools}
\usepackage{cancel}
\usepackage{subfigure}
\usepackage{comment}
\usepackage{float}
\usepackage[normalem]{ulem}
\usepackage{graphicx}
\usepackage{diagbox}
\usepackage{txfonts}
\begin{document}
   \title{A complete spectroscopic catalogue of local galaxies in the Northern spring sky - Gas properties and nuclear activity in different environments
   \thanks{Tables A.1 and B.1 are fully available in electronic form
   at the CDS via anonymous ftp to cdsarc.u-strasbg.fr (130.79.128.5)
   }}

   \subtitle{}

  \author{Federico Cattorini \inst{1,}\inst{2} \and Giuseppe Gavazzi \inst{3} \and Alessandro Boselli \inst{4} \and Matteo Fossati \inst{3}}
  \institute{Dipartimento di Scienza e Alta Tecnologia, Università degli Studi dell'Insubria, Via Valleggio 11, I-22100, Como, Italy
        \and
   INFN, Sezione di Milano-Bicocca, Piazza della Scienza 3, I-20126 Milano, Italy\\
   \email{fcattorini@uninsubria.it}
        \and
   Dipartimento di Fisica G. Occhialini, Università di Milano- Bicocca, Piazza della Scienza 3, I-20126 Milano, Italy\\
   \email{peppogavazzi@gmail.com} 
        \and
    Aix Marseille Universit\'{e}, CNRS, LAM (Laboratoire d’Astrophysique de Marseille), UMR 7326, F-13388, Marseille, France\\
   \email{alessandro.boselli@lam.fr} 
	    }

   \date{Received - Accepted}

 
  \abstract
  {With the aim of providing the complete demography of galaxies in the local Universe, including their nuclear properties, we present SPRING, a complete census of local galaxies limited to the spring quarter of the Northern sky ($10 \mathrm{h} < \mathrm{RA} < 16\mathrm{h}$;  $0^{\circ}< \mathrm{Dec} < 65^{\circ}$).
  The SPRING catalogue is a flux-  and volume-limited sample ($ r<17.7$ mag, $cz<10000 \ \mathrm{km \ s}^{-1}$) of 30597 galaxies, including the Virgo, Coma and A1367 clusters. Images and spectra were individually examined to clear the sample from unwanted entries. \\ 
  To inspect possible secular and environmental dependencies of the various nuclear excitation properties (SF vs. AGN), we perform a multidimensional analysis by dividing the total sample according to \textit{(i)} their position in the $(\mathrm{NUV}-i)$ vs. $M_{\rm star}$ diagram, \textit{(ii)} local galaxy density, \textit{(iii)} stellar-mass, \textit{(iv)} halo-mass of the group to which galaxies belong, and \textit{(v)} neutral Hydrogen content. We present a new calibration of the optical diameter-based H\thinspace{\scriptsize I}-deficiency parameter $\mathrm{H} \thinspace \scriptsize{\text{I}}_{\mathrm{def}}$ employing a reference sample of isolated galaxies extracted from SPRING. At intermediate distances between Virgo and Coma, we identify a ring-like structure of galaxies constituted by three large filaments, each with approximately 20$h^{-1}$ Mpc length, mostly composed of blue-cloud galaxies with stellar-mass $M_{\rm star} \lesssim 10^{10}$ M$_{\odot}$. The fraction of H\thinspace{\scriptsize I}-deficient galaxies within the filament ($\sim$30\%) suggests that filaments are a transitioning environment between field and cluster in terms of H\thinspace{\scriptsize I} content, as we find a clear progression from field galaxies to filament and cluster galaxies for increasing $\mathrm{H} \thinspace \scriptsize{\text{I}}_{\mathrm{def}}$ parameter. \\
  We classify the nuclear spectra according to the four-line BPT and the two-line WHAN diagnostic diagrams, and investigate the variation in the fraction of active nuclei hosts with stellar-mass, as well as their colours and environments.
  We observe that the fraction of LINERs is a steep function of stellar-mass, \textit{e.g.,} it is consistent with zero up to $M_{\rm star} \lesssim 10^{9.5}$ M$_{\odot}$ and becomes $\sim$ 40 \% for $M_{\rm star} \gtrsim 10^{10.5}$ M$_{\odot}$, whereas, for $ M_{\rm star} \lesssim 10^{9-9.5}$ M$_{\odot}$, almost the entire spectroscopic sample is constituted of galaxies with star-forming nuclei. 
  We investigate whether the nuclear-excitation fractions depend predominantly on the stellar-mass or, conversely, on the galaxy environment.
  In general, we observe that the mass-dependency of the fraction of Seyfert nuclei is little sensitive to the galaxy environment, whereas the fraction of star-forming nuclei is a steeper function of stellar-mass in lower-density environments and in blue-cloud galaxies. We find that the fraction of LINERs depends on galaxy colour and, for $M_{\rm star} \gtrsim 10^{9.5-10}$ M$_{\odot}$, increases in galaxies belonging to the green valley.}

  \keywords{Galaxies: active -- Galaxies: nuclei -- Galaxies: Seyfert}

\titlerunning{The complete census of local galaxies in the Northern spring sky - I. The catalogue}

\maketitle
\section{Introduction}
The spring sky, comprised between 10 and 16 hours of right ascension in the Northern hemisphere, has hosted most of the near-field cosmology studies in the history of astrophysics, due to several unique characteristics.
It contains \textit{(i)} the Northern Galactic Pole, the area less crowded by the Milky Way, which is thus less susceptible to dust extinction; \textit{(ii)} the Local Supercluster, including \textit{(iii)} the Virgo cluster, and \textit{(iv)} the Coma cluster (with the ``Great Wall'').

Starting from the 1950s, when the Palomar sky survey became available, a large number of astronomers interested in near-scale cosmology (\textit{e.g.,} Fritz Zwicky, Allan Sandage, Gerard de Vaucouleurs, and collaborators) focused their work in this region of the sky using photographic material taken at 1.5-2.5 meter-class telescopes  \cite[see, \textit{e.g.},][]{Zwicky-1961, Tully-1982, Einasto-1983, Binggeli-1985}. 
Their goal was to shed light onto the physical mechanisms that govern the structure of the Universe on local scale (within the distance of the Coma cluster, \textit{i.e.}, $\sim$100$h^{-1}$ Mpc).
Among the early studies, many efforts aimed to establish the scale relations that govern the evolution of galaxies, such as, for example, the ``downsizing'' law \cite[][]{Gavazzi-1996}.

At the turn of the millennium, the digital revolution took place with the announcement of the Sloan Digital Sky Survey \cite[SDSS,][]{York-2000}. The SDSS combined five-band ($u,g,r,i,z$) photometry from digital imaging with exquisite medium resolution ($R \approx 2000$) spectroscopy taken (except for few cases) in the nuclei of external galaxies brighter than  $17.7$ mag in the $r$-band.  However, instead of reforming the study of the Local Supercluster, the SDSS shifted the focus of near-field cosmology to a slightly higher redshift ($0.1 \lesssim z \lesssim 0.5$).
This turn was owing to the difficulty of extracting precise photometric parameters from digital images in areas crowded with extended galaxies using automatic algorithms because of significant overestimate of the sky background; another consequence of this effect is the so-called ``shredding'' that affects the SDSS photometry and spectroscopy of nearby galaxies \cite[][]{Blanton-2005a, Blanton-2005b, Blanton-2005c}.

Because of these difficulties, the vast majority of near-field-cosmology works based on the SDSS are at $0.1\lesssim z\lesssim 0.5$ \cite[see, \textit{e.g.,} ][]{Kauffmann-2003b, Gomez-2003, Blanton-2017, Peng-2010}. Alternatively, a number of studies employed versions of the SDSS  with improved background determination in case of very extended galaxies \cite[\textit{e.g.}, the NASA-Sloan Atlas catalogue of][]{Blanton-2011}, or with the limitation $z>0.01$ (which, anyway, excludes the Local Supercluster and the Virgo cluster).
This is, for example, the case of \cite{Yang-2007, Yang-2008, Yang-2009}, who, despite limiting their sample (in this case, the DR4) to $0.01<z<0.2$, retain more than 360.000 galaxies in it. This number is sufficiently large to ensure a satisfactory statistical coverage of the sky, but is too large to allow the firsthand inspection of all galaxies (images and spectra) in search for errors (\textit{e.g.,} false objects caused by diffraction spikes of bright stars, duplicates, spectroscopy not taken at the centre of galaxies). 
Such mistakes do not affect the sample statistically, but pollute the catalogue with unwanted individual objects.

Besides the SDSS, another outcome of the third millennium digital revolution was the availability of multi-wavelength photometry (from the ultraviolet to the far-infrared bands) based on space missions such as GALEX \cite[][]{GALEX, GALEX07} and WISE \cite[][]{WISE}. Multi-wavelength data have proven to be very effective in the investigation of the role of the environment in galaxy evolution \cite[see, \textit{e.g.},][]{Gavazzi-2013Ha3_a, Gavazzi-2013Ha3_b}. This is the case of the extensive work on the Virgo cluster carried out by \cite{Boselli-2014} \cite[built upon the pioneering work by][which, in spite of being a fully analogical work, represented a breakthrough for the study of the Virgo cluster]{Binggeli-1985}. The work of \cite{Boselli-2014} set up a new standard for the study of galaxy evolution in the Virgo cluster by dividing galaxies into red sequence, green valley, and blue cloud according to their $\mathrm{NUV}-i$ colour, where both $\mathrm{NUV}$ and $i$ magnitudes are corrected for extinction by the Milky Way and for dust attenuation taking advantage of $22\mu$m photometry from WISE \cite[][]{WISE}. This colour range contains all galaxies, with red galaxies separated by approximately 5 mag from blue ones. 
This new standard has opened a new method for studying the influence of the environment (\textit{e.g.,} ram pressure) on the transition between the blue cloud and the red sequence, passing through the green valley due to the progressive ablation of H\thinspace{\scriptsize I} gas necessary for the formation of new stars.
With the present work, we apply the new standards of \cite{Boselli-2014} to the entire spring sky, expanding from $\sim$1000 Virgo galaxies to $\sim$30000 galaxies analysed here. We aim to provide a census of galaxies in the local Universe, including their nuclear classification. 

It is possible to disentangle the variety of galactic nuclear activities in diverse ways: at optical wavelengths owing to a number of emission-line ratios, \textit{e.g.,} [NII]/$\rm H\alpha$ and [OIII]/$\rm H\beta$ \cite[see][]{Baldwin-1981, Cid_Fernandez-2010, Cid_Fernandez-2011},
\cite{Kauffmann-2003, Kewley-2001, Kewley-2006, Decarli-2007, Gavazzi-2011, Gavazzi-2013, Gavazzi-2018a, Reines-2013}; by the detection of nuclear radio continuum \cite[][]{Bregman-1990}; by X-ray sources \cite[][]{Dewangan-2008}; or by a combination of the two \cite[][]{Ballo-2012, Agostino-2019}. 

This work also aims to provide a bridge between the empirical analysis carried out by  \cite{Peng-2010}, who examined the statistical dependence of star-formation quenching on stellar-mass and environmental density, and other local studies of selected clusters, such as Virgo, where the perturbing mechanisms are identified using multi-frequency observations, tuned models, and simulations of representative galaxies \cite[see][]{Boselli-2014, Boselli-2022-REV}. 

Despite the impressive statistics of the SDSS ($\sim$1.5 $\times$ 10$^6$ galaxies extracted with photometric and spectroscopic data in the redshift range 0.02$<$ $z$ $<$ 0.085), which makes \cite{Peng-2010} a reference for years to come, the identification of the dominant perturbing mechanisms is hampered by the lack of multi-frequency data covering the whole electromagnetic spectrum necessary to quantify the effects of the perturbations on the different galaxy components (stars, gas, dust) and the star-formation process. Furthermore, the lack of multi-frequency data (such as X-rays data) necessary to describe the properties of the intracluster medium in high-density regions is also a major limit in identifying the dominant perturbing mechanisms. Despite its limited statistics, the sample analysed in the present work has a broad multi-frequency coverage (from the UV to the infrared). It includes well-known structures such as the Local Supercluster with Virgo, the most studied cluster in the sky, but also the Coma Supercluster \textendash ~including the two massive clusters Coma and A1367 \textendash ~and the Great Wall, the closest filamentary structure linking high-density regions in the Universe \cite[see, \textit{e.g.},][]{Gavazzi-2010, Cybulski-2014}.

\vspace{.3cm}
\noindent
The outline of the Paper is as follows.  
Section $\S$ \ref{sample} presents our flux-  and volume-limited sample ($ r<17.7$ mag, $cz<10000 \ \mathrm{km \ s}^{-1}$) and describes the galaxy characterization based on their photometry (\S \ref{sec:photometry}), nuclear spectra (\S \ref{sec:nuclearspectra}), and environment (\S \ref{sec:environment}).
The nuclear spectra are analysed and classified in Section $\S$ \ref{sec:results} using the BPT ($\S$\ref{sec:bpt}) and WHAN ($\S$\ref{sec:whan}) diagnostic schemes. This classification is adopted to investigate the relation between nuclear excitation properties, galaxy stellar-mass, and environment in Section $\S$ \ref{subsec:stmass-nuclear}.
A critical feature of the method we adopted is that we checked all $\sim$30000 images and spectra individually using the visualization tool provided by SDSS. This enabled us to disregard targets that are, in fact, duplicates of galaxies observed in peripheral positions (spiral arms or H\thinspace{\scriptsize II} regions) not corresponding to galaxy nuclei.
Finally, we summarise our results and conclude in Section $\S$ \ref{conclusion}. 
Throughout the Paper, we assume a $\Lambda$CDM cosmology with  $\Omega_{\mathrm{M}}=0.3$,  $\Omega_{\Lambda} = 0.7$, $H_0 = 73 h_{73} \  \mathrm{km \ s}^{-1}\mathrm{Mpc}^{-1}$.

To mitigate the incompleteness of the SDSS nuclear spectroscopy at the bright luminosity end, in 2014-2020 we took 302 nuclear spectra with the Loiano 1.5m telescope. Appendix \ref{appendix-Loiano} provides the spectra obtained at Loiano.  
Appendix \ref{appendix-springSample} contains a one-page sample of the SPRING dataset, containing the first 30 of 30597 galaxies. The full table is available electronically at the CDS.

\section{The sample}\label{sample}
The sample analysed in this work is based on the SDSS Portsmouth emission-line catalogue  \cite[``emissionLinesPort'', ELP henceforth, see ][]{Thomas-2013}, which was queried via \texttt{SQL} with the SDSS Catalogue Archive Server Jobs System (CasJobs)\footnote{\url{http://skyserver.sdss.org/CasJobs/}}. Photometric data were obtained cross-correlating our sample with the NASA-Sloan Atlas\footnote{\url{http://nsatlas.org}}  \cite[NSA,][]{Blanton-2011} that, being limited to $z>0.01$, excludes the Local Supercluster and the Virgo cluster.
Across this work, our matching criteria is $\delta\theta < 5^{''}$ and $c\delta z < 200$ km s$^{-1}$ unless stated otherwise. All non-unique associations have been visually inspected.
Our sample comprises ($i$) local galaxies in the spring quarter of the Northern sky ($10 \mathrm{h} < \mathrm{RA} < 16\mathrm{h}$;  $0^{\circ}< \mathrm{Dec} < 65^{\circ}$), ($ii$) within a distance of $cz<10000$ km s$^{-1}$, and ($iii$) selected to have Petrosian magnitudes $ r<17.7$ mag (\textit{i.e.}, the spectroscopic completeness limit of the SDSS).

\begin{figure*}
\begin{centering}
\begin{minipage}{.495\textwidth}
\includegraphics[width=\textwidth]{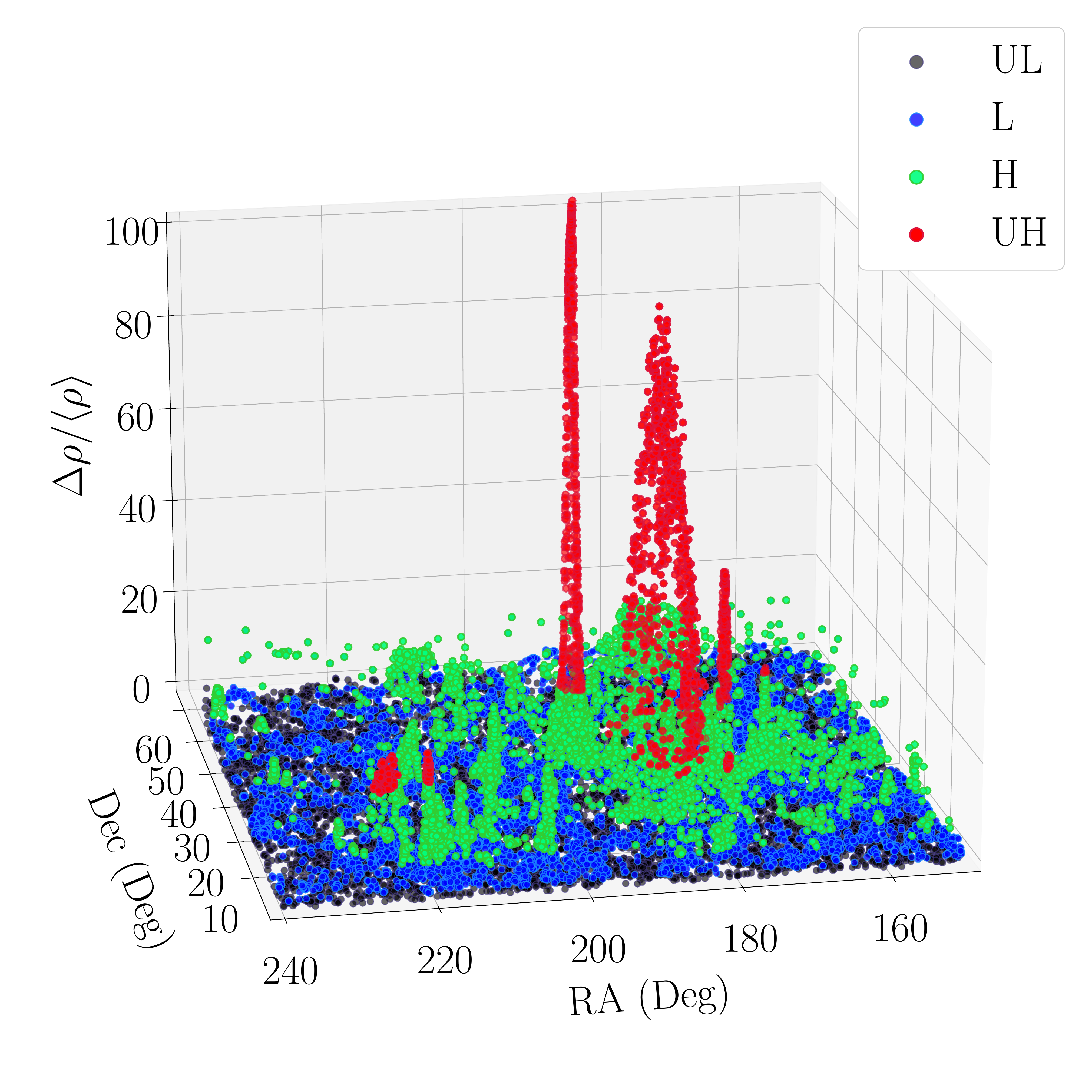}
\end{minipage}
\begin{minipage}{.495\textwidth}
\vspace{-0.35cm}
\includegraphics[width=\textwidth]{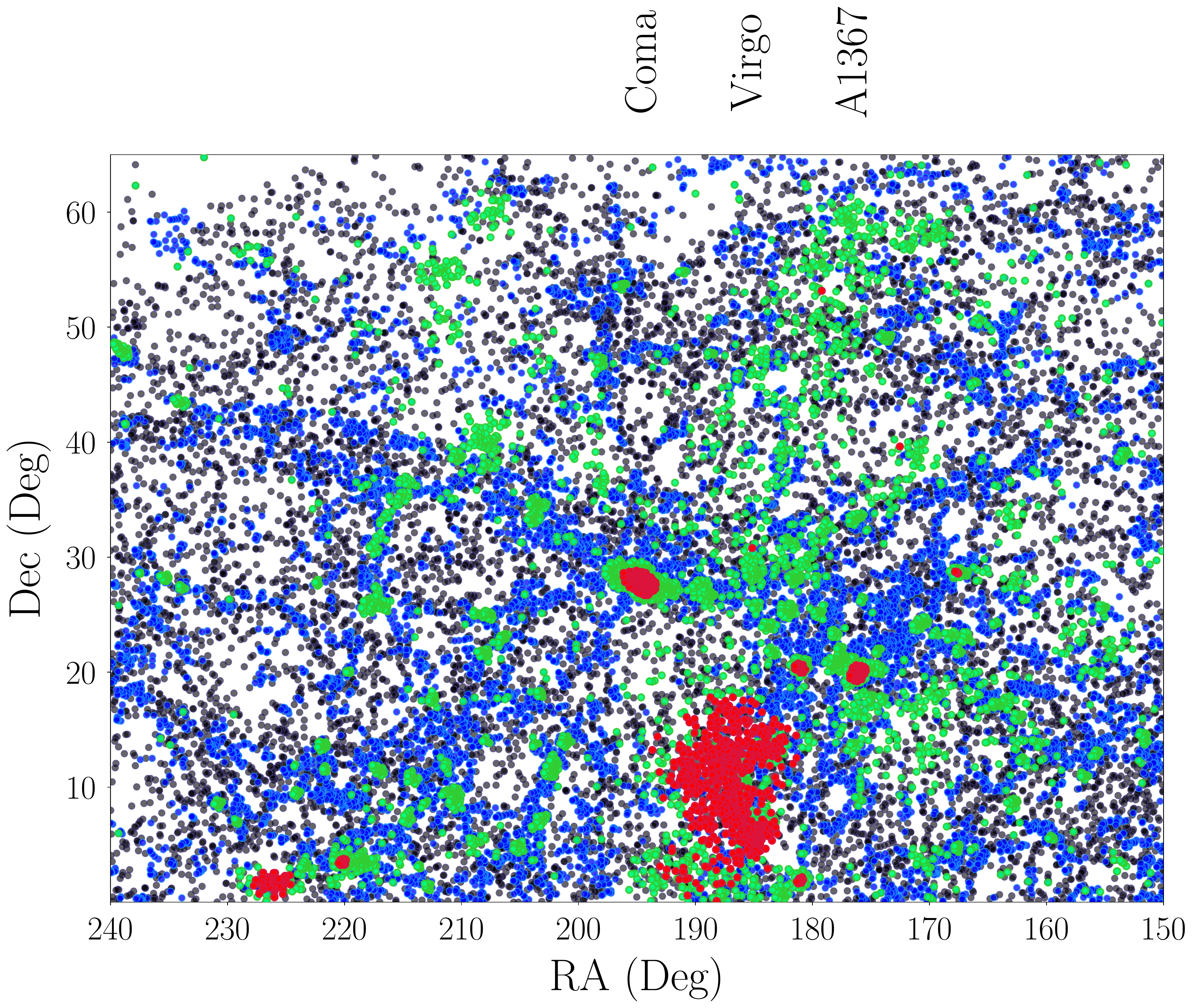}\end{minipage}
\caption{Celestial distribution of the 30597 galaxies analysed in the SPRING catalogue, distributed over nearly one-fourth of the Northern sky. They represent the totality of galaxies in the spring sky within $cz<10000 \ \mathrm{km \ s}^{-1}$ and $r<17.7$ mag. The colors represent the overdensity parameter $\Delta\rho / \langle\rho\rangle$ (defined in Sec. $\S$ \ref{subsec:overdensity}) for galaxies in four bins of local galaxy overdensity (red = UH, green = H, blue = L, black = UL). On the left panel, the third dimension emphasizes the overdensity parameter. The highest narrow peak represents the Coma cluster, the second narrow peak is A1367, and the broader peak is Virgo.}
\label{fig:density}
\end{centering}
\end{figure*}

We find 28190 galaxies meeting criteria ($i$)-($iii$) in the ELP table. However, adopting identical criteria to select galaxies in the NSA table of \cite{Blanton-2011} (who claims to have solved some of the incompleteness problems of bright galaxies) we find 28440 galaxies, among which 26966 are in both catalogues, 1474 are only in the NSA table and 1224 only in the ELP table. 
The differences between the NSA and the ELP may be due to a number of factors. The NSA was built on the basis of several catalogs: Sloan Digital Sky Server (DR8), NASA Extragalactic Database, Six-degree Field Galaxy Redshift Survey, Two-degree Field Galaxy Redshift Survey, ZCAT and ALFALFA, while the ELP table (DR12) is built from SDSS and BOSS galaxies.
Our analysis considers all (29590) targets found in either of the two catalogues. We further integrate this sample with the updated Zwicky catalogue \cite[UZC,][]{Zwicky-1961, Falco-1999}, the VCC catalogue \cite[][]{Binggeli-1985}, the UGC catalogue \cite[]{Nilson-1973}, the ALFALFA survey $\alpha.100$ \cite[]{Durbala-2020}, the optical spectroscopic survey by Ho, Filippenko and Sargent \cite[]{Ho-1993, Ho-1995, Ho-1997}, the spectra taken by us with the Loiano 1.5m telescope \cite[\textit{e.g.},][]{Gavazzi-2013}, and the spectra taken with the MUSE IFU by \cite{Consolandi-2017, Fossati-2019, Pedrini-2022}. We dub the database obtained in this way as the ``SPRING'' database, which amounts to 30597 unique objects. Its sky distribution is given in Fig. \ref{fig:density}.

We claim that our SPRING catalogue is a robust and exhaustive compilation of all galaxies in the Northern spring sky, including the brightest ones. For example, all 29 galaxies in the Messier catalogue projected in this region are listed in our database; also, we cross-correlated our catalogue with the Updated Zwicky Catalogue \cite[][]{Falco-1999} and found that in the window 150 < RA < 240, 0 < Dec < 65, cz < 10000, Falco lists 5904 galaxies, among which 5878 are found in the SPRING sample (99.5\%). We manually inspected the remaining 29 objects by means of the SDSS navigator tool, finding that they are constituted by fake detections such as diffraction spikes (7), faint and diffused local objects (14), objects which are not covered by the SDSS (8). Nontheless, this is not a suitable criteria for establishing completeness, since by construction our catalogue includes the UZC. Thus, we cross-matched our sample with the RC3 catalogue \cite[][]{RC3} which we did not use in the construction of our sample. We found 94 objects in RC3 that are missing in SPRING. We manually inspected the 94 entries, and found that only 5 objects are indeed missing galaxies, while the remaining are either targets with wrong coordinates, or faint (r $>$ 17.7 mag) objects, or errors in RC3.

\subsection{Photometry}\label{sec:photometry}
Photometric optical data are available for all galaxies in the sample. We take Petrosian ($u,g,r,i,z$) magnitudes from the SDSS. The GALEX ultraviolet (NUV and FUV) data \cite[][]{GALEX, GALEX07} are taken from \cite{Voyer-2014} for Virgo galaxies and from the Revised Catalog of GALEX Ultraviolet Sources \cite[][]{Bianchi-2017} otherwise; 25526 galaxies have NUV magnitudes, 24545 have FUV, and 24405 have both. For six galaxies, for which SDSS Petrosian $u, g, r, i, z$ magnitudes are not available, we take B and V magnitudes from the NASA/IPAC Extragalactic Database (NED).
Since a number of galaxies in the ELP/NSA tables have their photometry taken in oﬀ-nuclear positions, their given magnitudes are sometimes not representative of the whole galaxy. For those objects, we re-measure the $u, g, r, i, z$ magnitudes at the central position using the SDSS navigator.
For a small subsample of the SPRING catalogue (6764, 22\%), we collect information on the precise morphological type from \cite{Gavazzi-2003}, which were determined after the visual inspection of GOLDMine images.
\begin{figure}
\begin{centering}
\includegraphics[width=9cm]{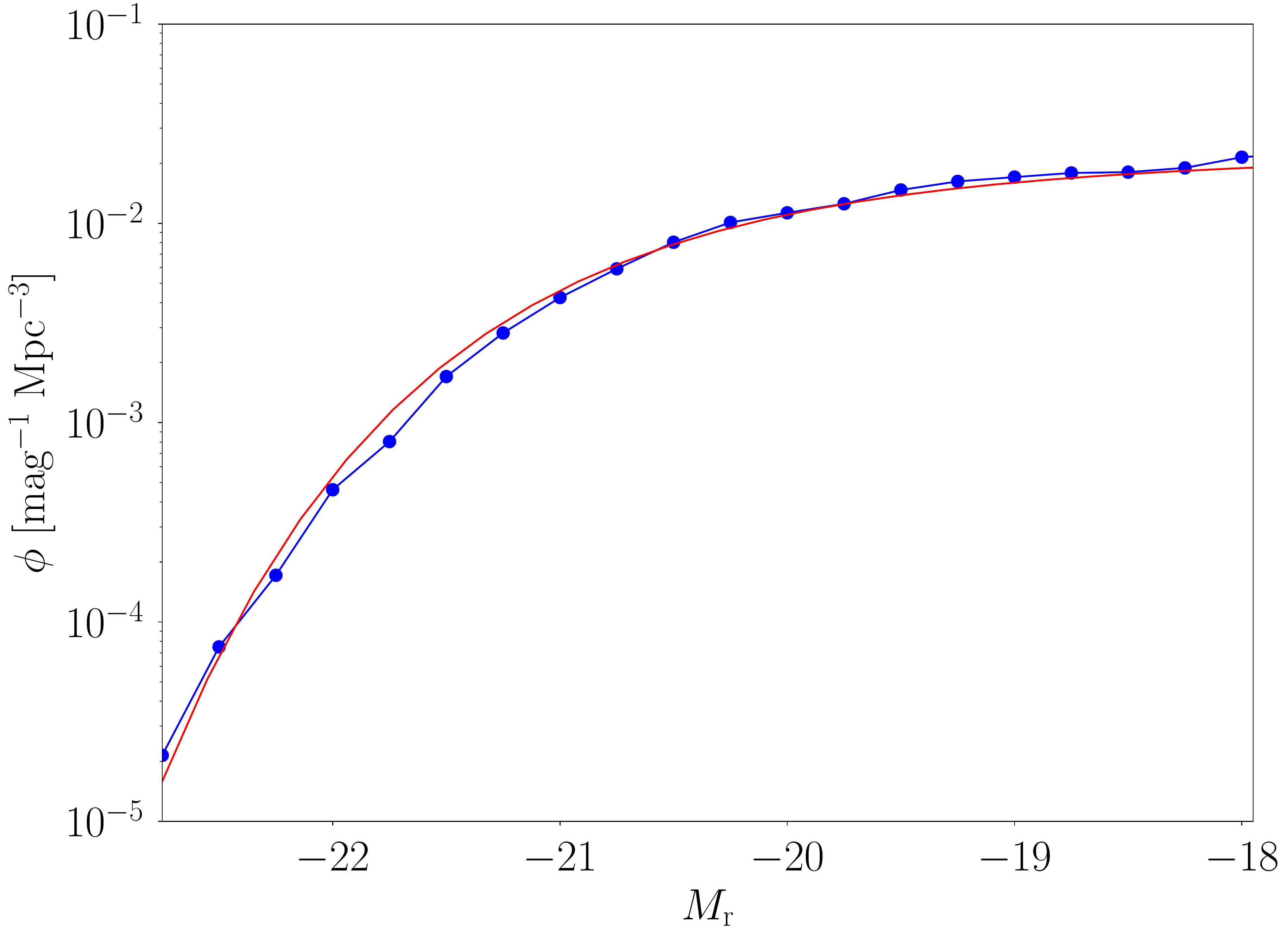}
\caption{Luminosity function of SPRING galaxies in the $r$-band (blue dots and line). Our best-fitting Schechter function (red line) agrees well with the data points.}
\label{fig:schechter}
\end{centering}
\end{figure}
The optical and UV data are corrected for Galactic extinction and dust attenuation. For this purpose, we cross-correlate the SPRING sample with the NASA’s Wide-field Infrared Survey Explorer \cite[WISE,][]{WISE} and extracted 22 $\mu$m W4 band data to quantify the attenuation in the FUV and NUV bands following the methods of \cite{Boselli-2014} \cite[see also][]{Hao-2011}. W4 band data are available for 30373 objects.
For galaxies without WISE 22$\mu$m data (224 objects), we estimated the internal attenuation using the mean value derived for objects of similar stellar mass using the relation:
\begin{equation}
A_{\mathrm{NUV}}  [\mathrm{mag}] = 0.111 \times (\log M_{\mathrm{star}})^2 - 1.555 \times \log M_{\mathrm{star}}  + 5.444
\end{equation}
\noindent
calibrated for the star forming galaxies on the rest of the sample. This relation was applied to the blue galaxies ($FUV-i$ $<$ 6) without WISE data, whereas we assumed $A(NUV)$ = 0 mag for the red (presumebly early-type) systems.

The adopted values of the stellar-mass are calculated by the $g$ and $i$ (corrected) magnitudes and the $i$-band luminosity assuming a Chabrier IMF \cite[][]{Chabrier-2003} and following the prescription of \cite{Zibetti-2009} as:
\begin{equation}
   \log (M_{\mathrm{star}}/\mathrm{M}_{\odot}) =-0.963+1.032(g-i)+\log l_I, 
\end{equation}
where $g$, $i$ are the corrected $g$-band and $i$-band magnitudes, $l_I$ is the $i$-band luminosity computed as $\log l_I=(I-4.56)/(-2.5)$, and $I$ is the absolute $i$-band magnitude (since our catalog is limited to very low redshift ($z\lesssim 0.03$, no $K$-correction is applied).
When the $g$ and $i$ magnitudes are not available, we employ the B and V magnitudes instead, and calculate the stellar-mass as
\begin{equation}
   \log (M_{\mathrm{star}}/\mathrm{M}_{\odot}) =-1.075+1.837(B-V)+\log l_V.
\end{equation}
The SPRING catalogue has been cross-matched with H\thinspace{\scriptsize I}-data from the GOLDMine database \cite[GM,][]{Gavazzi-2003}, the  Westerbork Coma Survey \cite[WCS,][]{Molnar-2022}, the ALFALFA survey \cite[AA,][]{Haynes-2011, Durbala-2020}, and the HyperLeda catalogue \cite[][]{Makarov-2014}. We cross-match our sample by coordinates separation with the criteria $c\delta z < 200$ km s$^{-1}$ (for all datasets), $\delta\theta < 5^{''}$ for GM, AA, and WSB and $\delta\theta < 10^{''}$ for LEDA. This choice was motivated by the lower accuracy of LEDA celestial coordinates; non-unique associations were visually inspected. H\thinspace{\scriptsize I} data are available for 9461 objects ($\sim$30\%).
If H\thinspace{\scriptsize I}-data from more than one source are available, we select data taken from GM over WCS, AA, and LEDA, respectively. This choice is motivated by \textit{(i)} the higher sensitivity of GM/WCS over AA, and \textit{(ii)} the inhomogeneity in LEDA sources, which were listed across a wide time span with diverse methodology.

\vspace{.3cm}
\noindent
A widely adopted method to identify galaxies at different evolutionary stages is the colour-stellar-mass relation \cite[][]{Boselli-2008a, Boselli-2014, Hughes-2009, Cortese-2009, Gavazzi-2010}. We use the $\mathrm{NUV}-i$ vs $M_{\mathrm{star}}$ relation \cite[][]{Kennicutt-1998} as a tracer of the mean age of the stellar population of galaxies in our sample, which we separate into three ``chromatic'' classes: \textit{(i)} a red sequence, composed of quiescent ETGs; \textit{(ii)} a blue cloud of star-forming LTGs \cite[][]{Gil-de-Paz-2007}; \textit{(iii)} and a green valley separating these two sequences \cite[][]{Martin-2007}.
We identify the red sequence as the region in the $(\mathrm{NUV}-i)$ vs $M_{\mathrm{star}}$ plane for which $\mathrm{NUV}-i>0.47\log  M_{\mathrm{star}}+0.1$; the blue cloud is the region for which $\mathrm{NUV}-i<0.47\log M_{\mathrm{star}}-1.0$; the green valley is the region between the former two. 
This representation, chosen by \cite{Boselli-2014} to study the Virgo cluster, is here extended to a much larger ($\sim30\times$) sample.

\subsection{Nuclear spectra}\label{sec:nuclearspectra}
Besides the SDSS, several other sources of nuclear spectroscopy are available in the literature; we integrate the SDSS spectroscopic data (available for 28971 targets) with two extensive spectral surveys: the one by \cite{Ho-1993, Ho-1995, Ho-1997} undertaken with the Palomar 200 inch telescope (224 spectra added), and the one by \cite{Falco-1999} using the Tillinghast 1.5m telescope (366 spectra added). 

On our side, since 2005 we undertook a spectroscopic project  using the $\mathrm{Cassini} 1.5 \mathrm{m}$ telescope of the Loiano  Observatory aiming at integrating the SDSS  with the nuclear spectroscopy of galaxies in the local Universe \cite[][]{Gavazzi-2011, Gavazzi-2013}. 
In 2014-2020, we obtained 302 new nuclear spectra (see Appendix \ref{appendix-Loiano}). These spectra are generally taken with the red-channel of the spectrograph, namely between 6000 and 8000 \text{\normalfont\AA}; only a dozen galaxies were observed also with the blue grism. In addition, five spectra were taken with the MUSE IFU by  \cite{Consolandi-2017, Fossati-2019, Pedrini-2022}. Altogether, the total number of galaxies with nuclear spectra observed independently from SDSS is 884.

All entries (images and spectra) in the SPRING catalogue were individually inspected in search of errors (\textit{e.g.,} false objects caused by diffraction spikes of bright stars, duplicates, misclassified objects). This initial iterative examination led us to clear $\sim$1\% of the objects in our sample; also, the direct inspection of all targets using the SDSS navigator tool (DR13 version) allowed us to eliminate 74 (less than 0.3 \%) targets with more than one spectrum available (\textit{i.e.}, galaxies with two or more nuclear spectra and galaxies with non-nuclear spectra). We leave two entries for the same object only in obvious post-merger targets where two nuclei are visible.

\begin{figure*}[!ht]
\begin{center} 
\includegraphics[width=8cm]{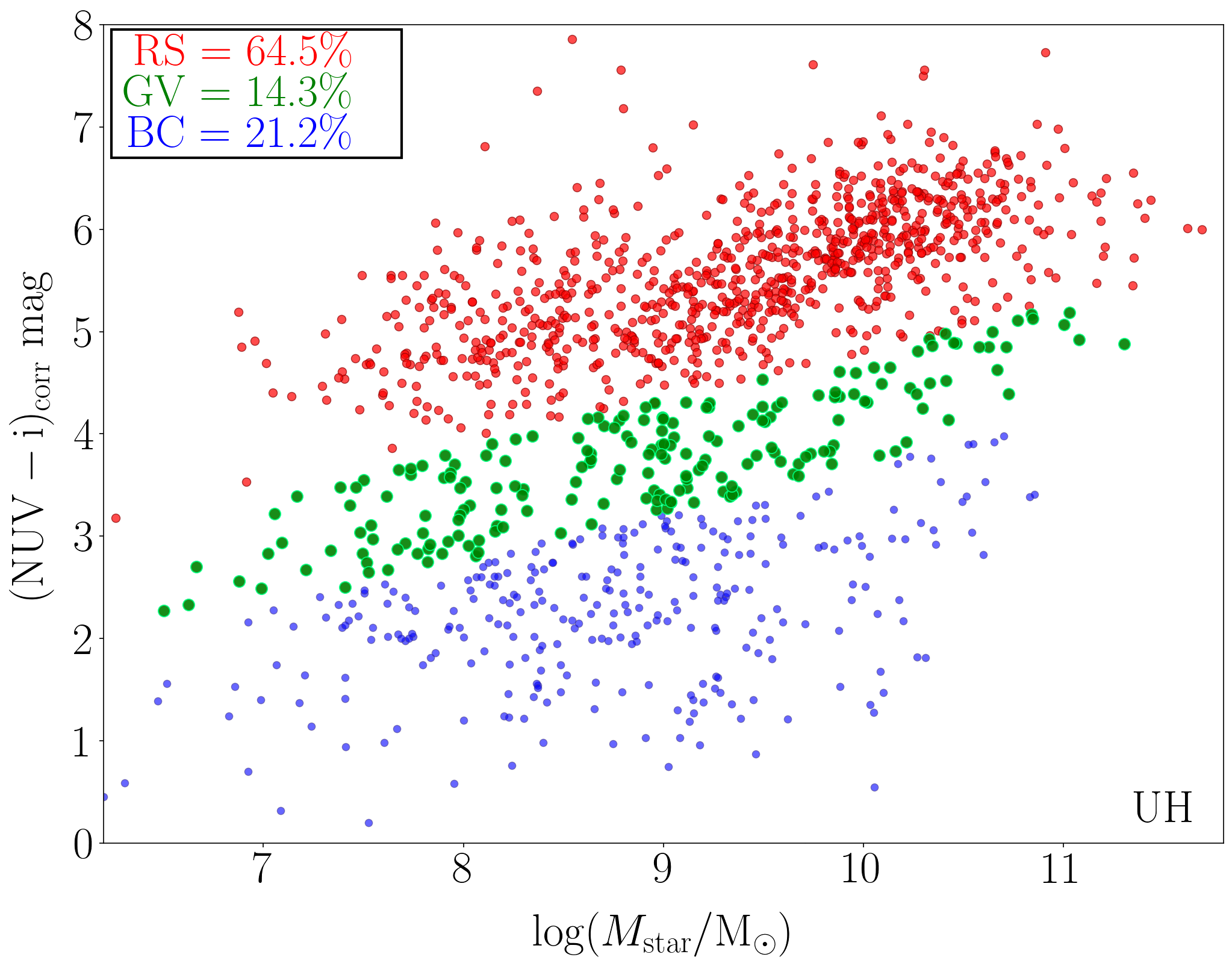}
\includegraphics[width=8cm]{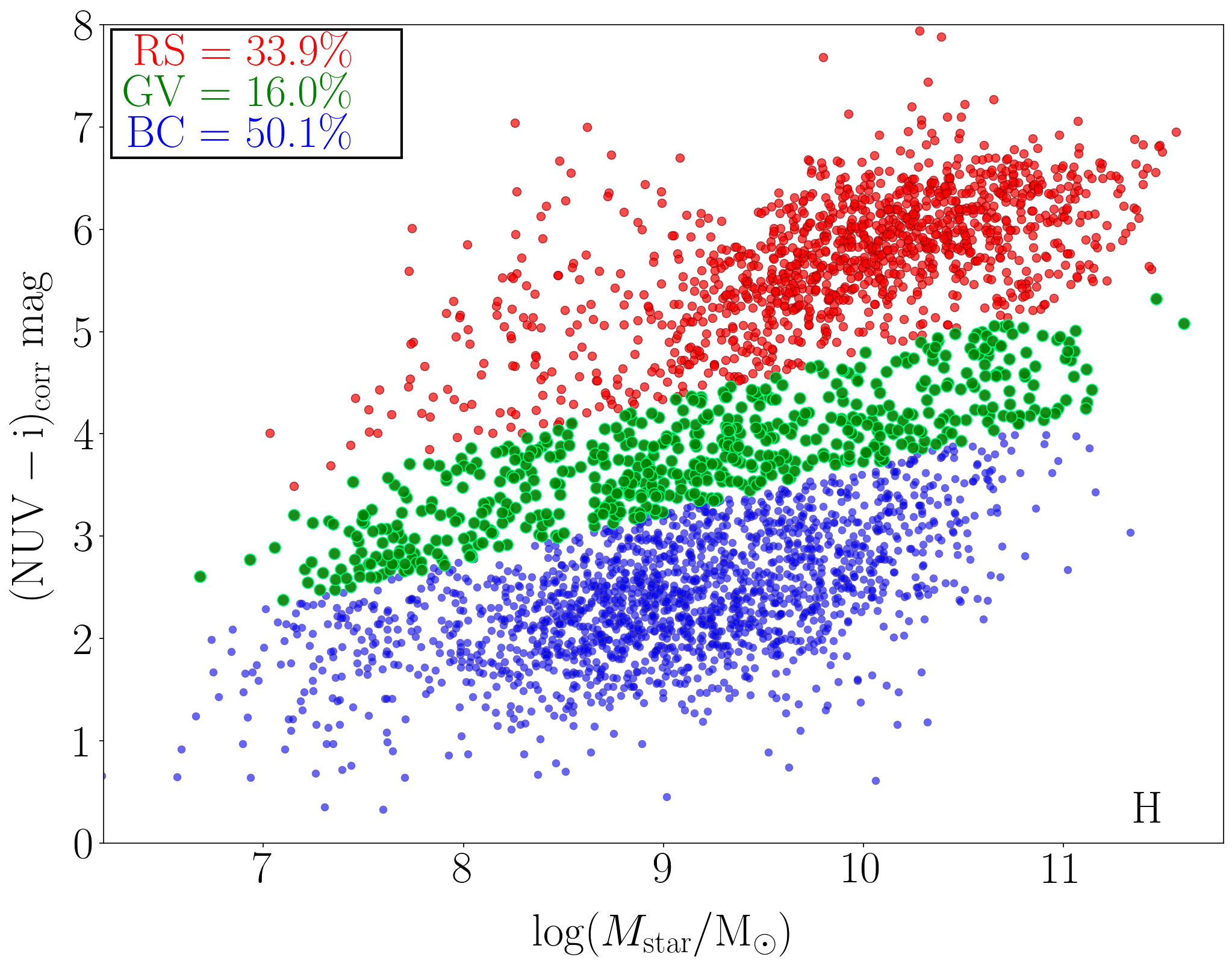}\\
\includegraphics[width=8cm]{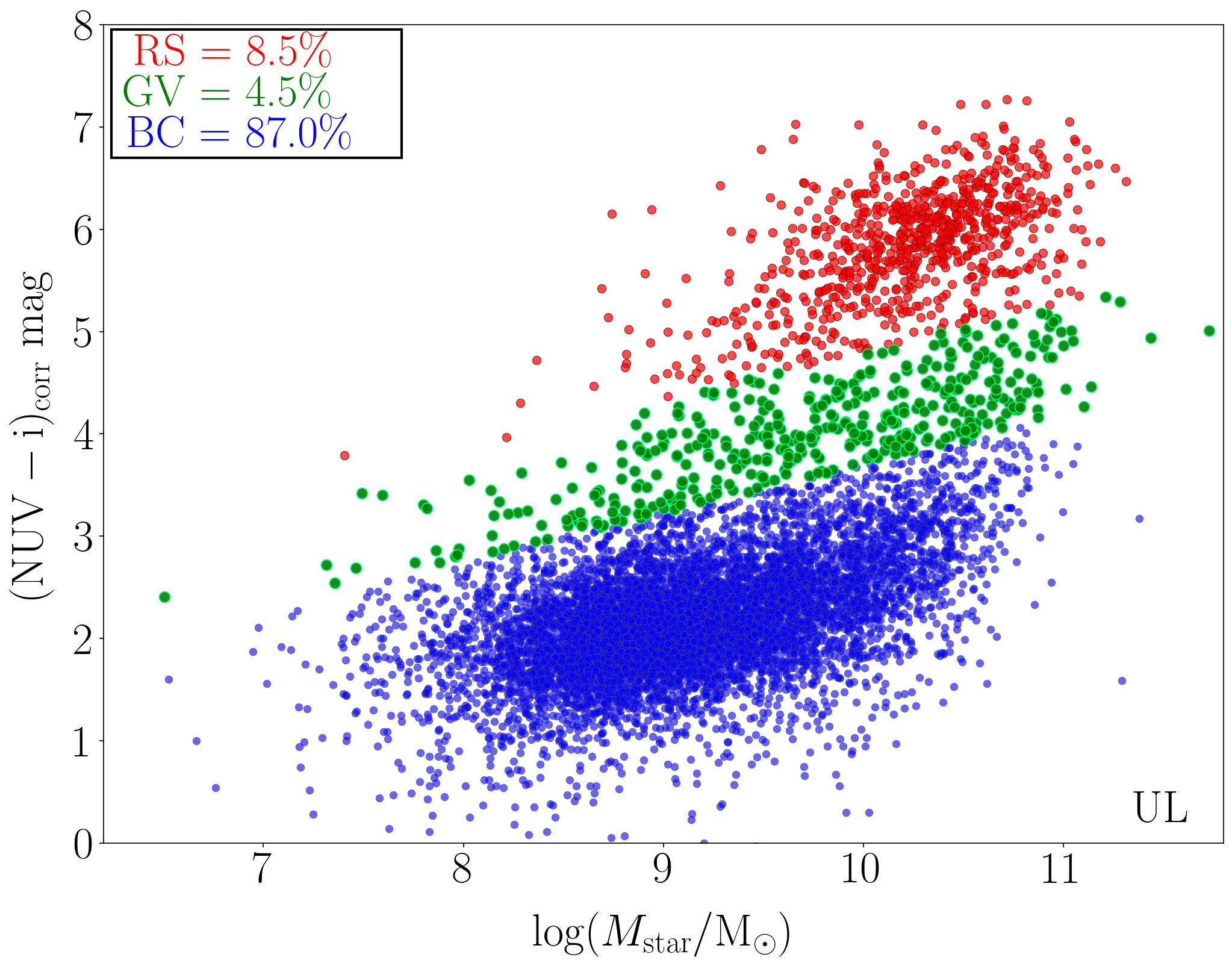}
\includegraphics[width=8cm]{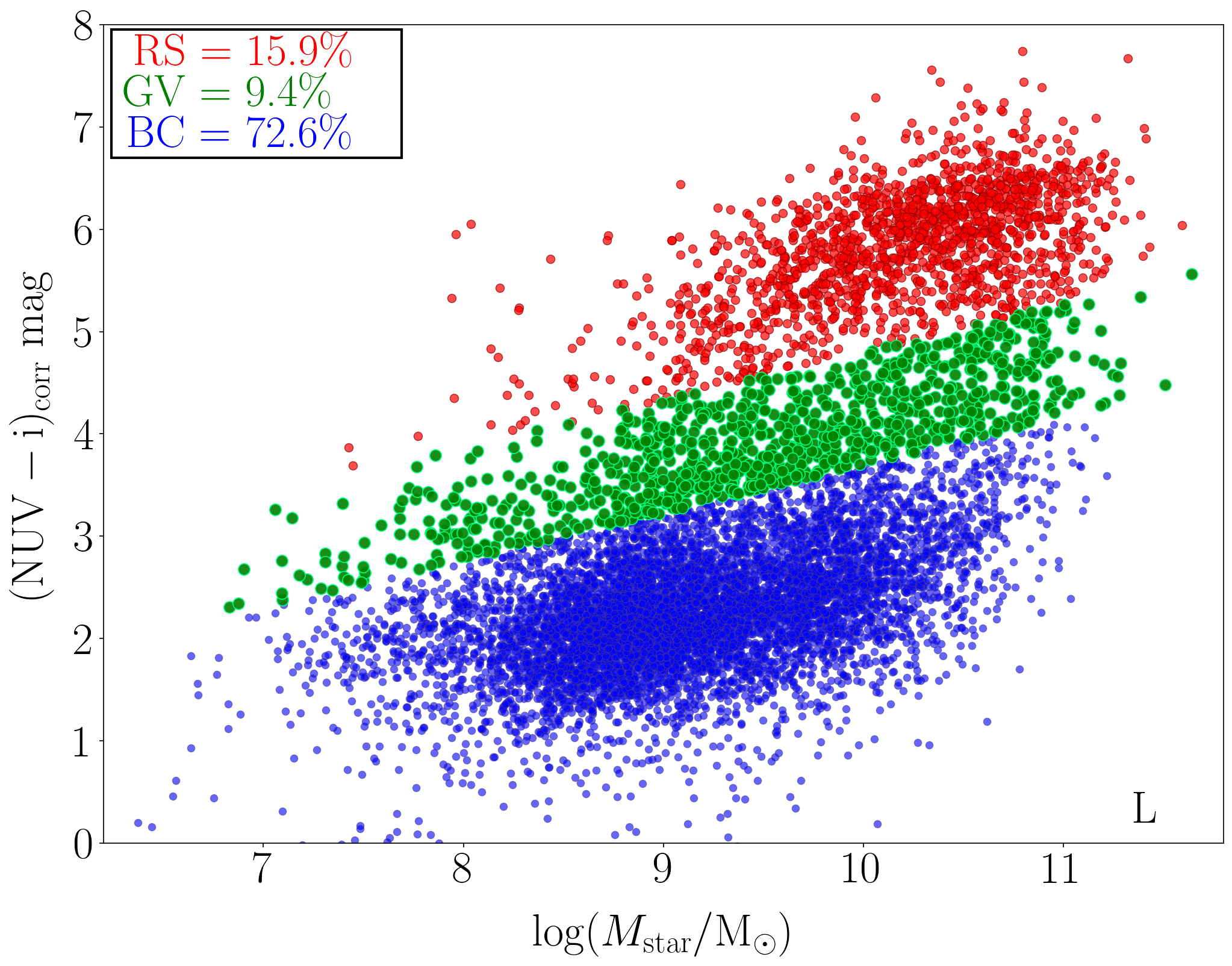}\\
\end{center}
\caption{Corrected $\mathrm{NUV}-i$ colour versus stellar-mass diagram relation
of all galaxies in this sample divided in blue cloud (BC), green valley (GV), and red sequence (RS), according to the chromatic criterion of \cite{Boselli-2014} (see Sec. $\S$ \ref{sec:photometry}), in four bins of galaxy overdensity (from UH to UL clockwise from top left). In each panel (corresponding to the four overdensities) are reported the fractions of galaxies in the three chromatic classes with respect to the total number of galaxies in the corresponding overdensity bin.}
\label{fig:colormass-overdensity}
\end{figure*}

\noindent
The SPRING catalogue is limited to $cz<$10000 km s$^{-1}$. Up to this redshift, corresponding to a distance of $\sim$137$h^{-1}$ Mpc and embracing a volume of 780000$h^{-3}$ Mpc${^3}$,  the sample is volume- and magnitude-limited, thus the number counts in bins of $M_{\mathrm{star}}< 10^9 \ \mathrm{M}_{\odot} $ (or $r<-17.7$ mag) can be converted into mass/luminosity functions. We verify that the $r$\textendash band luminosity function (LF) of our sample matches that derived by \cite{Blanton-2001} (including normalization, $M_{\mathrm{knee}}$ and slope). To this extent, we compare our galaxy distribution in the $r$-band magnitude space to a galaxy LF model using a maximum likelihood model (Fig. \ref{fig:schechter}). We adopt a Schechter function \cite[][]{Schechter-1976} and find the best-fit parameters $\alpha=-1.08, M^*=20.64, \phi_*=2.1\cdot10^{-2}$; the normalized mean residual between the fit and out data is $\sim 0.28$ dex.

\subsection{Probing the galaxy environment}\label{sec:environment}
Our analysis of the environmental properties of SPRING galaxies relies on differing techniques: in \S \ref{subsec:overdensity}, we reconstruct the 3D distribution of $\sim$30000 galaxies in 5400 square degrees of the spring sky and determine the density around each galaxy measuring the number of neighbors within fixed cylindrical volumes; in \S \ref{subsec:halomass}, we adopt the halo-mass as a tracer of the galaxy environment; in \S \ref{subsec:HIdef}, we employ the H\thinspace{\scriptsize I}-deficiency parameter to investigate the H\thinspace{\scriptsize I}-content distribution of galaxies.
\subsubsection{Galaxy density}\label{subsec:overdensity}
The local galaxy density around each object is quantified according to the method of \cite{Gavazzi-2010} which we briefly summarise in the following. 
After reducing the fingers of God of the three major clusters by collapsing the individual galaxy velocities to the average cluster velocity $\langle cz \rangle$, the density $\rho$ was determined in cylinders of 1$h^{-1}$ Mpc radius and 1000 km s$^{-1}$ half-length. Given an average density $\langle \rho \rangle$ of 0.04 gal $h^{3}$ $\text{Mpc}^{-3}$, the resulting overdensity parameter $(\rho -\langle\rho\rangle)/\langle\rho\rangle \equiv \Delta \rho/\langle \rho \rangle$ is estimated. This procedure is applied to all galaxies with a distance greater than 15$h^{-1}$ Mpc, to exclude very local galaxies which can be subject to large distance uncertainties due to peculiar velocities; 879 local galaxies are thus excluded from our overdensity determination. Following \cite{Gavazzi-2010}, we define four arbitrary overdensity thresholds which are chosen to highlight increasing levels of aggregation:
\begin{enumerate}
    \item[] Ultra low (UL):  $\ \ \ \ \ \ \ \ \ \ \ \ \ \ \ \ \Delta\rho/\langle \rho \rangle \leq 0 $
    \item[] Low (L): $\ \ \ \ \ \ \ \ \ \ \ \ \ \ \ \ \ \ \ \ 0 < \Delta \rho/\langle \rho \rangle \leq 4$
    \item[] High (H): $\ \ \ \ \ \ \ \ \ \ \ \ \ \ \ \ \ \ \ 4 < \Delta \rho/\langle \rho \rangle \leq 20$
    \item[] Ultra High (UH): $\ \ \ \ \ \ \ \ \ \ \ \ \ \ \Delta \rho/\langle \rho \rangle >20$.
\end{enumerate}
The UL density bin describes the underlying cosmic web; the L density bin is for the loose groups and the filaments in the Great Wall; the H bin includes the significant groups and cluster outskirts; the UH bin represents the cores of the clusters.
Figure \ref{fig:density} shows the overdensity parameter distribution in the whole sample.

To check the qualitative consistency with the morphology-segregation effect observed by \cite{Dressler-1980}, we derive that the fraction of  galaxies in the red sequence over the total in the four different density bins is 7\% in UL, 14\% in L, 29\% in H and  59\% in UH. 
In addition, to check consistency with \cite{Gavazzi-2010} (see Fig. 8 therein), we use the colour-stellar-mass relation to study the variation of the chromatic distribution with increasing overdensity parameter.
Figure \ref{fig:colormass-overdensity} displays four NUV-$i$ vs stellar-mass diagrams, highlighting the chromatic distributions of galaxies in the four overdensity bins. For each panel, the fractions of galaxies in the blue cloud, green valley, and red sequence are reported.

We observe that low-mass galaxies in the blue cloud dominate the population of galaxies in the UL bin ($\sim$86\%); still, there is a significant contribution of galaxies in the red sequence (and, to a lesser extent, in the green valley) which is, however, restricted to the larger ($M_{\rm star}>10^{9.5}$ M$_{\odot}$) masses. Conversely, in the UH bin, there is a minimal contribution of blue-cloud galaxies altogether ($\sim$20\%), and the red sequence includes a significant population of dwarf galaxies.

\subsubsection{Halo-mass}\label{subsec:halomass}
An alternative, widely used method to evaluate a galaxy's environment is by the halo-mass of the group to which it belongs \cite[see, \textit{e.g.}, ][]{Muldrew-2012, Fossati-2015}. The halo-mass is obtained from abundance matching techniques between the halo-mass function and the stellar-mass (or luminosity) of the most-massive galaxy identified within the group \cite[see \textit{e.g.},][]{Yang-2007, Yang-2008, Yang-2009}. We correlate our $cz<10000$ km s$^{-1}$ catalogue with the one by \cite{Yang-2007}, which extends much further out than our local sample, however excluding the Local Supercluster and the Virgo cluster ($cz>3000$ km s$^{-1}$).
We find 21608 matches in the shared redshift range between the two catalogues. 

Only 427 galaxies in the Yang catalogue have no counterpart in our database, and a visual inspection revealed them to be primarily extranuclear H\thinspace{\scriptsize II} regions, not galactic nuclei.
For the galaxies in common, we take the group halo-mass estimate obtained from the total stellar-mass in the group. For the Virgo cluster, which is not present in \cite{Yang-2007} and thus does not have an estimate of the virial mass, we use the estimate ($M_{200}=10^{14.4} \rm{M}_\odot$) reported by \cite{Boselli-2006}. 
\begin{figure}
\begin{centering}
\includegraphics[width=9cm]{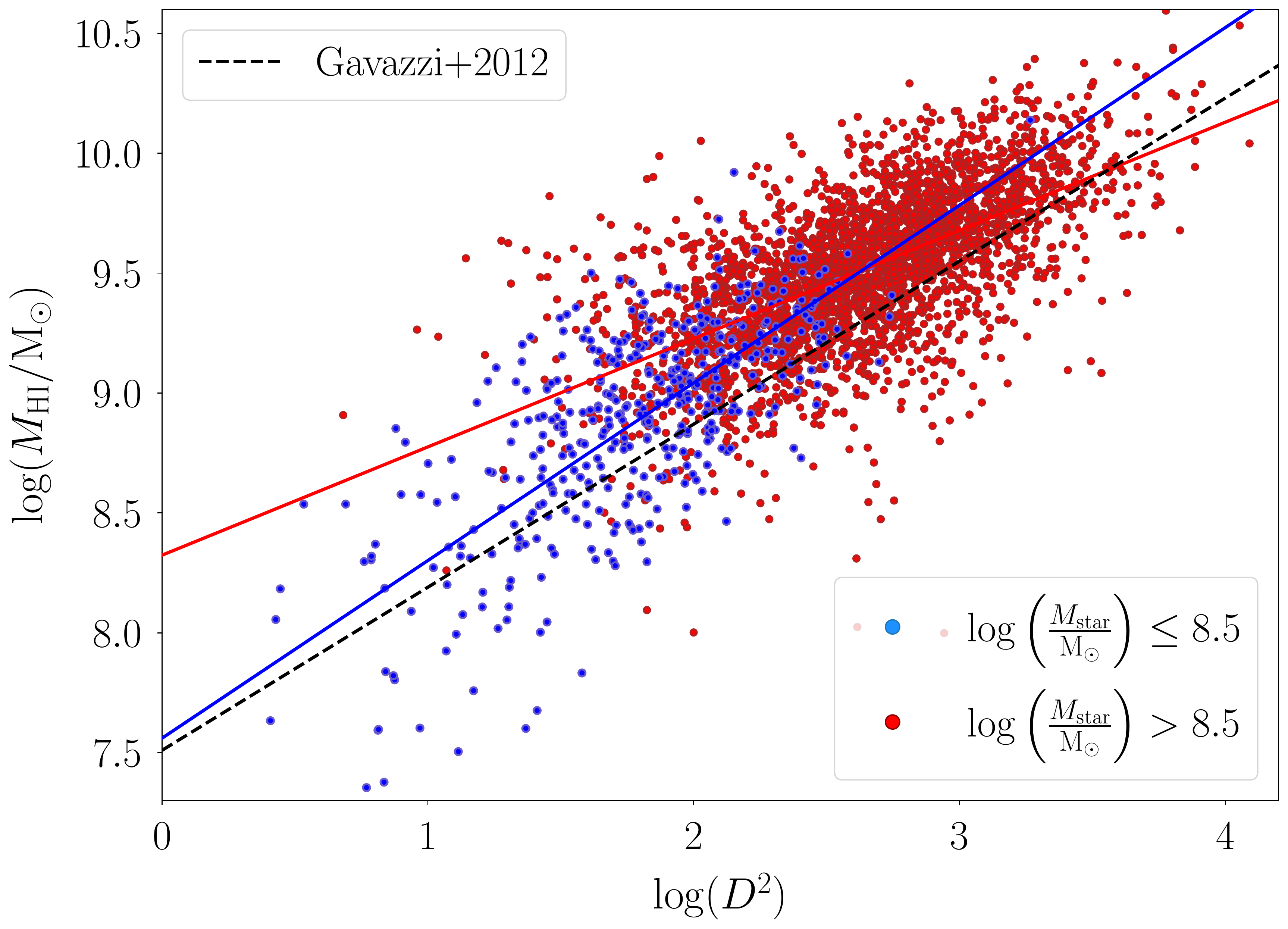}
\caption{$\log(M(\mathrm{H} \thinspace \scriptsize{\text{I}}))$ vs $\log(D^2)$ distribution of galaxies belonging to the UL overdensity bin and with $\log(M_{\mathrm{star}}/\mathrm{M}_{\odot}) \leq 8.5$ (blue dots), and $\log(M_{\mathrm{star}}/\mathrm{M}_{\odot}) > 8.5$ (red dots). The colour straight lines denote the best-fit linear regression in each mass bin, whose coefficients are given in Table \ref{tab:coeff}. The black dashed line denotes the linear fit given in \cite{Gavazzi-2013Ha3_a}.}
\label{fig:calib}
\end{centering}
\end{figure}
\begin{figure}
\begin{centering}
\includegraphics[width=9cm]{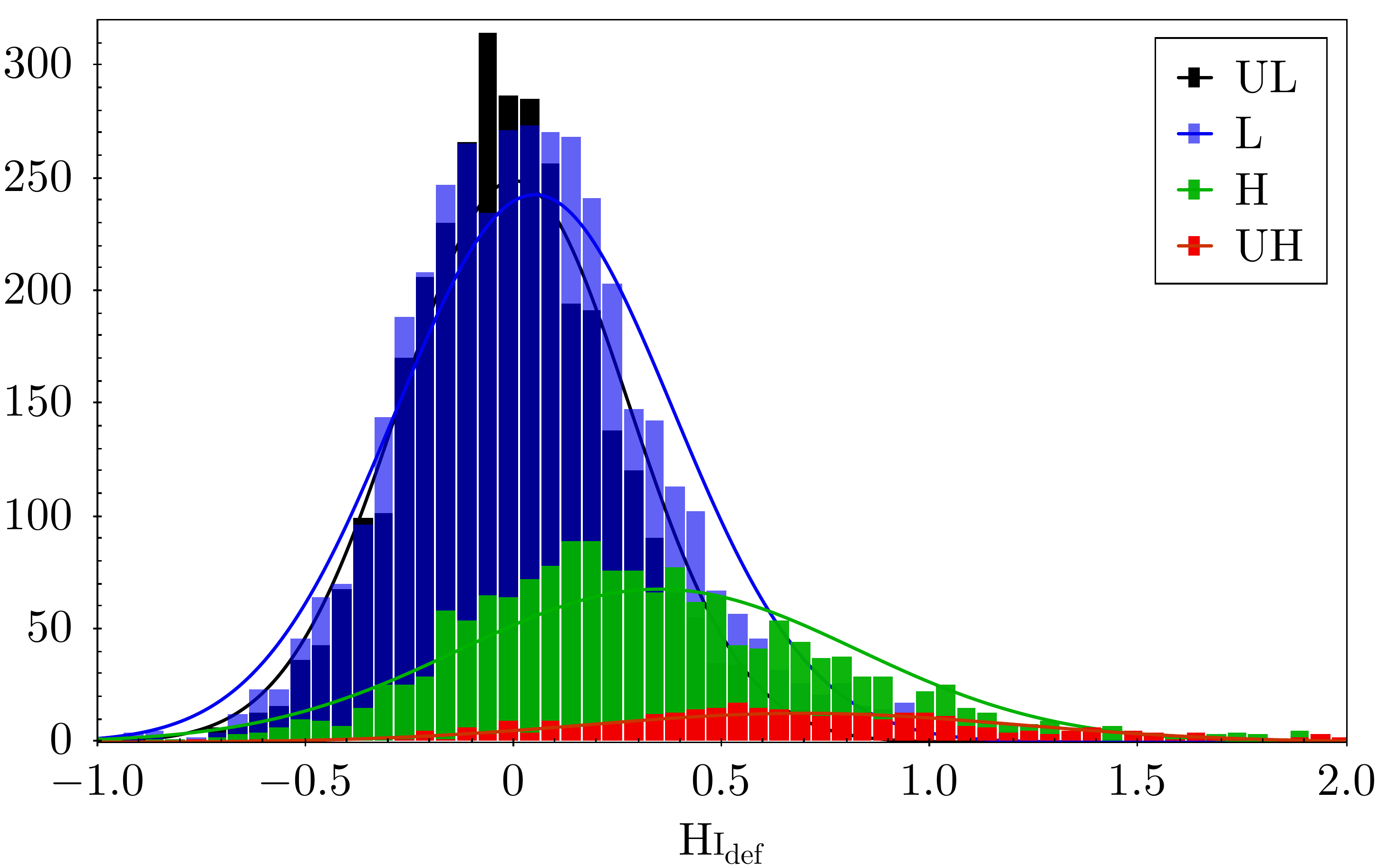}
\caption{Distribution of the newly calibrated $\mathrm{H} \thinspace \scriptsize{\text{I}}_{\mathrm{def}}$ parameter in four bins of galaxy overdensity. 
The mean values and standard deviations for each overdensity bin are given in Table \ref{tab:hist_params}.}
\label{fig:HIdef_hist}
\end{centering}
\end{figure}

\noindent
In analogy with the overdensity parameter, we define 3 halo-mass ranges:
\begin{enumerate}
    \item[] Low (L): $\ \ \ \ \ \ \ \ \ \ \ \ \ \ \ \ \ \ \ \ \ \ \ \ \ \ \ \ \ \ \ \ \log (M_{\mathrm{halo}}/\mathrm{M}_{\odot}) \leq 12.5$
    \item[] Intermediate (M): $ \ \ \ \ \ \ 12.5 < \log (M_{\mathrm{halo}}/\mathrm{M}_{\odot}) \leq 13.5$
    \item[] High (H): $\ \ \ \ \ \ \ \ \ \ \ \ \ \ \ \ \ \ \ \ \ \ \ \ \ \ \ \ \ \ \ \log (M_{\mathrm{halo}}/\mathrm{M}_{\odot}) > 13.5$ 
\end{enumerate}

\subsubsection{H\thinspace{\scriptsize I}-deficiency}\label{subsec:HIdef}
\begin{figure}
\begin{centering}
\includegraphics[width=9cm]{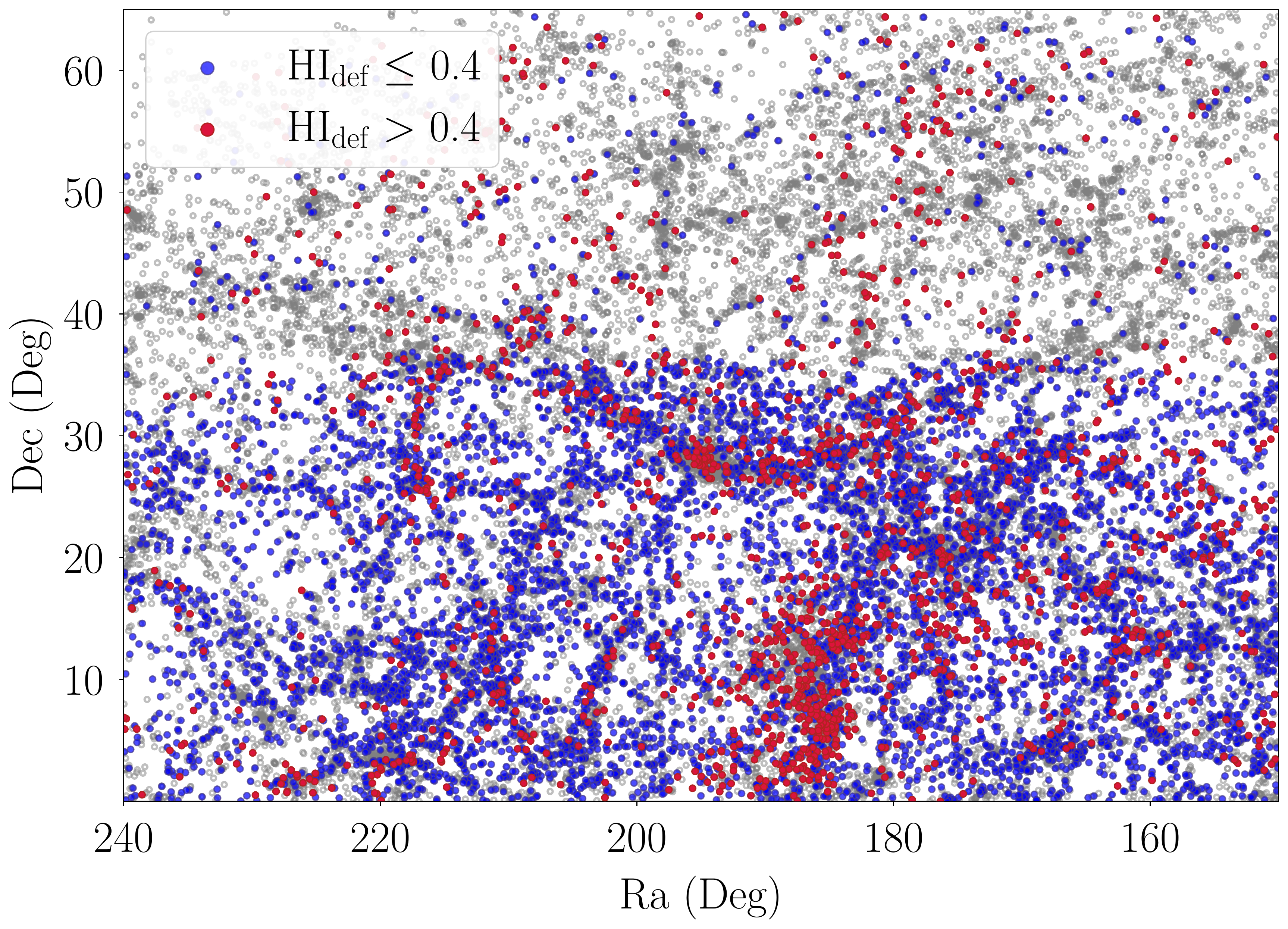}
\caption{Sky distribution of galaxies in the SPRING sample in classes of HI-deficiency parameter. The horizontal line at $\mathrm{Dec}  = 36 \ \mathrm{deg}$ traces the limiting pointing declination of the Arecibo dish. Grey circles stand for galaxies for which the H\thinspace{\scriptsize I} content is unknown. Blue dots denote galaxies with H\thinspace{\scriptsize I}$_{\mathrm{def}}\leq 0.4$, while red dots are for galaxies with H\thinspace{\scriptsize I}$_{\mathrm{def}}> 0.4$. The latter class is distributed consistently around (in) the Coma cluster and the Virgo cluster, with a small number of entries belonging to the A1367 cluster.
It is noticeable that our knowledge of the H\thinspace{\scriptsize I} properties of local galaxies in such an extended 
stretch of the sky depends on the availability of Arecibo data, which are limited to Declination $\lesssim$ 36$^{\circ}$.}
\label{fig:HIdef}
\end{centering}
\end{figure}
\begin{figure*}
\begin{center}
\includegraphics[width=8cm]{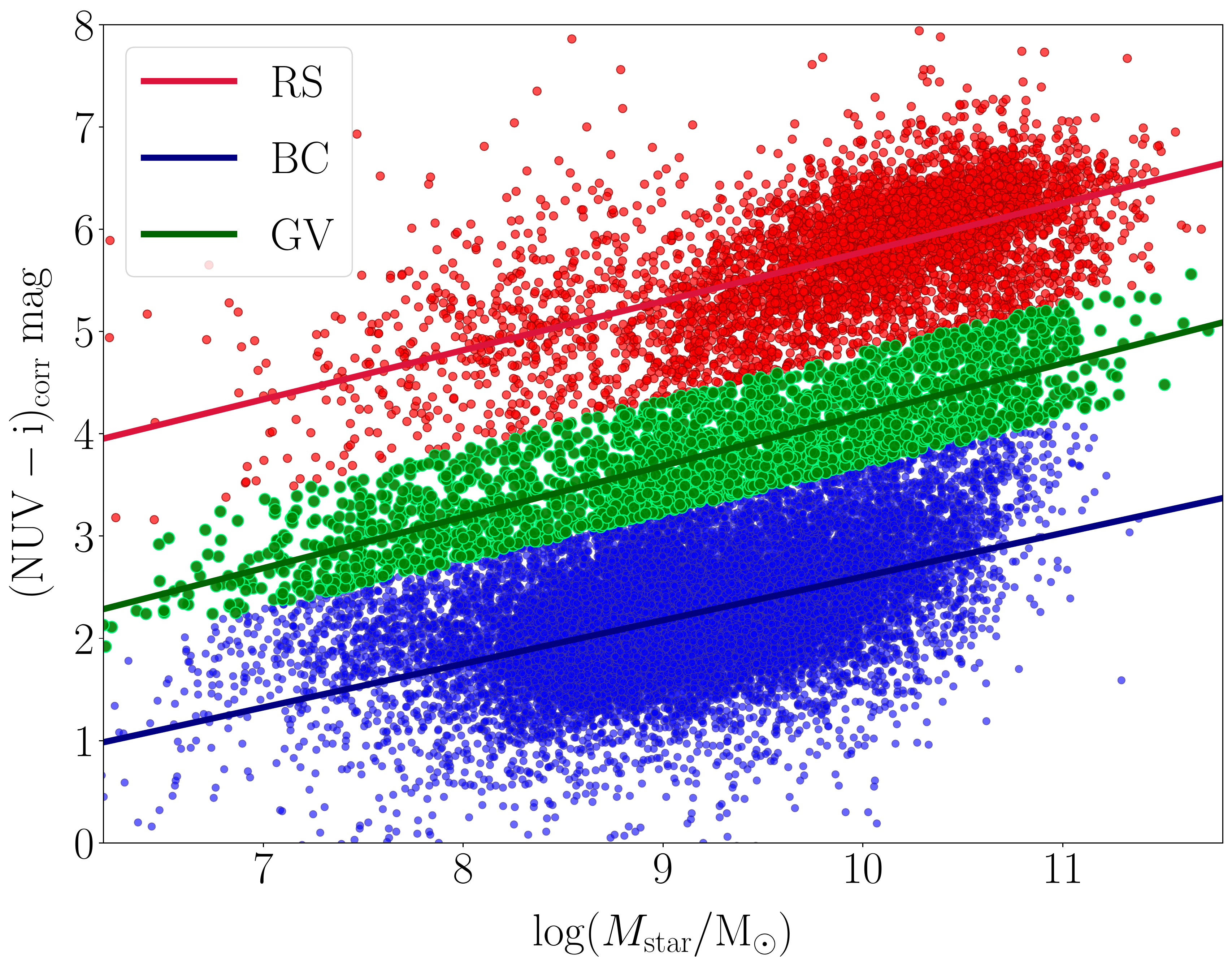}
\includegraphics[width=8cm]{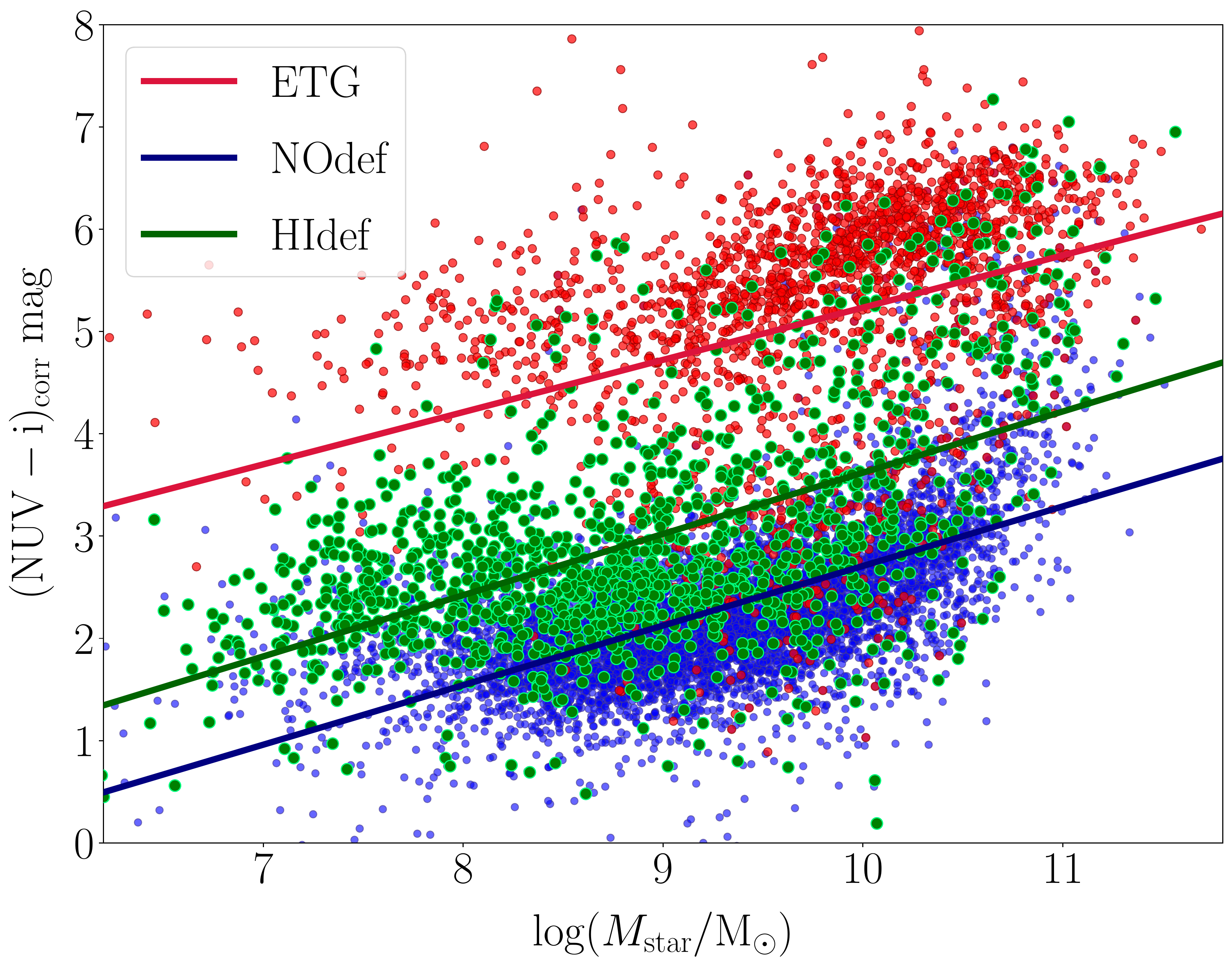}
\caption{Left: corrected $\mathrm{NUV}-i$ colour versus stellar-mass diagram relation
of all galaxies in this sample divided according to the chromatic criterion of \cite{Boselli-2014}  (see Fig. 2 therein) whether they belong to the red sequence (RS), the green valley (GV) or the blue cloud (BC). For each class, the linear regression analysis is given.
Right: the same relation for the subsample of ETGs with detailed morphological classification available (red); of deficient (H\thinspace{\scriptsize I}$> 0.4$, green) and non-deficient galaxies (H\thinspace{\scriptsize I}$\leq 0.4$, blue). The slopes of the linear regressions for the ETGs and the H\thinspace{\scriptsize I}-deficiency classes are consistent with the ones corresponding to the three chromatic classes.
The NUV and $i$ magnitudes are corrected for dust attenuation \cite[see ][]{Boselli-2014}.}
\label{fig:colormass}
\end{center}
\end{figure*}
\begin{figure*}
\begin{center}
\includegraphics[width=8cm]{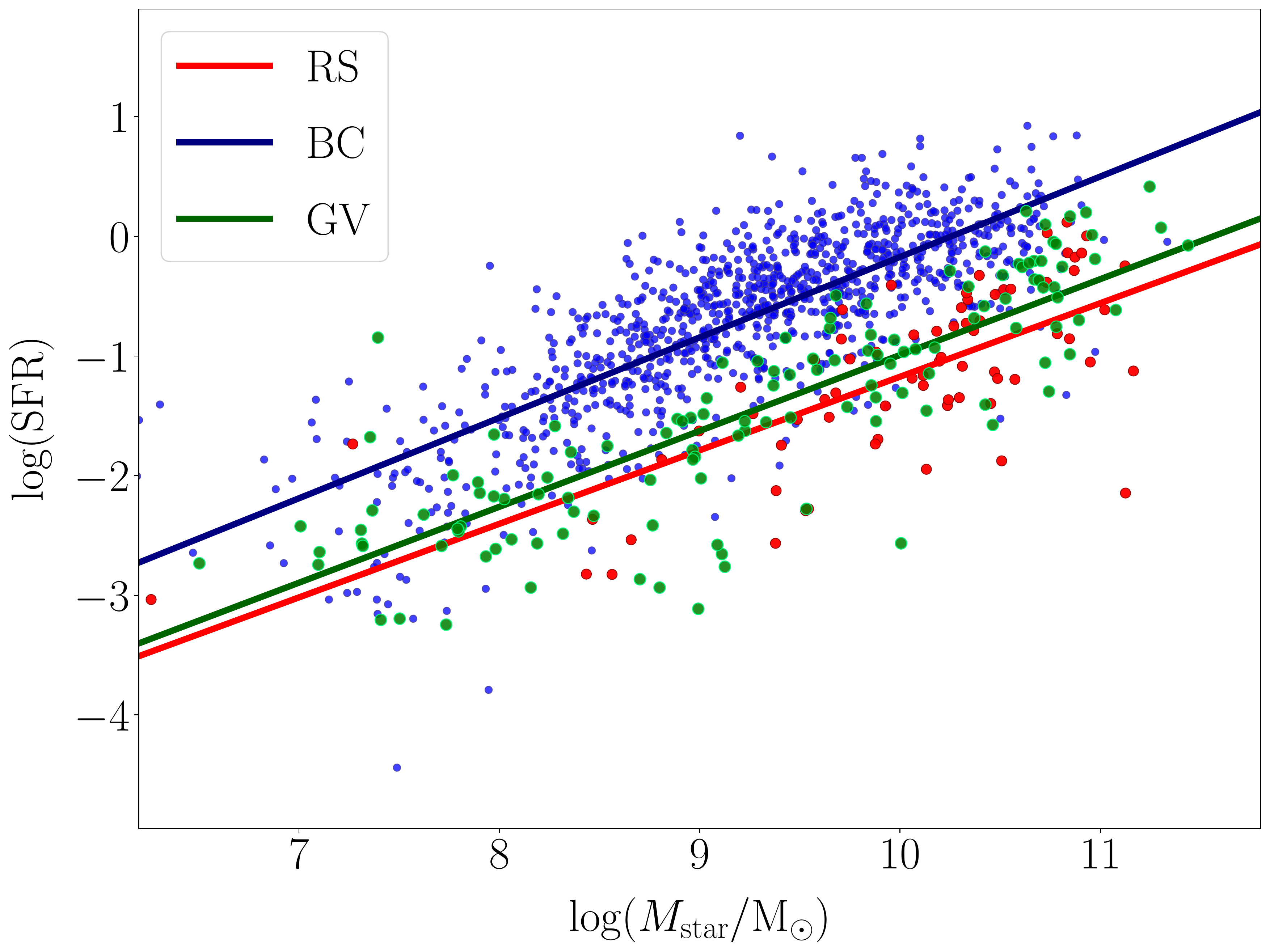}
\includegraphics[width=8cm]{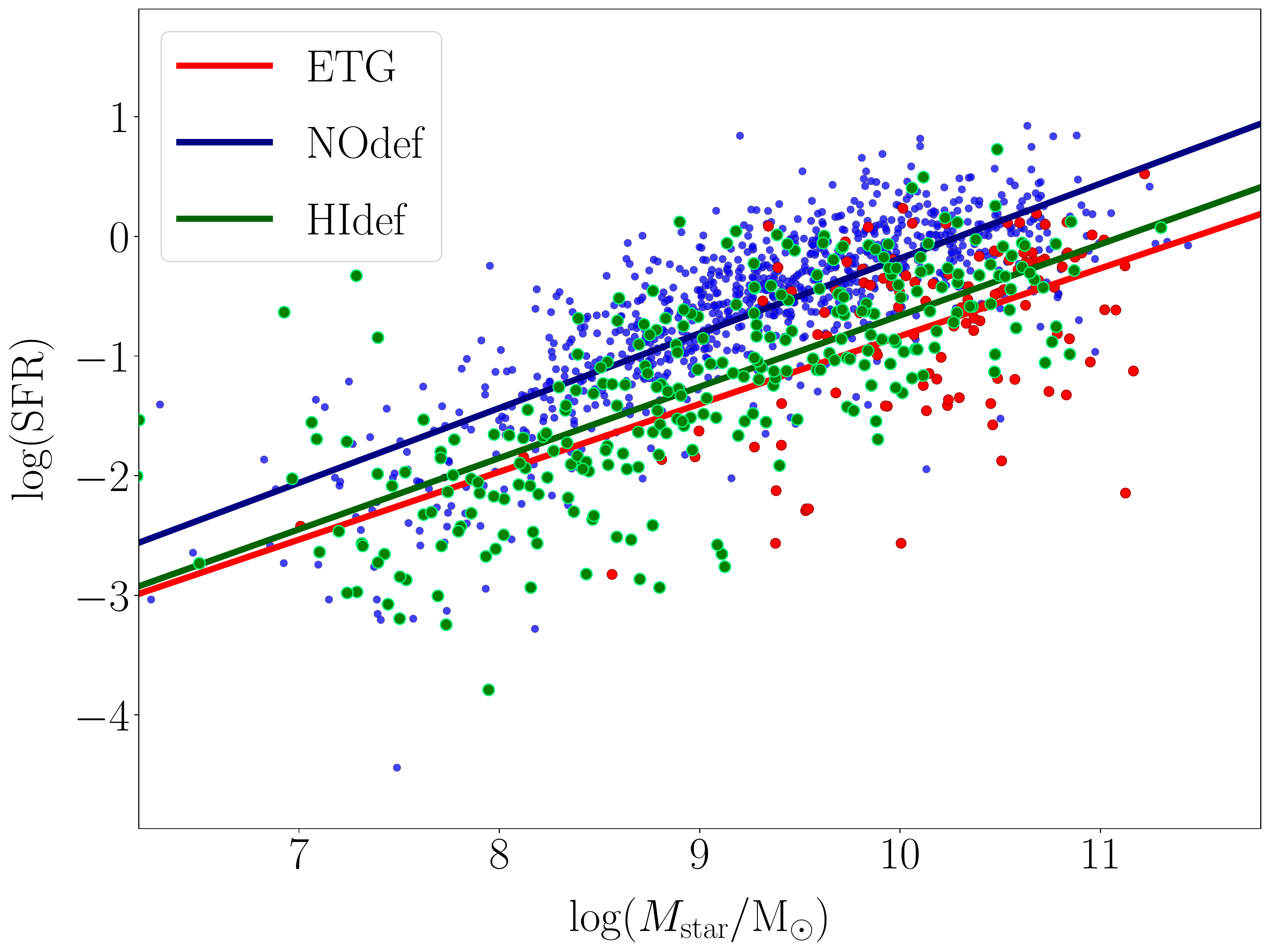}
\caption{Left: main sequence relation
of all galaxies in this sample divided according to the chromatic criterion of \cite{Boselli-2014}  (see Fig. 2 therein) whether they belong to the red sequence (RS), the green valley (GV) or the blue cloud (BC). For each class, the linear regression analysis is given.
Right: the same relation for the subsample of ETGs with detailed morphological classification available (red); of deficient (H\thinspace{\scriptsize I}$> 0.4$, green) and non-deficient galaxies (H\thinspace{\scriptsize I}$\leq 0.4$, blue). The slopes of the linear regressions for the ETGs and the H\thinspace{\scriptsize I}-deficiency classes are consistent with the ones corresponding to the three chromatic classes.}
\label{fig:SFRmass}
\end{center}
\end{figure*}

A third method for investigating environmental effects is studying the atomic Hydrogen (H\thinspace{\scriptsize I}) content of galaxies, because galaxies in clusters are H\thinspace{\scriptsize I}-deficient compared to galaxies in lower density environments.
H\thinspace{\scriptsize I}-observations provide powerful diagnostics of the role of the environment in regulating the evolution of late-type galaxies in the local Universe \cite[][]{Gavazzi-2012Ha3, Gavazzi-2013Ha3_a, Gavazzi-2013Ha3_b, Gavazzi-2013Ha3_c, Gavazzi-2015Ha3_a, Gavazzi-2015Ha3_b}.
In the present work, we employ ALFALFA \cite[AA,][]{Durbala-2020}, GOLDMine \cite[GM,][]{Gavazzi-2003}, the Westerbork Coma Survey \cite[WCS,][]{Molnar-2022}, and the HyperLeda catalogue \cite[LEDA,][]{Makarov-2014} H\thinspace{\scriptsize I} 21 cm line data to estimate the H\thinspace{\scriptsize I}-deficiency parameter $\mathrm{H} \thinspace \scriptsize{\text{I}}_{\mathrm{def}}$, defined as 
\cite[]{Haynes-1984, Giovanelli-1985, Gavazzi-2003}
\begin{equation}
    \mathrm{H} \thinspace
\scriptsize{\text{I}}_{\mathrm{def}} = \log M(\mathrm{H} \thinspace \scriptsize{\text{I}})_D - \log M(\mathrm{H} \thinspace \scriptsize{\text{I}})_{\mathrm{obs}},
\end{equation}
where  $\log M(\mathrm{H} \thinspace \scriptsize{\text{I}})_D$ is the H\thinspace{\scriptsize I}-mass computed as a function of the optical linear diameter $D$ \cite[][]{Meyer-2017} as
\begin{equation}
    \log M(\mathrm{H} \thinspace \scriptsize{\text{I}})_D = C_1 + C_2 \log(D^2).
\end{equation}
The coefficients $C_1$ and $C_2$ are empirically determined by studying a control sample of isolated objects \cite[see, \textit{e.g.},][]{Haynes-1984, Gavazzi-2013Ha3_a}. The SPRING catalogue includes 9461  galaxies with H\thinspace{\scriptsize I} data from GM, WCS, AA, and LEDA. When data from more than one catalogue are available, we choose GM data over WCS, AA, and LEDA, respectively. The diameters of this subset of galaxies are retrieved from NED. Since our HI data are taken from surveys with different limiting sensitivities, our HI-mass determination only includes detections and no upper limits.

As discussed in \cite{Cortese-2021}, there are different methods which can be employed to quantify the H\thinspace{\scriptsize I}-deficiency parameter.
We perform a recalibration of the optical diameter-based H\thinspace{\scriptsize I}-deficiency parameter employing a reference sample of isolated galaxies corresponding to the UL overdensity bin of our catalogue containing 11051 objects, among which 3416 have H\thinspace{\scriptsize I} data.
We find that the slope of the $\log(M(\mathrm{H} \thinspace \scriptsize{\text{I}}))$ vs $\log(D^2)$ relation is strongly dependent on the galaxy stellar-mass; in contrast, no significant discrepancy in the linear fits for different morphological types is observed. To further investigate the stellar-mass dependence, we divide the galaxies in our sample in six bins of stellar-mass (dubbed as M1,...,M6) defined as
\begin{itemize}
    \item \ \ M1: \ \ \ \ \ \ \ \ \ \ \  log($M_{\mathrm{star}}/\mathrm{M}_{\odot}) \leq 8.5$;
    \item \ \ M2: \ 8.5 $<$ log($M_{\mathrm{star}}/\mathrm{M}_{\odot}) \leq 9$;
    \item \ \ M3: \ \ \ \ 9 $<$ log($M_{\mathrm{star}}/\mathrm{M}_{\odot}) \leq 9.5$;
    \item \ \ M4: \ 9.5 $<$ log($M_{\mathrm{star}}/\mathrm{M}_{\odot}) \leq 10$;
    \item \ \ M5: \ \  10 $<$ log($M_{\mathrm{star}}/\mathrm{M}_{\odot}) \leq 10.5$;
    \item \ \ M6: \ \ \ \ \ \ \ \ \ \ \  log($M_{\mathrm{star}}/\mathrm{M}_{\odot}) > 10.5$.
\end{itemize}
For each stellar-mass bin, we study individually the $\log(M(\mathrm{H} \thinspace \scriptsize{\text{I}}))$ vs $\log(D^2)$ relation and find that the slope of M1 galaxies ($\log\left(M_{\mathrm{star}}/\mathrm{M}_{\odot}\right) \leq 8.5$) is significantly steeper than that of galaxies in larger stellar-mass bins (see Fig. \ref{fig:calib}). The values of the best-linear fit coefficients $C_1$ and $C_2$ in the two mass bins are reported in Table \ref{tab:coeff}.
The histogram in Fig. \ref{fig:HIdef_hist} displays the distribution of our parametrization for the H\thinspace{\scriptsize I}-deficiency parameter of the galaxies in our sample divided in overdensity bins. In Table \ref{tab:hist_params} we give the mean values and the standard deviations of Gaussian fits to the $\mathrm{H} \thinspace \scriptsize{\text{I}}_{\mathrm{def}}$ distribution in each overdensity bin. We observe that the mean value of $\mathrm{H} \thinspace \scriptsize{\text{I}}_{\mathrm{def}}$ increases as we move from isolated galaxies (UL bin, $\langle \mathrm{H} \thinspace \scriptsize{\text{I}}_{\mathrm{def}} \rangle \sim 0$) to galaxies in the centre of clusters (UH bin, $\langle \mathrm{H} \thinspace \scriptsize{\text{I}}_{\mathrm{def}} \rangle \sim 0.7$). 
\begin{figure*}
\begin{centering}
\begin{minipage}{.48\textwidth}
\includegraphics[width=\textwidth]{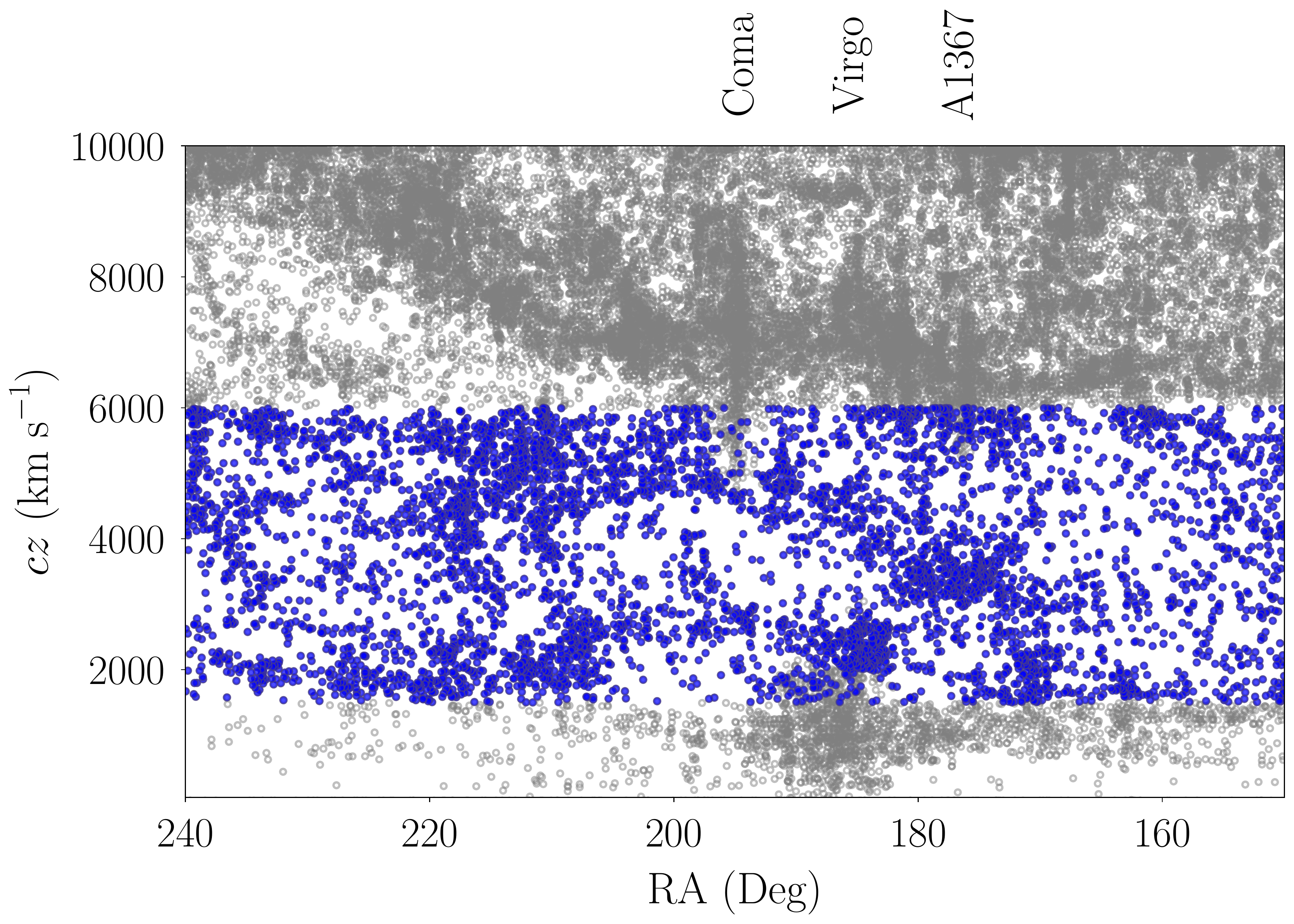}
\end{minipage}
\hspace{.5cm}
\begin{minipage}{.48\textwidth}
\includegraphics[width=\textwidth]{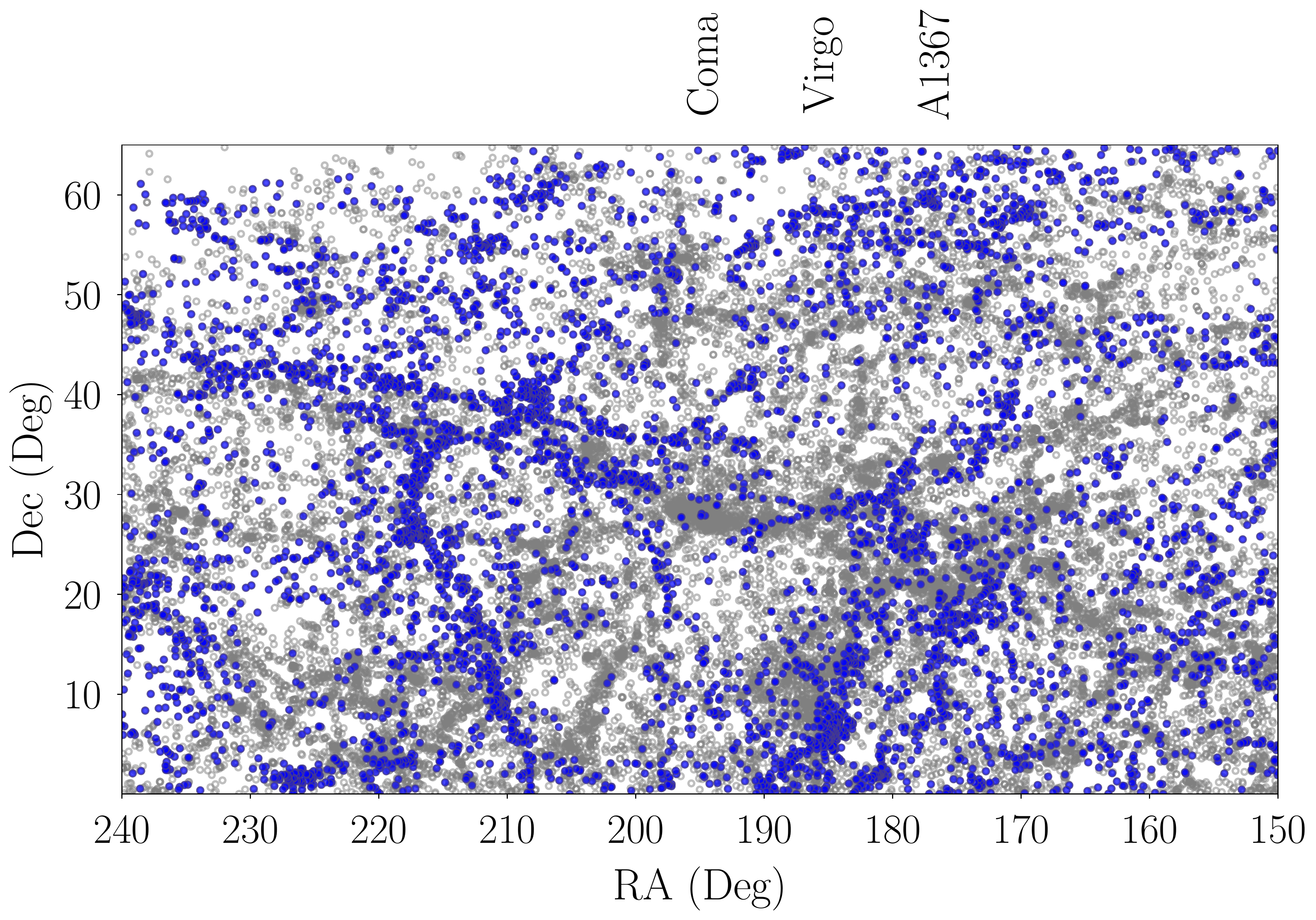}
\end{minipage}
\caption{Left: Distribution of the Hubble velocity $cz$ as a function of RA in the spring sky to illustrate the criterion discussed in Sec. \S \ref{sec:springRing} (1500 < $cz/\mathrm{km \ s}^{-1}$< 6000, excluding objects belonging to the fingers of God of either Coma or Virgo).  Right: sky distribution of galaxies in the 1500 < $cz/\mathrm{km \ s}^{-1}$< 6000 range displaying the existence of a filamentary structure that we dub the Spring Ring.}
\label{fig:ringdistro}
\end{centering}
\end{figure*}

\begin{figure}
\begin{centering}
\includegraphics[width=9cm]{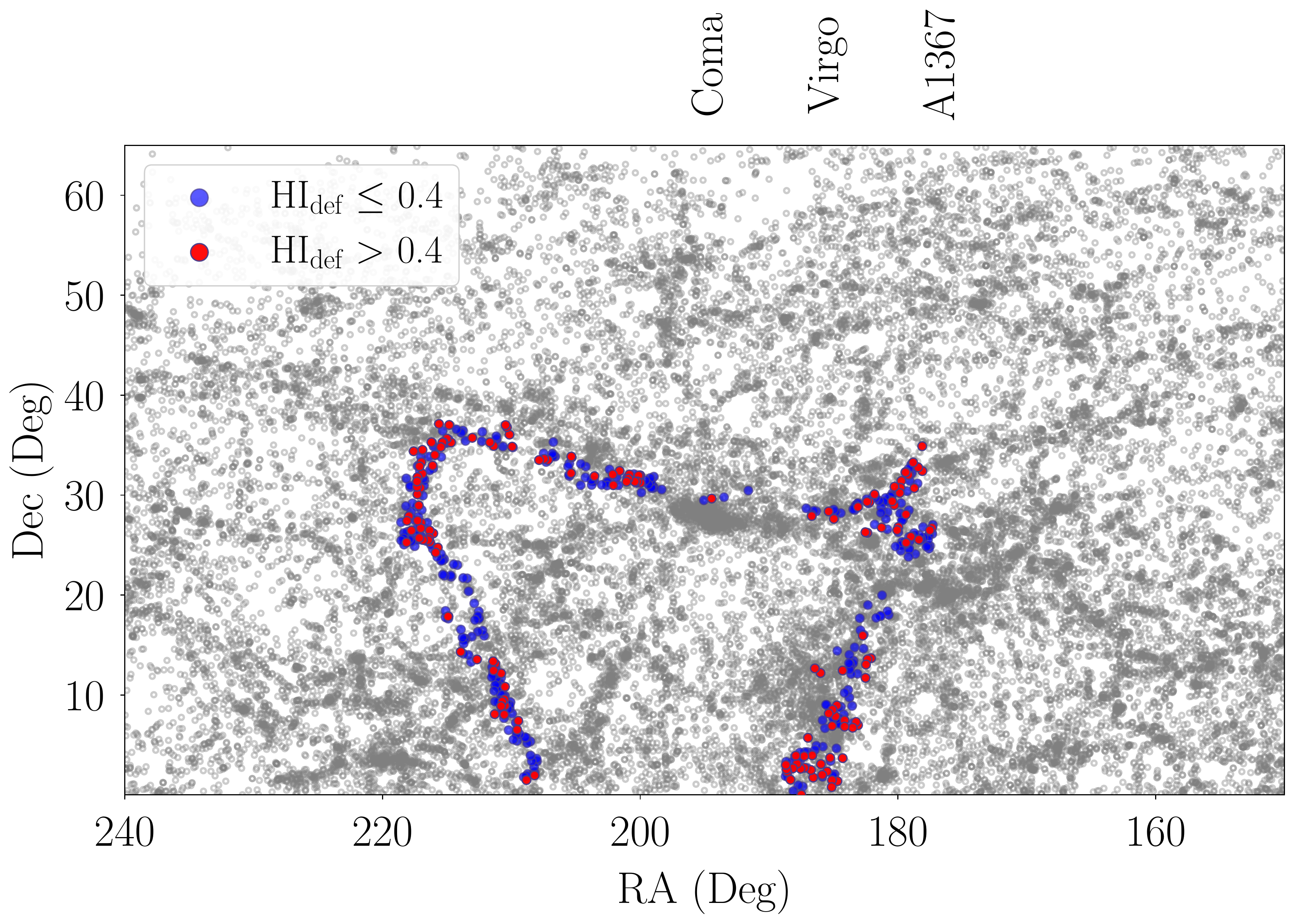}
\caption{The sky distribution of galaxies belonging to the Spring Ring in two bins of H\thinspace{\scriptsize I} deficiency parameter $\mathrm{H} \thinspace \scriptsize{\text{I}}_{\mathrm{def}} \leq 0.4$ (blue) and $\mathrm{H} \thinspace \scriptsize{\text{I}}_{\mathrm{def}} > 0.4$ (red).}
\label{fig:HIdef_ring}
\end{centering}
\end{figure}

\begin{figure}
       \begin{center}
       \includegraphics[width=9cm]{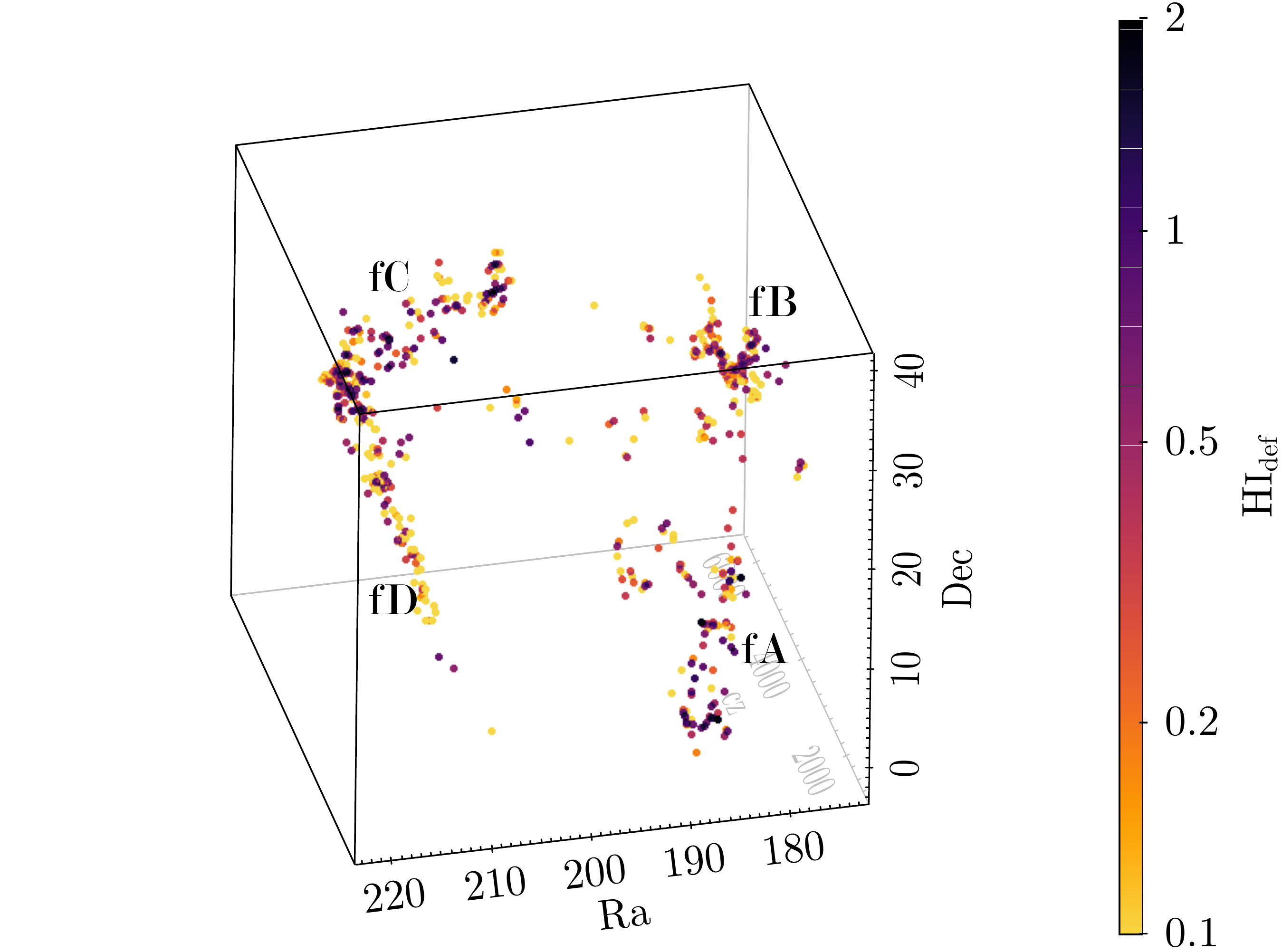}
       \end{center}
       \caption{The sky distribution of the 515 galaxies belonging to the Spring ring and having  H\thinspace{\scriptsize I} data. The third dimension is given by the Hubble velocity $cz$. The color map displays the $\mathrm{H} \thinspace \scriptsize{\text{I}}_{\mathrm{def}}$ parameter.}
       \label{fig:ring3d}
       \end{figure}

\begin{table}
\small
\centering
\caption{Coefficients of the $\log(M(\mathrm{H} \thinspace \scriptsize{\text{I}}))$ vs $\log(D^2)$ relation for the two stellar-mass bins M1 ($\log\left(M_{\mathrm{star}}/\mathrm{M}_{\odot}\right) \leq 8.5$) and M2-M6 ($\log\left(M_{\mathrm{star}}/\mathrm{M}_{\odot}\right) > 8.5$).}\label{tab:coeff}
 \begin{tabular}{lrcc}
\hline
\hline
  Stellar-mass &  &$C_1$ & $C_2$ \\
\hline
  $\log\left(M_{\mathrm{star}}/\mathrm{M}_{\odot}\right) \leq 8.5$ &  & 5.10   & 0.73 \\
\hline
  $\log\left(M_{\mathrm{star}}/\mathrm{M}_{\odot}\right) > 8.5$ &  & 6.98   & 0.42 \\
\hline
\hline
 \end{tabular}
\end{table}
We classify galaxies as deficient (non-deficient) if $\mathrm{H} \thinspace \scriptsize{\text{I}}_{\mathrm{def}}> 0.4$ ($\mathrm{H} \thinspace \scriptsize{\text{I}}_{\mathrm{def}}\leq 0.4$).
The $\mathrm{H} \thinspace \scriptsize{\text{I}}_{\mathrm{def}}$ parameter is a proxy for the influence of the environment on the neutral Hydrogen content of galaxies (in fact, it significantly correlates with our overdensity parameter $\Delta\rho/\langle\rho\rangle$).
\begin{table}
\small
\centering
\caption{Mean value and standard deviation of the Gaussian fits for the H\thinspace{\scriptsize I}-deficiency distributions in each overdensity bin.}\label{tab:hist_params}
 \begin{tabular}{lrcc}
\hline
\hline
  Overdensity &  & Mean & $\sigma$ \\
\hline
  UL &  & 0.00   & 0.26 \\
\hline
  L &  & 0.04   & 0.32 \\
\hline
  H &  & 0.24   & 0.41 \\
\hline
  UH &  & 0.64   & 0.52 \\
\hline
\hline
 \end{tabular}
\end{table}
Our sample contains three rich clusters (Coma, A1367 and Virgo, see Figure \ref{fig:density}), thus it includes a significant number of H\thinspace{\scriptsize I}-deficient objects (see Figure \ref{fig:HIdef}). Since the deficiency parameter depends on the distance, AA detects galaxies with $\mathrm{H} \thinspace \scriptsize{\text{I}}_{\mathrm{def}}\gtrsim 0.3$ in the Local Supercluster, but it only manages to detect galaxies with normal H\thinspace{\scriptsize I} content at the distance of Coma. The detection of deficient galaxies in Coma and A1367 requires more sensitive observations than those by AA, such as those by GM or WCS.
\begin{figure}
\begin{centering}
\includegraphics[width=8cm]{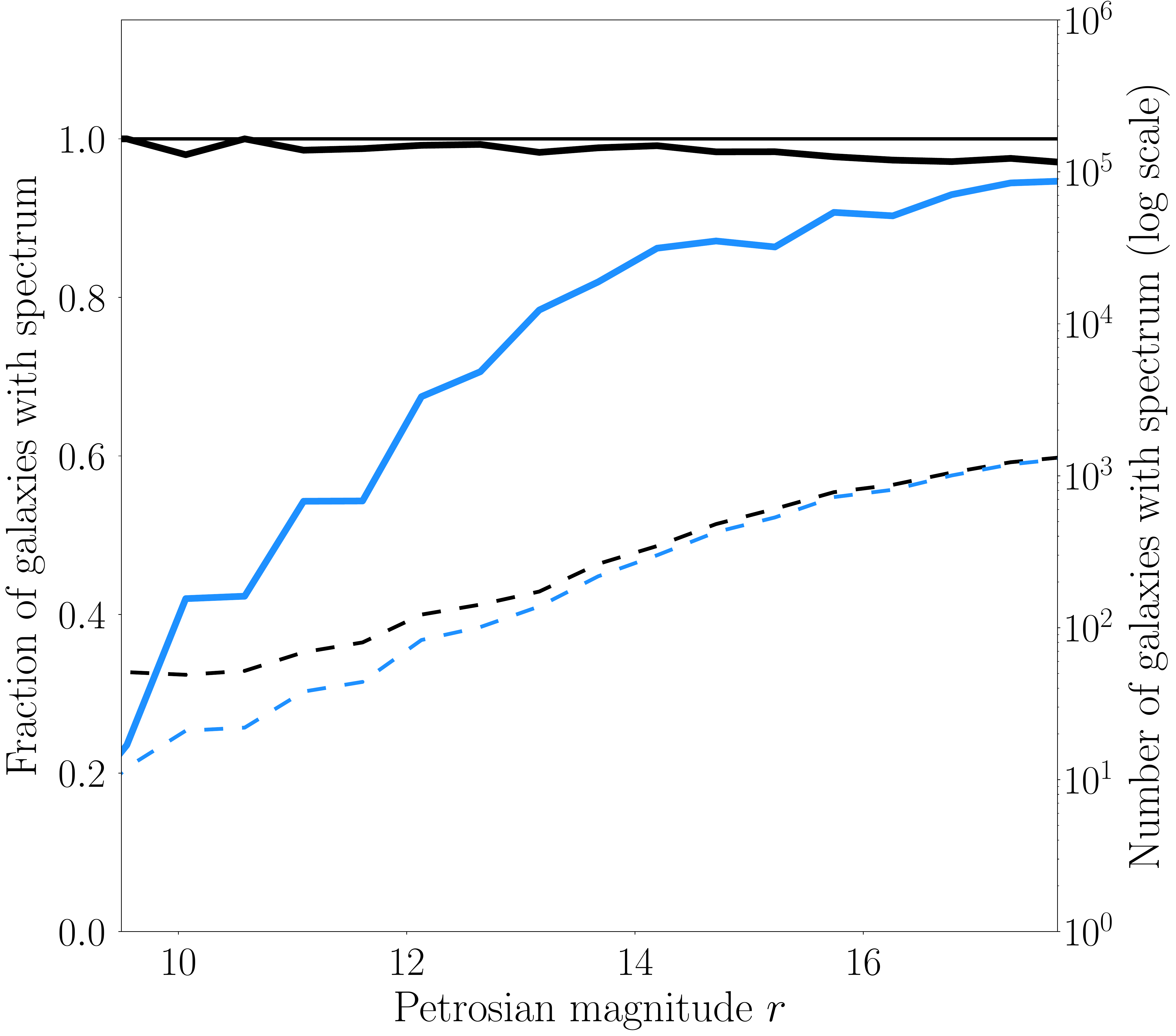}
\includegraphics[width=8cm]{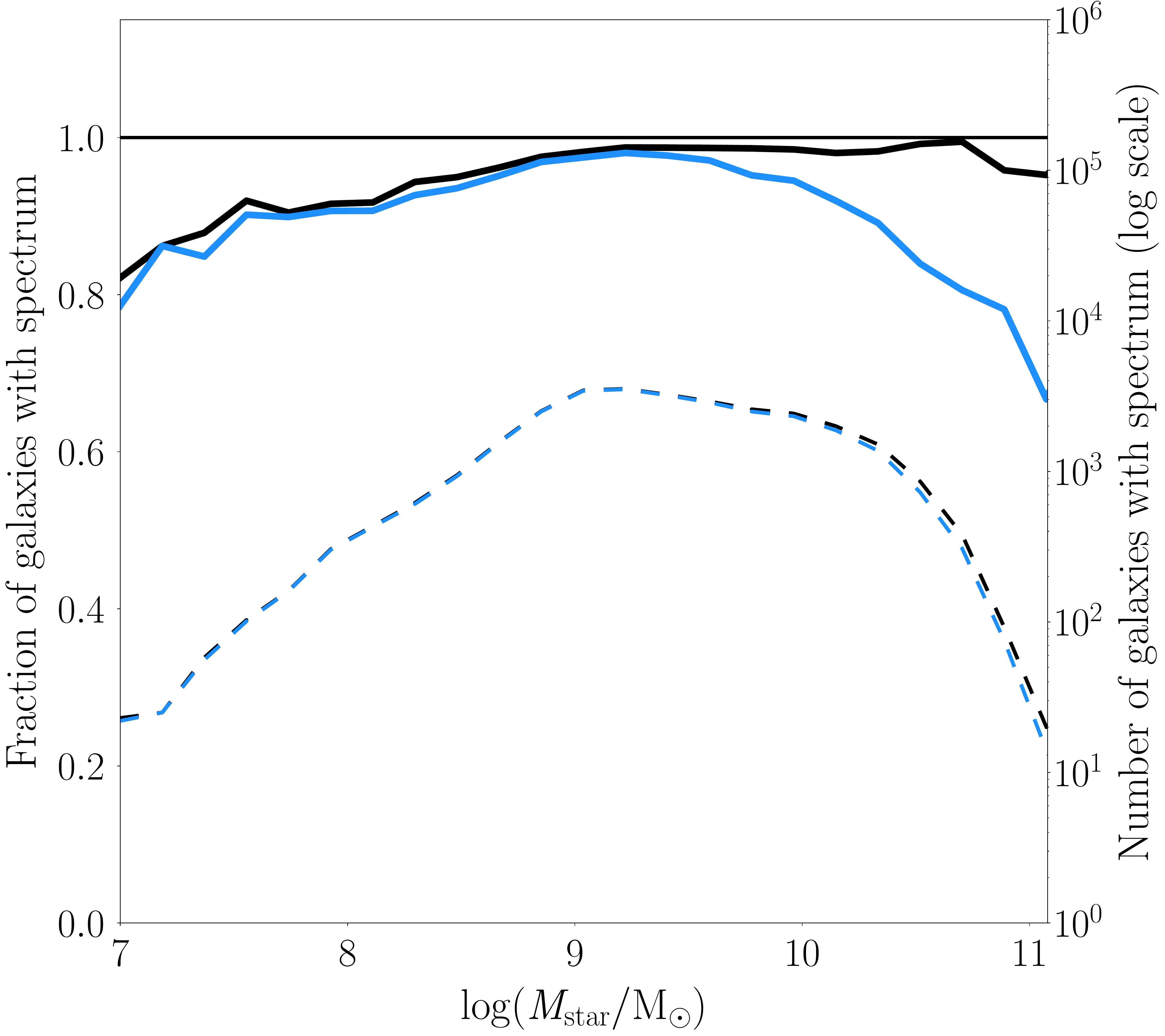}
\caption{Completeness (defined as the fraction of galaxies with nuclear spectra over the total number of galaxies, left axis and thick straight lines) and actual number of galaxies with nuclear spectra (right axis and dashed lines) as a function of $r$ apparent magnitude (top) and as a function of stellar-mass (bottom).
The blue lines represent SDSS-only spectra, the black lines include spectra taken from the SDSS and other sources.}
\label{Fig:comple}
\end{centering}
\end{figure}
Using different symbols (and colours) for the non-deficient galaxies
($\mathrm{H} \thinspace \scriptsize{\text{I}}_{\mathrm{def}}\leq 0.4$, blue), for the deficient galaxies ($\mathrm{H} \thinspace \scriptsize{\text{I}}_{\mathrm{def}}>0.4$, green) and for early-type galaxies whose detailed morphological information is available from \cite{Gavazzi-2003} (red), we show the dependence of the $\mathrm{NUV}-i$ colour on the stellar-mass (see Fig. \ref{fig:colormass}, right panel). 

Along with the subsets of galaxies coded by H\thinspace{\scriptsize I}-content, we chose to display ETGs to highlight the correspondence between the three classes (ETGs, HIdef, non-HIdef) and the chromatic classification (red sequence, green valley, blue cloud).

We observe that our chromatic classification is closely related to the content of H\thinspace{\scriptsize I}, as the slopes of the H\thinspace{\scriptsize I}-deficient and H\thinspace{\scriptsize I}-normal galaxies in the color-mass diagram (Fig. \ref{fig:colormass}, right) parallel those of the three colour classes representation (Fig. \ref{fig:colormass}, left) which was chosen by \cite{Boselli-2014} to study the Virgo cluster. The similarity between the classes of H\thinspace{\scriptsize I}-content and the BC and GV chromatic classes is visible also in the main sequence (\textit{i.e.}, SFR vs stellar-mass) relation. We cross-match our catalogue with the $H\alpha 3$ catalogue \cite[][1396 galaxies in common]{Gavazzi-2012Ha3}, from which we extract the star formation rate which is derived from observed, integrated $H\alpha$ flux after applying correction for (\textit{i}) Galactic extinction \cite[assuming the extinction law by][]{Cardelli-1989}, (\textit{ii}) [NII] contamination in the narrow-band imaging filter, and (\textit{iii}) internal extinction, this last estimated as described in \cite{Gavazzi-2012Ha3} \cite[see also][]{Gavazzi-2013Ha3_c, Gavazzi-2015Ha3_a, Gavazzi-2015Ha3_b}.
Figure \ref{fig:SFRmass} shows the main sequence relation for this subset of galaxies coded according to their H\thinspace{\scriptsize I}-content (right) and their chromatic classification (left). We observe that H\thinspace{\scriptsize I}-deficient galaxies display lower SFR than H\thinspace{\scriptsize I}-normal objects of the same stellar-mass, suggesting that the lack of H\thinspace{\scriptsize I} gas leads to a reduction in the activity of star formation \cite[][]{Boselli-2022a}.

\subsection{The Spring Ring}\label{sec:springRing}
As shown in Fig. \ref{fig:HIdef}, deficient objects are mostly distributed in clusters. The fraction of objects with $\mathrm{H} \thinspace \scriptsize{\text{I}}_{\mathrm{def}} > 0.4$ in the three clusters Virgo, Coma, and A1367 is $\sim62.4\%$. In addition, we observe that a consistent fraction of deficient galaxies is distributed in filamentary structures. To further investigate the significance of this finding, we select galaxies at intermediate distances (velocities) between Virgo and Coma (1500 < $cz/\mathrm{km \ s}^{-1}$< 6000), excluding objects belonging to the fingers of God of either Coma or Virgo (see Fig. \ref{fig:ringdistro}, left panel). A total of 6095 objects are selected this way. In addition, we remove from this selection 194 objects belonging to the highest (UH) overdensity bin.
The celestial distribution of 5911 galaxies selected with this criterion is given in Fig. \ref{fig:ringdistro} (right panel), showing the existence of a definite filamentary structure which we dub the ``Spring Ring'' (SR).
We manually isolate galaxies in the ring to study their H\thinspace{\scriptsize I}-content. Two- and three-dimensional sky distributions of this structure are given in Figs. \ref{fig:HIdef_ring}-\ref{fig:ring3d}.
We identify four main branches within the SR, denoted with fA, fB, fC, and fD in the figures. The fA branch displays evidence of gravitational influence from the Virgo cluster; fB is made by galaxies which show hints of gravitational attraction from A1367; fC comprises galaxies infalling toward the Coma cluster; fD does not exhibit signs of influence from one of the three clusters considered in the present work.

The Spring Ring is mostly composed of blue-cloud galaxies ($\sim80\%$) with stellar-mass $M_{\rm star} \lesssim 10^{10}$ M$_{\odot}$. Remarkably, when we divide galaxies with H\thinspace{\scriptsize I} data belonging to the SR (515 entries) among H\thinspace{\scriptsize I}-deficient and H\thinspace{\scriptsize I}-normal objects, we find that 158 ($\sim$30.7\%) of them are H\thinspace{\scriptsize I}-deficient (see Figs. \ref{fig:HIdef_ring}-\ref{fig:ring3d}). Conversely, the fraction of H\thinspace{\scriptsize I}-deficient objects among field galaxies (\textit{i.e.}, SPRING galaxies that do not belong to the UH overdensity bin) is $\sim$12\%. We cross-matched these 515 SR galaxies with the recent catalogue by \cite{Castignani-2022b}. We find that 143 galaxies in the SR are reported also in \cite{Castignani-2022b}. Among these, the most noticeable fractions are found among galaxies belonging to the W-M sheet (57 entries), the NGC5353\_4 filament (41), and the Coma Berenices filament (29).
It is noticeable that the majority of the left branch of the SR is not identified in \cite{Castignani-2022b}.

To accurately compare the fraction of H\thinspace{\scriptsize I}-deficient objects in the filament to the general population, we randomly select a sample of 515 objects from the galaxies in the SPRING sample which have H\thinspace{\scriptsize I} data, and count the fraction of H\thinspace{\scriptsize I}-deficient objects. We repeat this procedure ten thousand times, and find that the fraction of $\mathrm{H} \thinspace \scriptsize{\text{I}}_{\mathrm{def}}$ galaxies is (16.7 $\pm$ 1.6)\%.

In addition, we investigate whether the fraction of H\thinspace{\scriptsize I}-deficient objects in the SR is related to the distribution of galaxy overdensity within the filamentary structure; to this end, we randomly select ten thousand overdensity-matched samples of 515 galaxies with H\thinspace{\scriptsize I} data (\textit{i.e.}, the overdensity parameter distribution of each random sample matches the overdensity distribution within the SR). We find that (20.2 $\pm$ 1.6)\% galaxies in the matched-samples are H\thinspace{\scriptsize I}-deficient. This value is larger than the fraction found in the random samples with no restriction on the overdensity distribution (20.2\% vs 16.7\%). Still, it is considerably smaller than the fraction of deficient objects found within the SR (30.7\%).
This result suggests that \textit{(i)} local galaxies in filamentary structures are more H\thinspace{\scriptsize I}-deficient than field galaxies, and \textit{(ii)} their H\thinspace{\scriptsize I}-content does not depend (only) on their local galaxy overdensity.

In the literature, there is controversial evidence regarding the H\thinspace{\scriptsize I}-content of filament galaxies. \cite{Lee-2021} claims that the H\thinspace{\scriptsize I}-content of galaxies in filaments is normal, whereas \cite{Castignani-2022a} finds a higher incidence of H\thinspace{\scriptsize I}-deficient galaxies in filaments \cite[see, \textit{e.g.}, Fig. 17 in][]{Castignani-2022a}.
Since galaxies in the SR are mostly low-mass, they have low binding energies and thus can be easily affected by perturbations as they move along the filament. This scenario is consistent with \cite{Odekon-2018}, who examined a sample of $\sim$4000 LTGs with H\thinspace{\scriptsize I} data and found that galaxy H\thinspace{\scriptsize I}-content decreases with distance from a filament spine. Multiple mechanisms (\textit{e.g.}, gas-stripping, tidal interaction) may affect filament galaxies before they enter higher-density regions; in Fig. \ref{fig:ring3d}, we see that most SR galaxies at the early stage of gas stripping (yellow dots) are found in the fD branch, while galaxies with higher H\thinspace{\scriptsize I}-deficiency (purple dots) mostly appear to be falling towards the core of a cluster. In addition, a high fraction of deficient objects is found in the proximity of the intersection between the fD and fC branches.

\section{Nuclear Spectral classification}\label{sec:results}
\begin{figure*}
\begin{center} 
\includegraphics[width=0.30\textwidth]{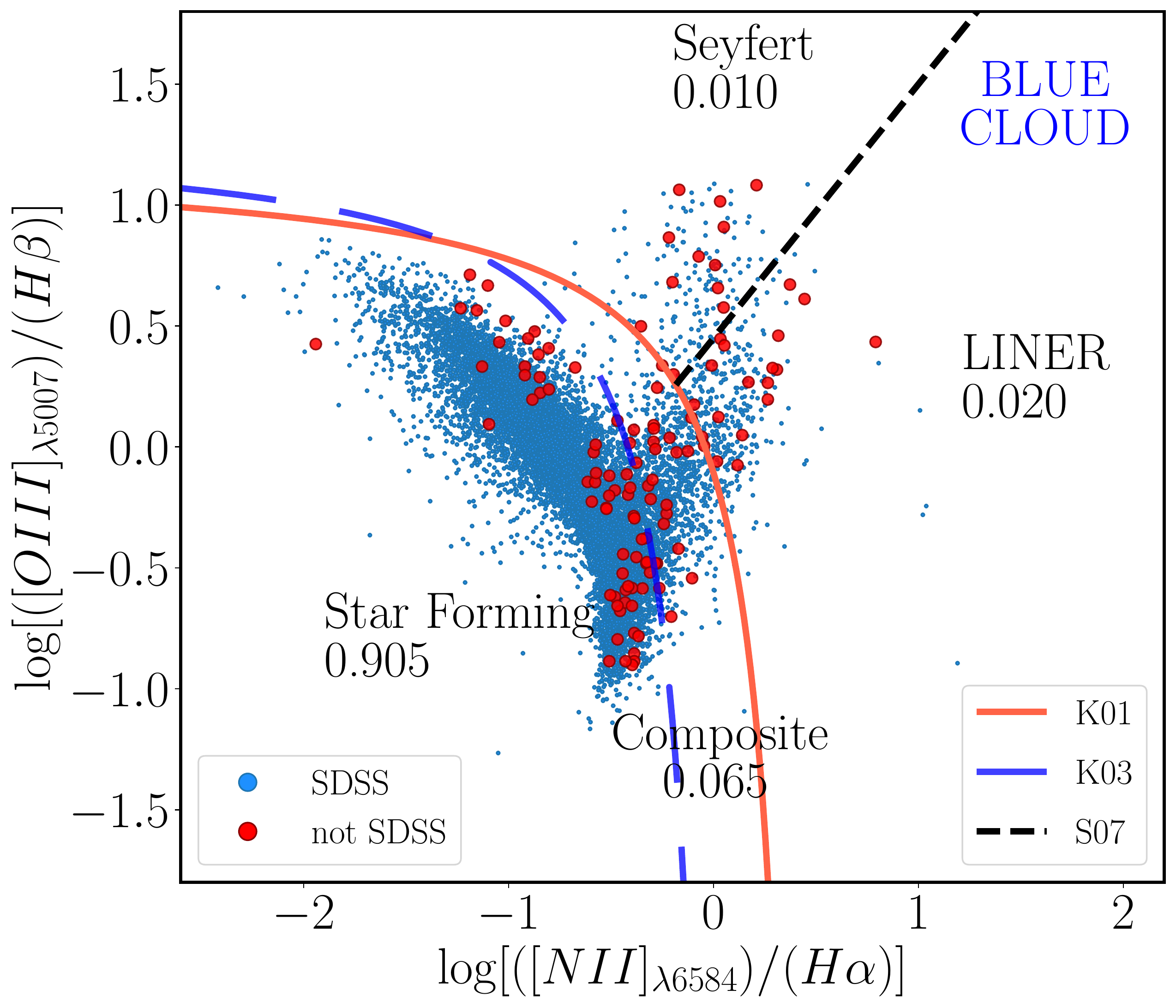}
\includegraphics[width=0.30\textwidth]{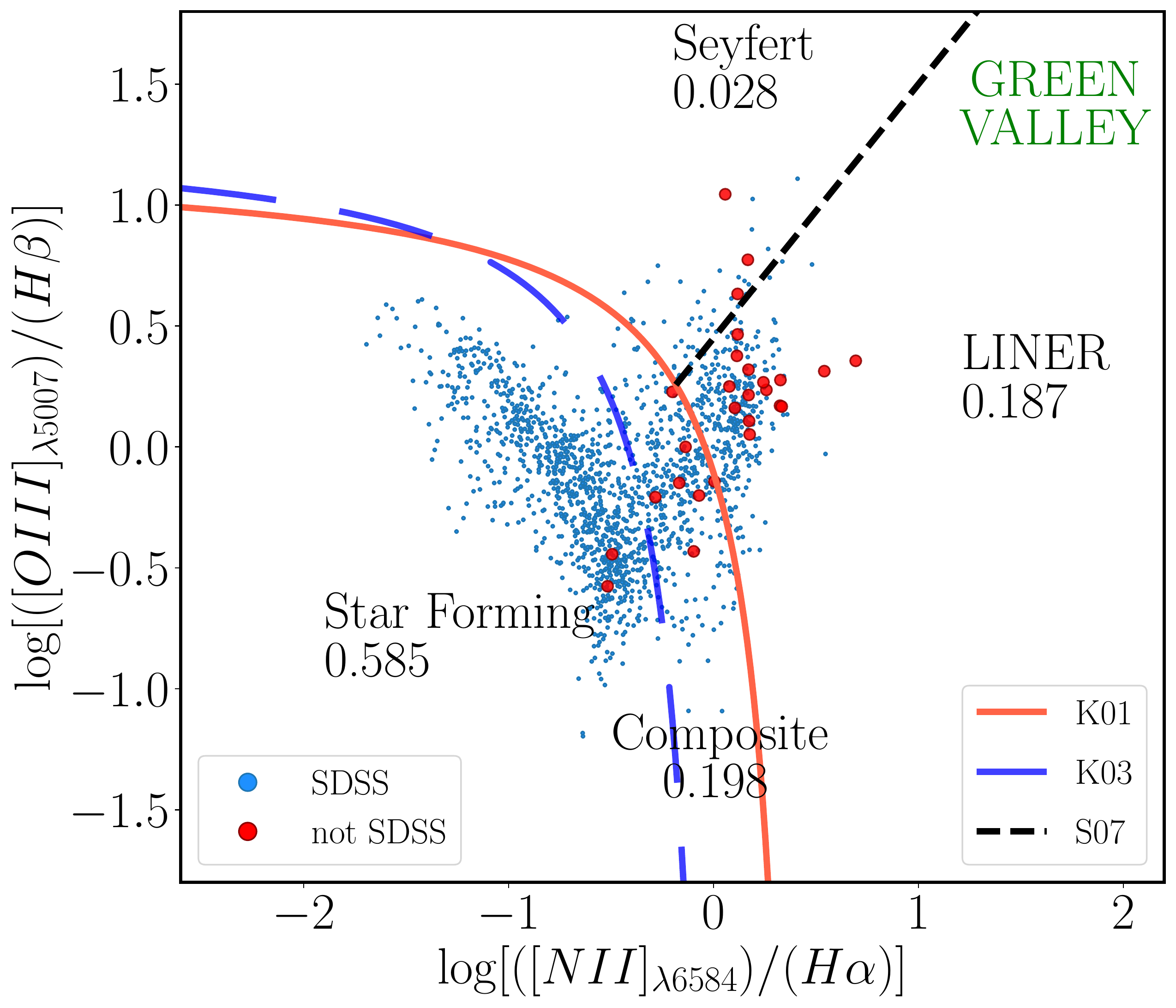}
\includegraphics[width=0.30\textwidth]{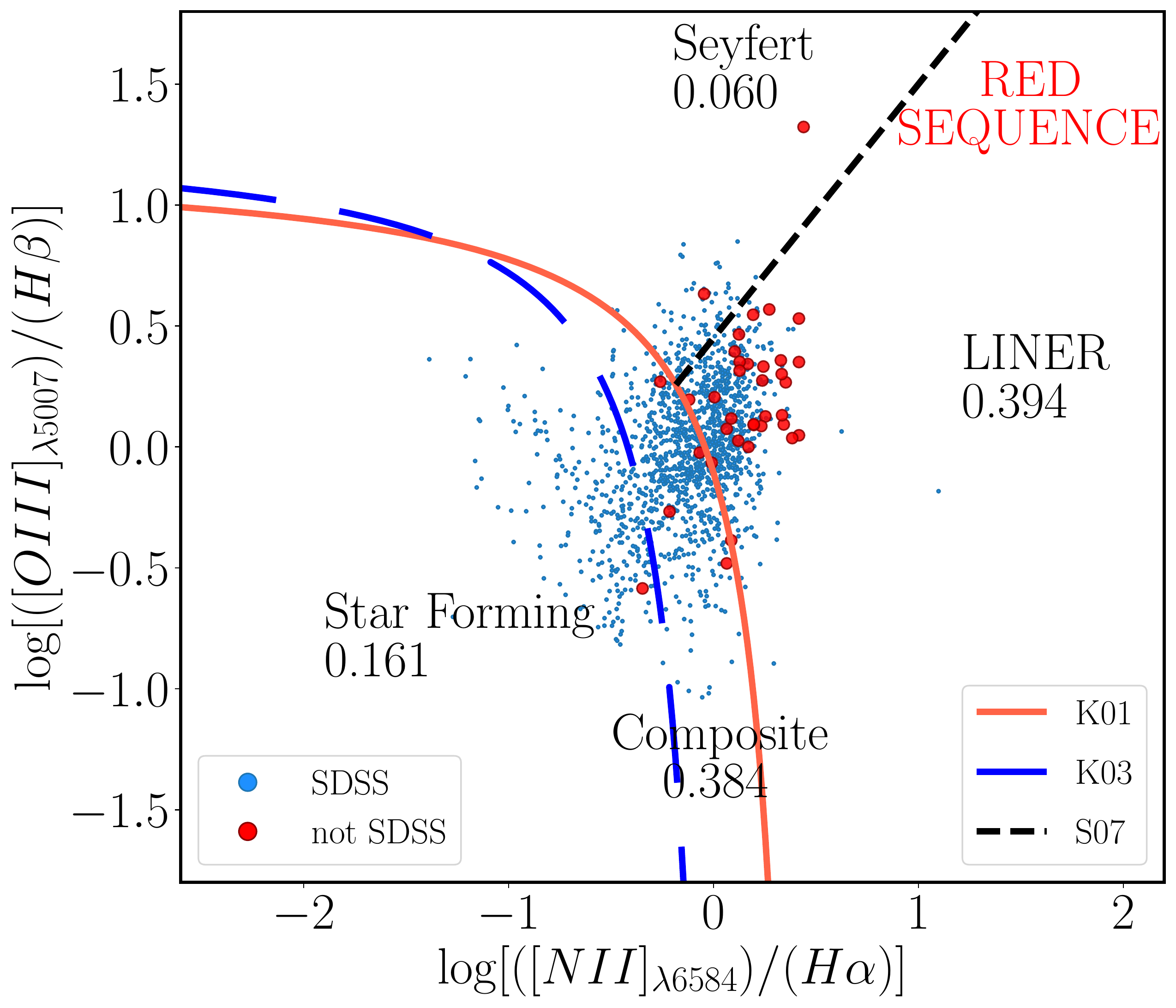}\\
\includegraphics[width=0.30\textwidth]{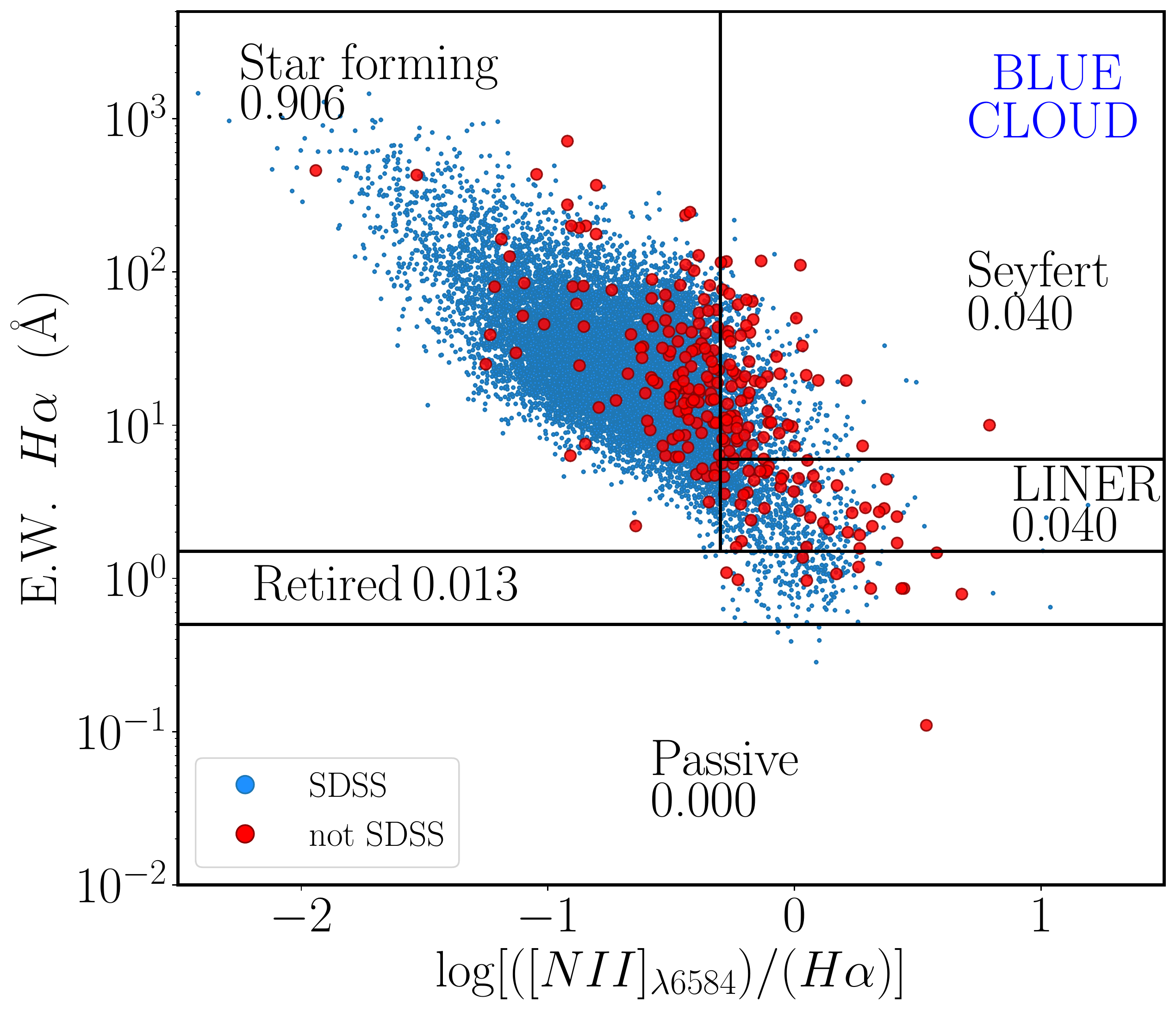}
\includegraphics[width=0.30\textwidth]{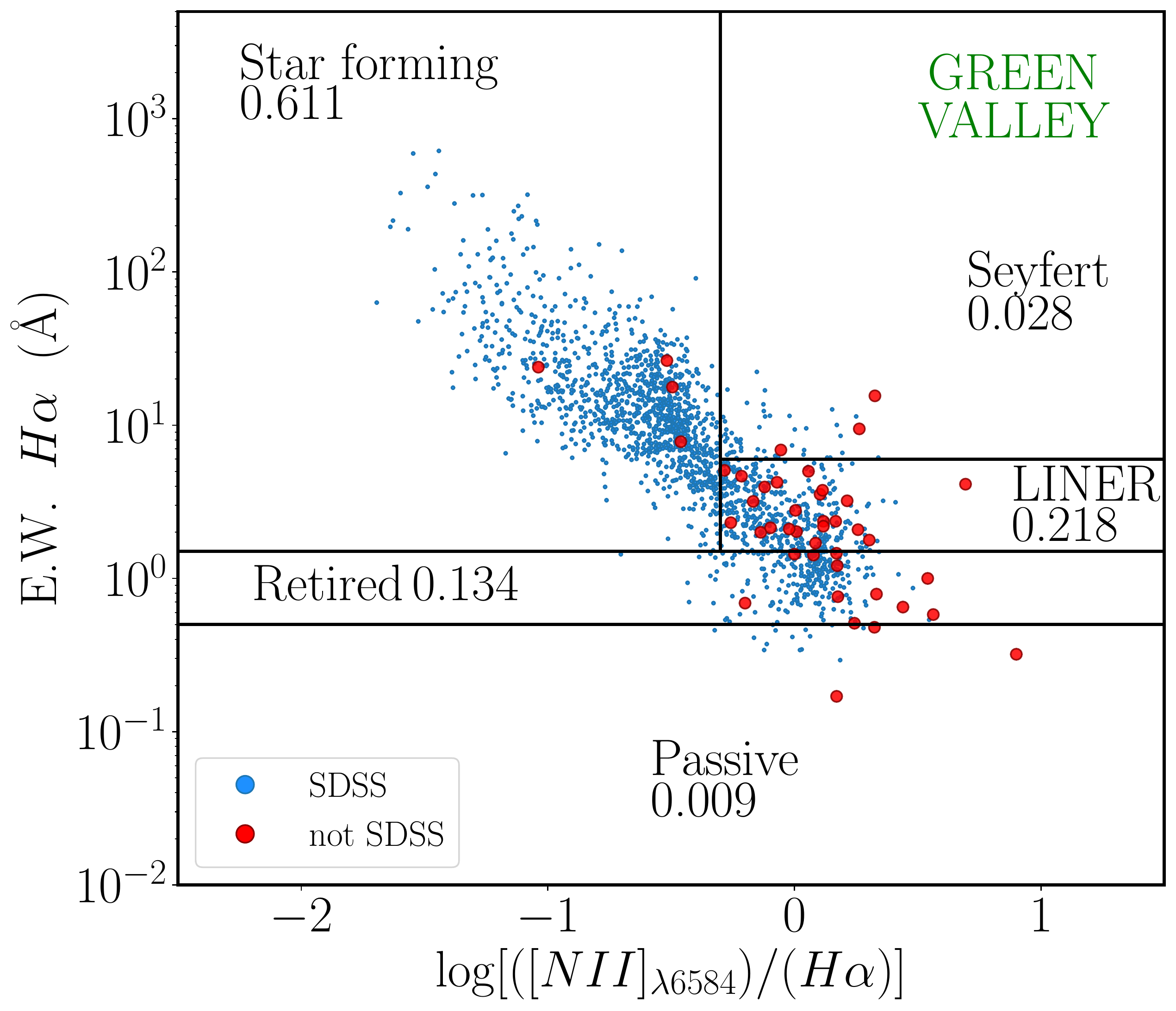}
\includegraphics[width=0.30\textwidth]{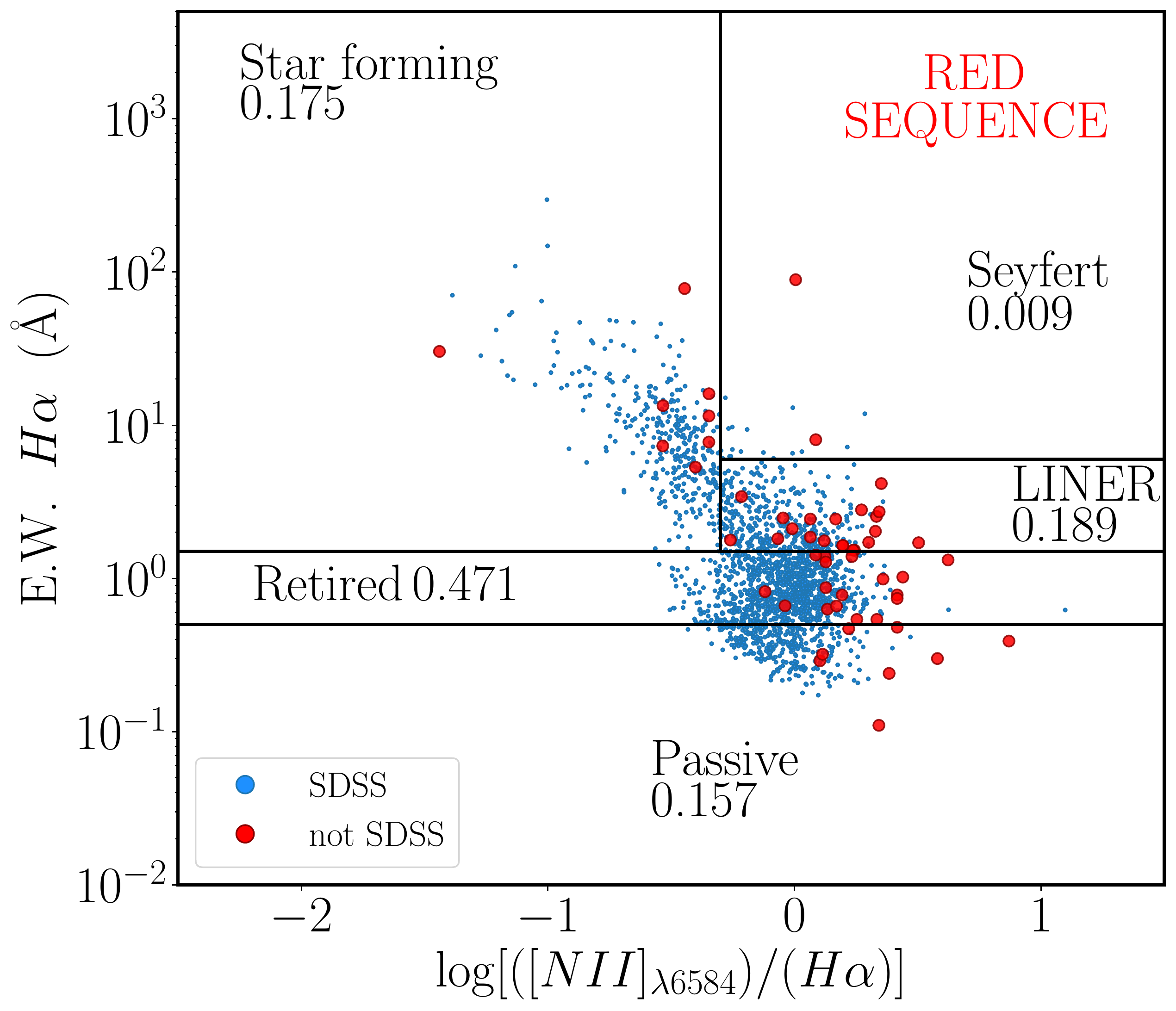}\\
\end{center}
\caption{(Top row): The BPT diagnostic diagram for galactic nuclei in the blue cloud, green valley and red sequence (left, centre and right respectively). The SDSS galaxies are given with blue points. Measurements from other sources are given with larger red dots. For each spectral category is given the associated percentage of galaxies.
The red straight line represents the model of \cite{Kewley-2001} to identify AGN. The blue long-dashed line gives the model of \cite{Kauffmann-2000} above which AGN are found. Composite nuclei fall between these two models. The black dashed line (\cite{Schawinski-2007}) separates Seyfert from LINERs.
(Bottom row): The WHAN diagnostic diagram for galactic nuclei in the blue cloud, green valley and red sequence (left, centre and right respectively). The SDSS galaxies are given with blue points. Measurements from other sources are given with larger red dots. For each spectral category is given the associated percentage of galaxies.}
\label{allmass-BPT-WHAN}
\end{figure*}

In this Section, we aim to link the photometric and environmental properties of the galaxies in the SPRING sample with the spectral features of their nuclei. 
We provide nuclear-emission classification by optical spectra based on two ionisation diagnostic diagrams: \textit{(i)} the Baldwin-Phillips-Terlevich diagnostic \cite[BPT diagram,][]{Baldwin-1981}, and the EWH$\alpha$ vs. [NII]/H$\alpha$ diagram \cite[WHAN,][]{Cid_Fernandez-2010, Cid_Fernandez-2011}. These diagnostics aim to identify the main ionizing mechanisms in galactic nuclei (\textit{e.g.}, young/old stars, AGN).
Out of the total of 30597 entries, we restrict our classification of nuclear activity to spectra having all lines with signal to noise ratio $>3$: we classify 22564 (74\%) galaxies of the SPRING catalogue using the BPT, and 24559 galaxies (80\%) using the WHAN. 
To investigate the impact of the inclusion of non-SDSS spectra to our sample, we define the spectral completeness as the fraction of galaxies with nuclear spectra over the total number of galaxies. In Fig. \ref{Fig:comple}, we show the completeness as a function of the $r$-band magnitude (top panel) and of the stellar-mass  $\log(M_{\mathrm{star}}/\mathrm{M}_{\odot})$ (bottom panel). The blue lines represent SDSS-only spectra, the black lines include spectra taken from the SDSS and other sources. Along with the completeness of the SPRING catalogue compared to SDSS-only data (thick straight lines, left $y$-axis), the actual numbers of galaxies with nuclear spectra for both samples is displayed as well (dashed lines, right $y$-axis). 

The addition of non-SDSS spectra to our flux- and volume-limited sample dramatically increases the completeness at the high-mass end of the catalogue (or, similarly, at the bright end), which is known to suffer a residual incompleteness in the SDSS database.
Only 361 targets ($\sim$1\%) in the SPRING catalogue remain without nuclear spectra\footnote{We point out that all galaxies in our sample have spectroscopic redshift, also those with no nuclear spectrum, for which the values of $z_{\mathrm{spec}}$ were taken from NED}.
\subsection{BPT Diagnostic}\label{sec:bpt}
The BPT diagnostic relies on four optical emission-line intensity ratios ([OIII]$\lambdaup$5007/H$\beta$ vs. [NII]$\lambdaup$6584/H$\alpha$) to classify galaxy nuclear spectra into four independent classes of nuclear activity:
\begin{itemize}
    \item Star-Forming nuclei (SF) are characterised by intense star-formation activity;
    \item Seyfert (SEY) or strong AGN contain a powerful galactic nucleus, likely triggered by a supermassive black hole;
    \item LINERs (Low-Ionisation Nuclear Emitting Regions, LIN) or weak AGN display a milder nuclear activity than SEY, possibly triggered by a mixture of stellar and true AGN processes \cite[][]{Cid-Fernandes-2004};
    \item Composite regions (star-formation + AGN) are transition objects with spectra intermediate between ``pure'' SF and ``pure'' SEY/LIN.
\end{itemize}
\begin{figure*}[ht!]
\begin{center}
\includegraphics[width=8cm]{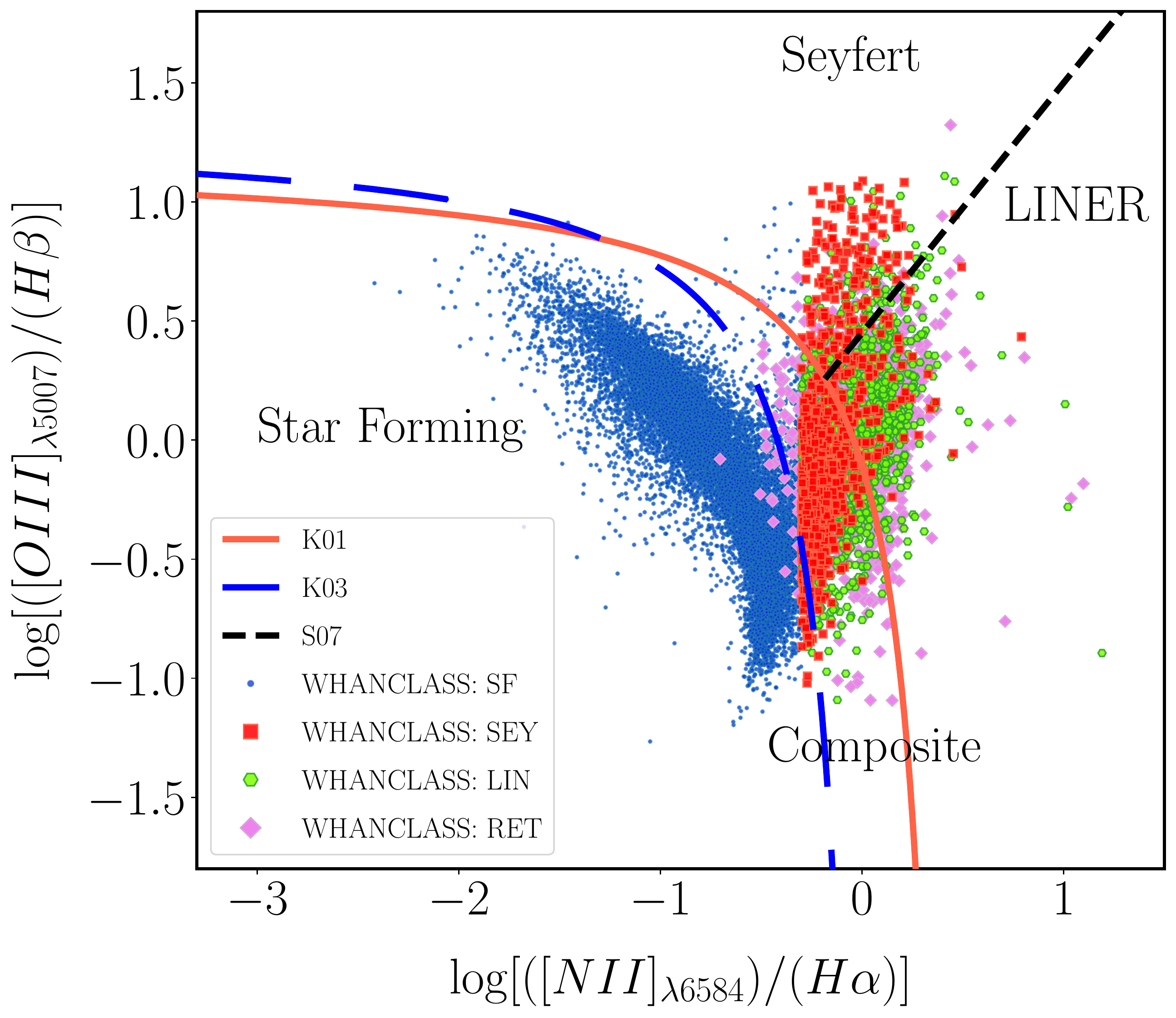}
\includegraphics[width=8cm]{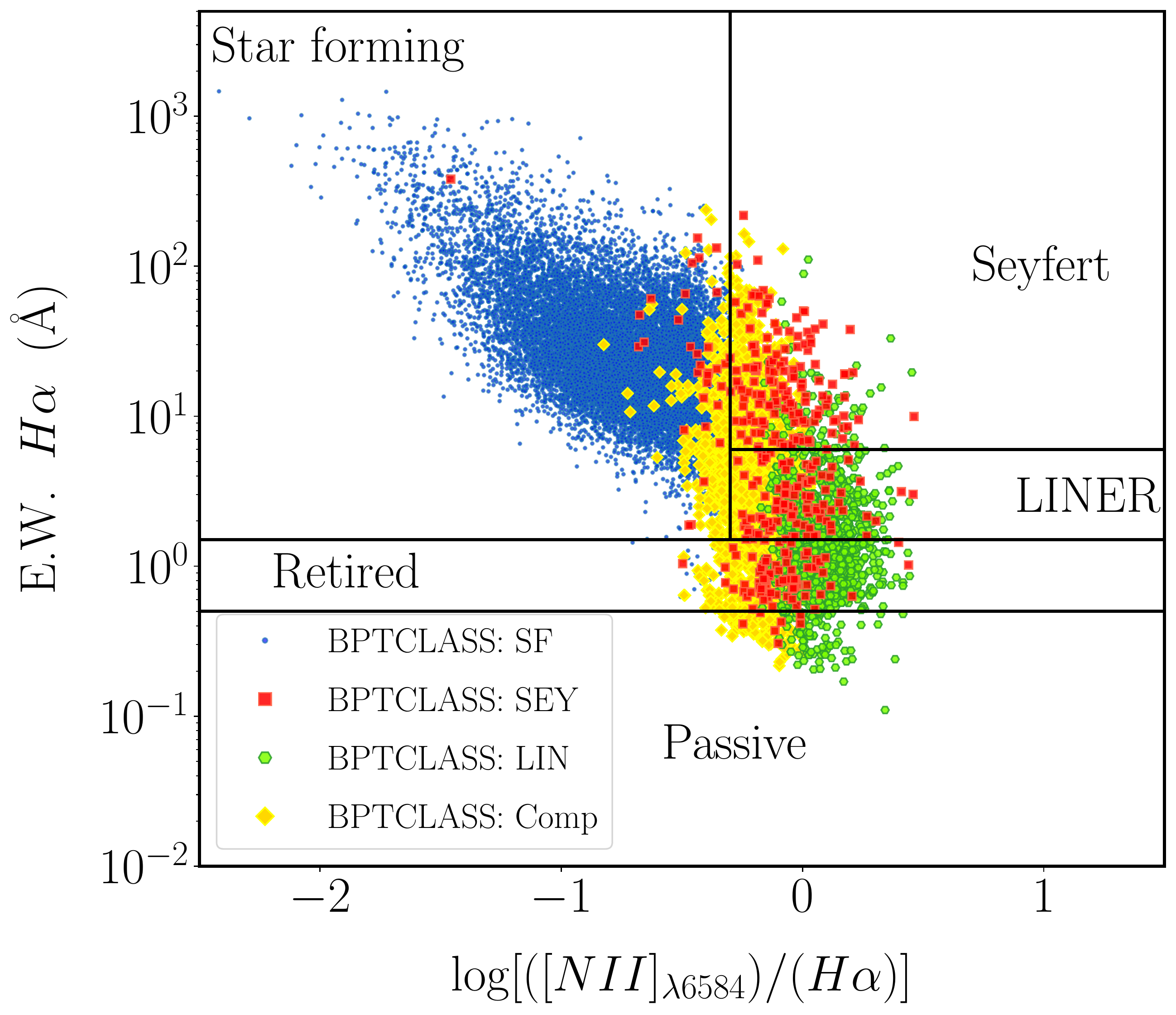}
\end{center}
\caption{Left: BPT diagram of galaxies grouped according to their WHAN classification (blue=SF, red==SEY, green= LIN, violet=RET).
Right: WHAN diagram of galaxies grouped according to their BPT classification (blue=SF, red==SEY, green= LIN, yellow=Comp).
}
\label{fig:BPT_WHAN_mix}
\end{figure*}
\begin{figure*}
       \begin{center}
       \includegraphics[width=18cm]{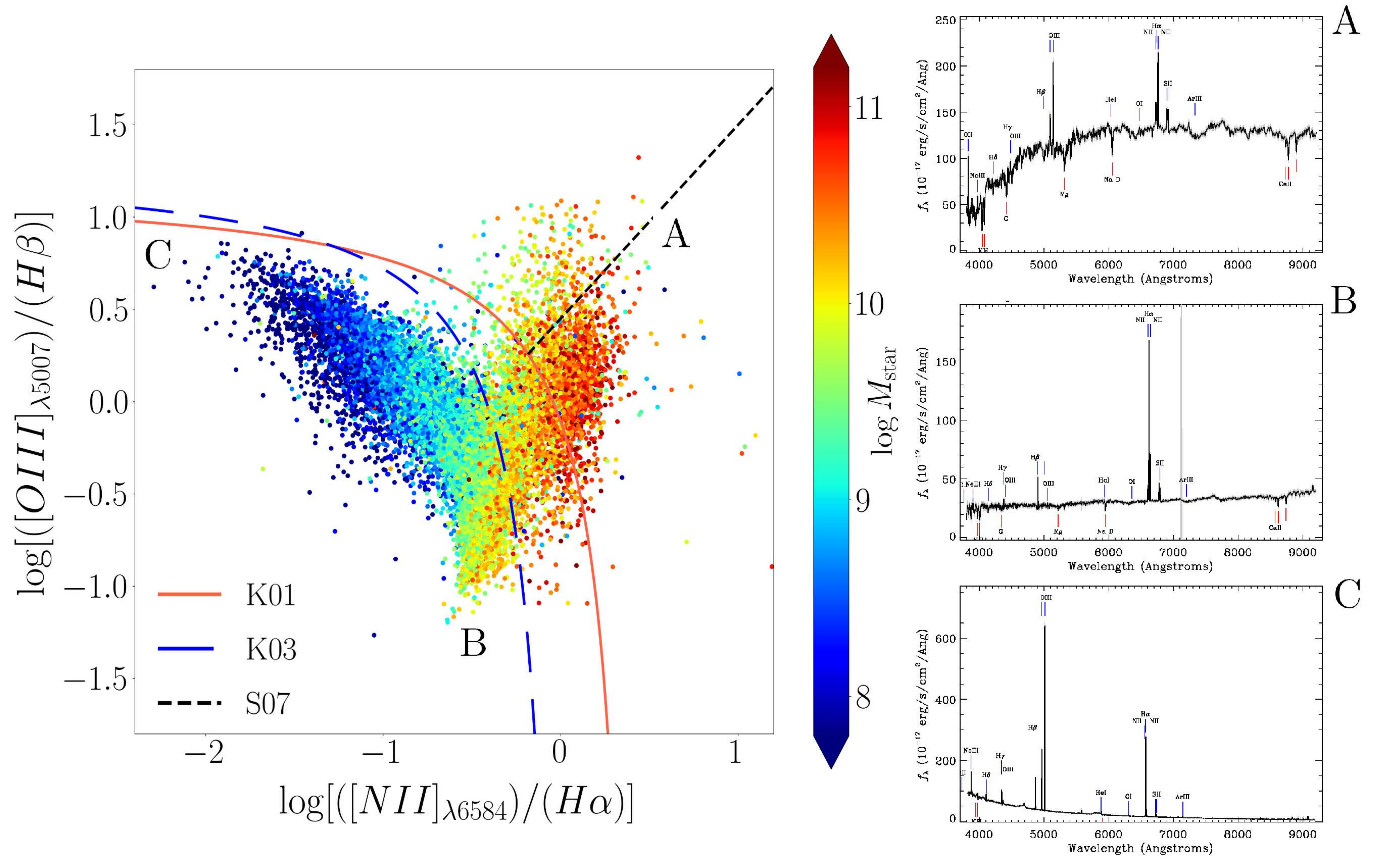}
       \end{center}
       \caption{The BPT diagnostic diagrams for galaxies in the SPRING catalogue. Three examples of spectra are given. Colours provide the logarithm of $M_{\mathrm{star}}$ in solar-mass units.}
       \label{fig:BPTcolor_v2}
       \end{figure*}
The prominent left wing of the BPT is traced by pure SF nuclei, lying below the empirical line included in the BPT scheme by \cite{Kauffmann-2003} (K03, red straight line in Fig. \ref{allmass-BPT-WHAN}, top panels). Active nuclei (SEY+LIN) are defined as those galaxies which lie on the right wing above the ``maximum starburst line'' determined by \cite{Kewley-2001} (K01, blue long-dashed line). Composite nuclei lie in between AGN and SF nuclei. Seyfert and LINERs are divided by the empirical separation line defined by \cite{Schawinski-2007} (S07, black dashed line).

In the top row of Fig. \ref{allmass-BPT-WHAN}, we display the BPT diagnostic diagrams separately for the three chromatic classes (blue cloud, green valley, and red sequence) defined in Sec. $\S$ \ref{sec:photometry}. 
\begin{itemize}
    \item In the blue cloud, most galactic nuclei are classified as star-forming ($\sim$90\%). Approximately 7\% objects are composite, and only few percent are LINERs and Seyfert nuclei ($\sim$2\% LIN and $\sim$1\% SEY).
    \item In the green valley, the majority of targets is still found in the SF wing of the BPT, although fewer than in the blue cloud ($\sim$60\%). A substantial amount of nuclei falls in the composite and LINERs branches ($\sim$20\% and $\sim$16\%, respectively). As in the blue cloud, only a small fraction of objects is classified as SEY ($\sim$ 2\%).
    \item Star-forming nuclei in the red sequence decrease to less than a fifth of the total ($\sim$19\%), whereas composite and LINERs account for more than half of the RS nuclei ($\sim$38\% each). Seyfert nuclei are $\sim$6\%. 
\end{itemize}
We point out that our nuclear classification of galaxies belonging to the SPRING sample considers only emission-line galaxies whose nuclear spectra have all four lines with signal-to-noise ratio $>$ 2. This selection introduces a noticeable bias in our classification as far as the RS concerns, since only $\sim$30\% of galaxies in the RS (1572 out of 5258) satisfies our criteria, opposite to the BC and the GV, whose majority of galaxies show all four lines in emission, and therefore are classified with the BPT diagram ($\sim$90\% and $\sim$70\%, respectively).

The mass metallicity relation \cite[cf. Fig. 3 in ][]{Thomas-2019} is partly responsible for the pattern in the star-formation arm of the BPT. From $12+\log(\mathrm{O/H}) = 7.7$ to $12+\log(\mathrm{O/H}) = 9.2$, most star-forming galaxies with stellar-mass between $10^{8.5}$M$_{\odot}$ and $10^{11}$M$_{\odot}$ are found, due to the well known correlation between $12+\log (\mathrm{O/H})$ and $\log \mathrm{([NII]/H}\alpha)$ \cite[see, \textit{e.g.},][]{Yates-2012}. 
However, it does not fully explain the two orders of magnitude decrease in log[OIII]/H$\beta$ over the entire SF branch, a fact that is hinting to a dependence of the ionisation properties on stellar-mass. With increasing stellar-mass, galaxies have decreasing fraction of OB associations in their arms. OB stars are the only stars capable of ionizing Hydrogen and their fraction is a steep function of mass \cite[][]{Kennicut-1994, Kennicutt-1998, Boselli-2009}. 
Since OB stars have the shortest lives of all stars this implies that low-mass galaxies have ``younger'' stellar population than higher mass. This is consistent with the picture of \cite{Gavazzi-2002}.
\subsection{WHAN diagnostic}\label{sec:whan}
In addition to the BPT diagnostics, the classification of the nuclear activity based on optical emission-lines is also performed following the two-line WHAN classification scheme. It is a less demanding diagnostic than the BPT, as it only requires the intensities of two lines (H$\alpha$ and [NII]$\lambdaup$6584), and is based on the $\rm [NII]/H\alpha$ ratio combined with the strength of the $\rm H\alpha$ line corrected\footnote{Since $\rm H\alpha$ and [NII] are close one another and since we use fluxes normalised to the continuum under $\rm H\alpha$, the same correction for stellar absorption is applied to the fluxes as to the EWHs.} for underlying stellar absorption.
Owing to this two-line diagnostic, even the red-channel spectra such as the majority of those obtained at Loiano \cite[][]{Gavazzi-2011, Gavazzi-2013, Gavazzi-2018a} contribute to increasing the number of known AGN, especially those associated with nearby bright galaxies, most affected by the residual incompleteness of the SDSS spectral database.

The WHAN diagnostic was introduced by \cite{Cid_Fernandez-2010, Cid_Fernandez-2011} to disentangle ``true'' active nuclei, believed to be triggered by supermassive black holes, from ``fake AGN'' (retired galaxies), \textit{i.e.}, galaxies with low SFR that are ionised by the hot low-mass evolved stars (HOLMES) within them \cite[][]{Trinchieri-1991, Binette-1994, Macchetto-1996, Cid-Fernandes-2004, Stasinska-2008, Sarzi-2010, Capetti-2011}. We identify five classes of nuclear activity adopting the modified WHAN classification of \cite[][]{Gavazzi-2011}:

\begin{itemize}
    \item Star-Forming nuclei (SF): $\log\mathrm{[NII]}/\mathrm{H}\alpha < -0.3$ and EWH$\alpha>1.5$ \text{\normalfont\AA};
    \item Seyfert (SEY): $\log\mathrm{[NII]}/\mathrm{H}\alpha > -0.3$ and EWH$\alpha>6$ \text{\normalfont\AA};
    \item LINERs (LIN): $\log\mathrm{[NII]}/\mathrm{H}\alpha > -0.3$ and $6<$ EWH$\alpha$/\text{\normalfont\AA}$<1.5$;
    \item Retired galaxies (RET): EWH$\alpha < 1.5$\text{\normalfont\AA};
    \item Passive galaxies (PAS): EWH$\alpha$, EW[NII] $<0.5$\text{\normalfont\AA}.
\end{itemize}

\begin{table}[h!]
\small
\centering
\caption{Fractions of the nuclear excitation classes classified with the WHAN diagram (SF, LIN, SEY, RET) calculated in four samples defining the BPT-classified nuclear activity (SF, LIN, SEY, Comp). }\label{tab:BPT_WHAN}
\begin{tabular}{|l||*{4}{c|}}\hline
\backslashbox{BPT}{WHAN}
&\makebox[3em]{SF (\%)}&\makebox[3em]{LIN (\%)}&\makebox[3em]{SEY (\%)}
&\makebox[3em]{RET (\%)} \\
\hline\hline
SF & 99.3  & 0.1  & 0.3  &  0.1 \\
\hline
LIN & 0.0  & 34.6  & 6.3  & 48.6  \\
\hline
SEY & 8.9  & 25.0  &  40.3 &  19.7 \\
\hline
Comp & 16.1  & 38.3  & 21.8  & 17.1  \\
\hline
\end{tabular}
\end{table}
\begin{table}[h!]
\small
\centering
\caption{Fractions of the nuclear excitation classes classified with the BPT diagram (SF, LIN, SEY, Comp) calculated in four samples defining the WHAN-classified nuclear activity (SF, LIN, SEY, RET). }\label{tab:WHAN_BPT}
\begin{tabular}{|l||*{4}{c|}}\hline
\backslashbox{WHAN}{BPT}
&\makebox[3em]{SF (\%)}&\makebox[3em]{LIN (\%)}&\makebox[3em]{SEY (\%)}
&\makebox[3em]{Comp (\%)} \\
\hline\hline
SF & 93.1  & 0.0  & 0.2  &  2.2 \\
\hline
LIN & 1.1  & 26.5  & 4.8  & 55.6  \\
\hline
SEY & 5.6  & 10.0  &  15.9 &  65.7 \\
\hline
RET & 12.2  & 40.8  & 4.1  & 27.1  \\
\hline
\end{tabular}
\end{table}

Analogously to the BPT, we display the WHAN diagnostic diagrams separately for the three chromatic classes (Fig. \ref{allmass-BPT-WHAN}, bottom panels).

\begin{itemize}
    \item Star-forming nuclei are the majority of objects in the blue cloud (90\%). The fraction of AGN is larger than in the BPT-classification, as both LIN and SEY account for $\sim$4\% of the total each. Retired objects are $\sim$1\%. Less than 1\% of nuclei are passive.
    \item Approximately 64\% nuclei in the green valley are star-forming. LINERs increase to $\sim$20\%, whereas the fraction of Seyfert nuclei mildly dereases ($\sim$3\%). Retired objects are $\sim$10\%. Again, less than 1\% of nuclei are passive.
    \item In the red sequence, star-forming nuclei drop to $\sim$20\%. The fraction of LINERs is $\sim$20\%.
    The fractions of retired and passive objects both increase to $\sim$45\% and $\sim$16\%, respectively. Finally, SEY only account for $\sim1\%$ of nuclei. 
\end{itemize}
In general, there is a satisfactory agreement between the BPT and the WHAN classification among Star-Forming systems. A visual comparison between the two classification schemes is given in Fig. \ref{fig:BPT_WHAN_mix}, displaying the BPT (WHAN) diagram of galaxies classified according to the WHAN (BPT).

We observe that the two diagnostics display some noticeable deviations in the classification of non-star-forming nuclei (see Tables \ref{tab:BPT_WHAN}-\ref{tab:WHAN_BPT}): there are significant overlaps between the classification of Seyfert and LINERs; \textit{e.g.}, most BPT-classified LINERs are identified as Retired by the WHAN diagnostic (48.6\%); similarly, most WHAN-classified LINERs are classified as composite nuclei by the BPT (55.6\%); the majority of galaxies classified as Seyfert by the BPT are marked the same way by the WHAN (40.3\%); conversely, WHAN-classified Seyfert are mostly classified as composite nuclei (65.7\%) by the BPT. 

The disparity between the BPT and WHAN classifications of AGN (and, especially, LINERs) is not surprising. The two schemes employ a different number of emission-lines, and were proposed following different theoretical and empirical criteria \cite[][]{Baldwin-1981, Kewley-2001, Kauffmann-2003, Cid_Fernandez-2011}. In addition, there is no clear consensus yet on the ionisation source of LINERs. \cite{Heckman-1980} proposed that gas shocks are mainly responsible for LINER-like emission. \cite{Halpern-1983} and \cite{Ferland-1983} suggested that the primary mechanism behind the ionisation of LINERs is the accretion of gas onto a central massive black hole. Later, \cite{Terlevich-1985} and \cite{Shields-1992} proposed that the excitation of these nuclei is produced by photoionisation due to hot stars. Recently, \cite{Oliveira-2022} argued that a sample of 43 LINERs is probably ionised by post-AGB (asymptotyc giant branch) stars.
The uncertainty on the mechanisms driving LINER-like emission could be a major factor for the differences between the BPT and WHAN classifications.

\begin{figure}
\begin{center}
\includegraphics[width=8cm]{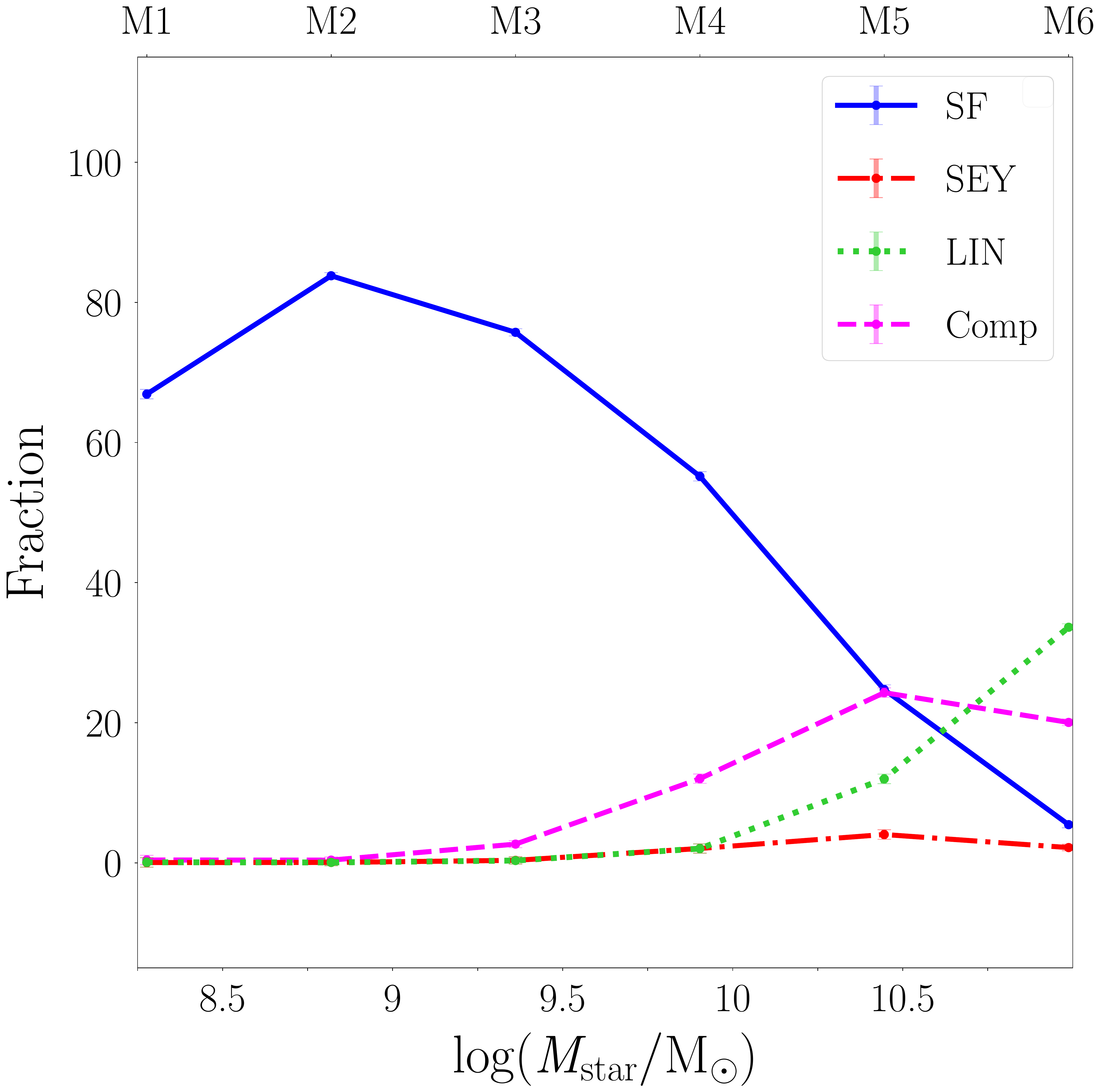}\\
\includegraphics[width=8cm]{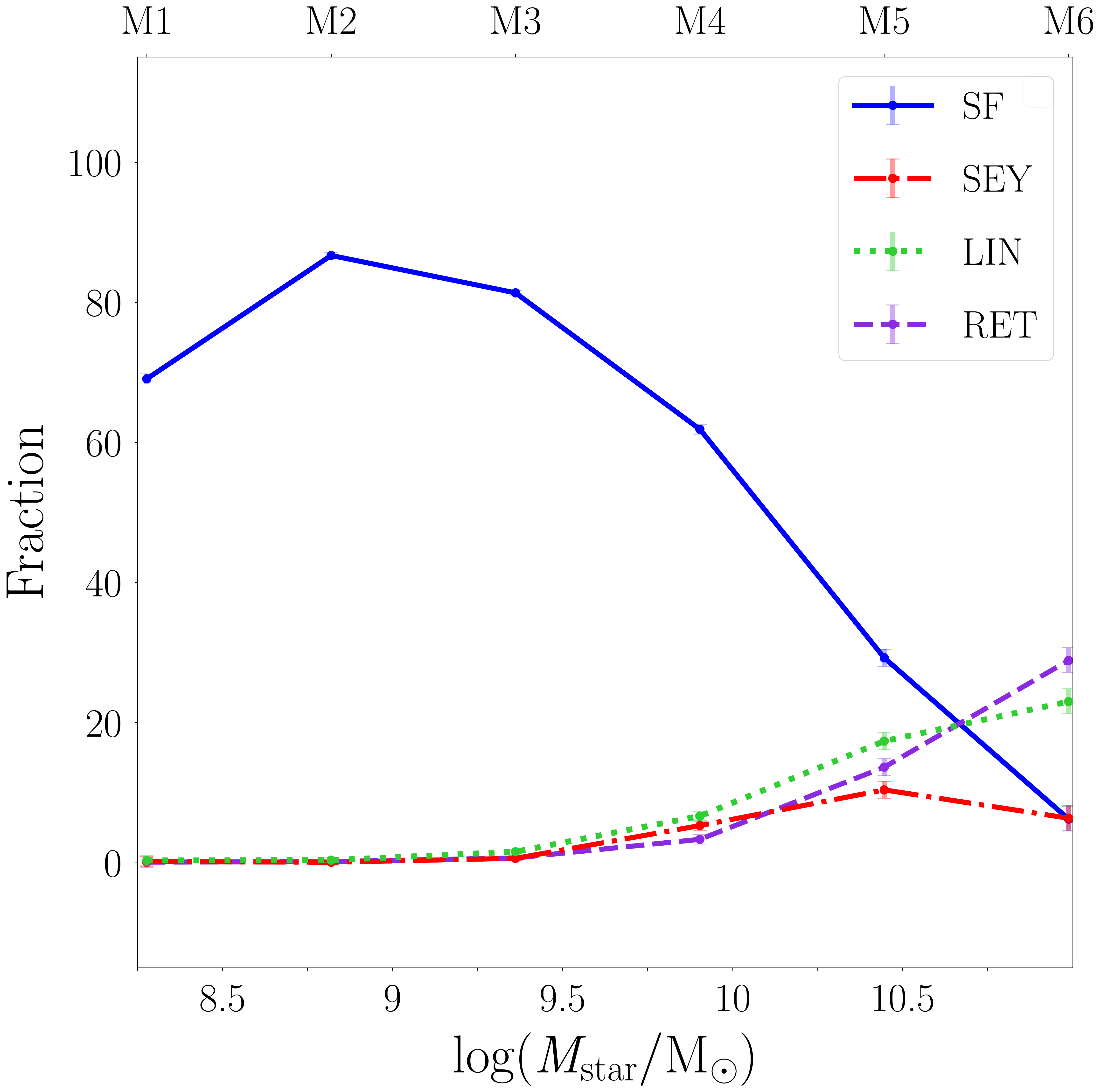}
\caption{Frequency of BPT (top) and WHAN (bottom) excitation classes as a function of $\log(M_{\mathrm{star}}/\mathrm{M}_{\odot})$ (represented by the six bins M1, ... , M6). The binomial errors are computed following \cite{Wilson-1927}.}\label{fig:FIG5ALL}     
\end{center}
\end{figure}
\subsection{Stellar-mass and environment vs nuclear activity}\label{subsec:stmass-nuclear}
The ionisation diagnostics considered here are sensitive to the stellar-mass of the targets \cite[see, \textit{e.g.},][]{Kauffmann-2003}. To explore the relation between nuclear activity and stellar-mass, we classify the galaxy nuclear properties with the BPT and WHAN diagnostics separately in each stellar-mass bin (M1, ..., M6). In Fig. \ref{fig:BPTcolor_v2}, we show the BPT diagnostic diagram for the emission-line galaxies in our catalogue. Using different colours to represent different mass bins, we see the characteristic dependence of the nuclear excitation properties on the stellar-mass: in accordance with \cite{Kauffmann-2003}, the nuclei of lowest-mass galaxies are SF regions, progressively filling the left-wing of the BPT from top-left to bottom-centre with decreasing ionisation strength (top-bottom) and increasing metallicity (left-right). For three representative regions of the BPT diagram, a typical spectrum is also reported showing the progressive decrease of H$\alpha$ over [NII], and of [OIII] over H$\beta$. 
Only in the two highest stellar-mass bins (\textit{i.e.}, $10^{10}$ to $\gtrsim10^{10.5} \ M_{\mathrm{star}}/\mathrm{M}_{\odot}$) a higher fraction of ionised nuclei (LINERs, and, to a lower extent, Seyfert nuclei) is found.

In Fig. \ref{fig:FIG5ALL}, we show the incidence\footnote{We emphasise that the fractions of emission-line galaxies (ELGs) computed throughout this paper (\textit{e.g.}, $f_{\mathrm{SEY}}/f_{\mathrm{TOT}}$) are calculated with respect to the whole galaxy sample in each stellar-mass bin; \textit{i.e.}, $f_{\mathrm{TOT}}$ consists of both ELGs and passive galaxies. This choice allows for a more accurate determination of the relative number of each class of ELG in each stellar-mass bin.} of the BPT- and WHAN-classified  SPRING emission-line galaxies as a function of their stellar-mass.
We see that up to $M_{\mathrm{star}}=10^{9-9.5} \ \mathrm{M}_{\odot}$ (mass bin M3), almost the entire spectroscopic sample is constituted of galaxies with SF nuclei, whose incidence is highest in the M2 bin.
The fraction of LINERs becomes non-negligible in the M4 bin ($M_{\mathrm{star}}=10^{9.5-10} \ \mathrm{M}_{\odot}$) and turns out to be dominant over the SF fraction ($\sim30\%$ over $\sim10\%$, respectively) for $M_{\mathrm{star}} \gtrsim 10^{10.5} \ \mathrm{M}_{\odot}$ (M6 bin) \cite[see][]{Kauffmann-2003}.

\begin{figure*}
\begin{center} 
\includegraphics[width=0.24\textwidth]{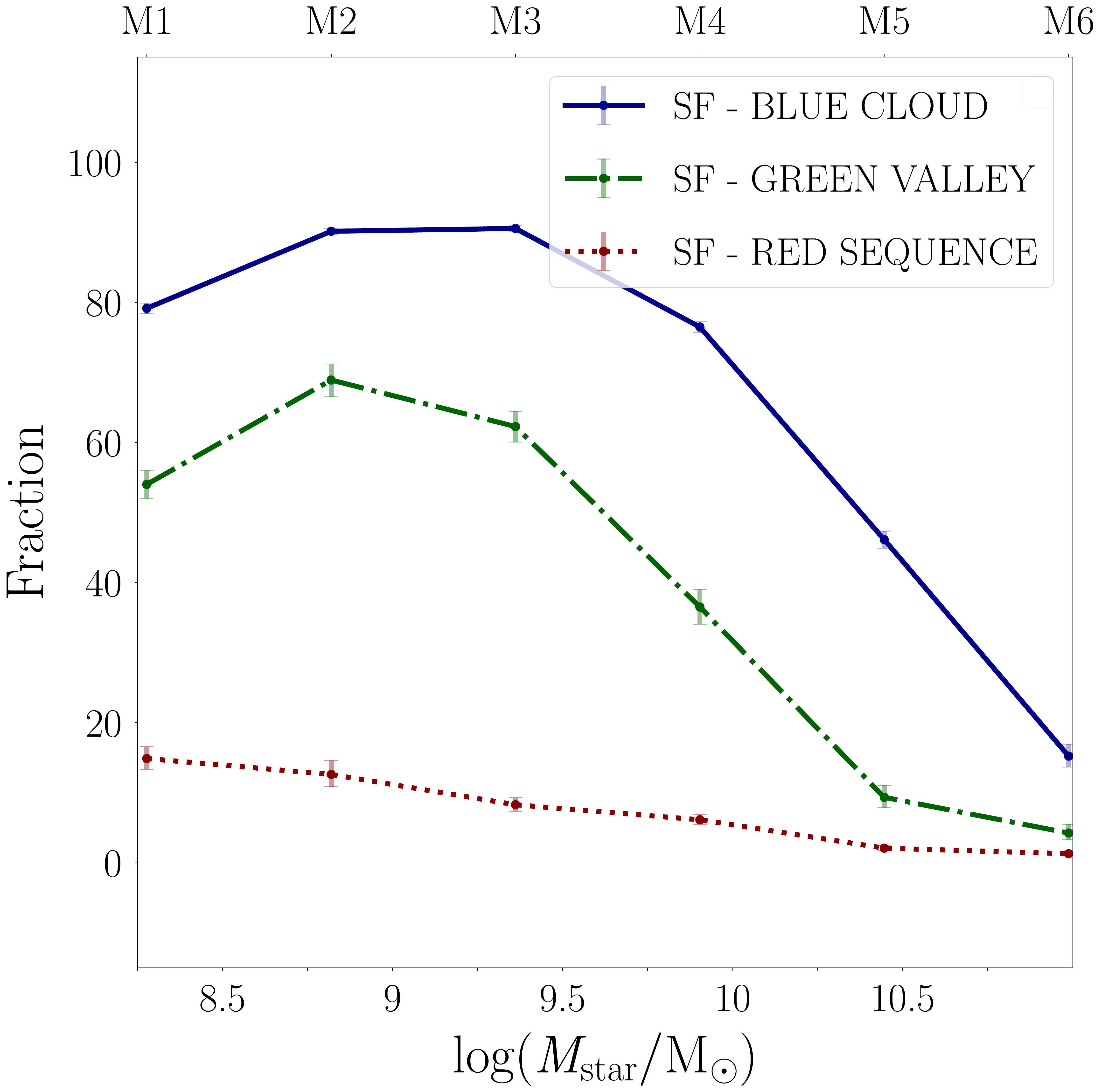}
\includegraphics[width=0.24\textwidth]{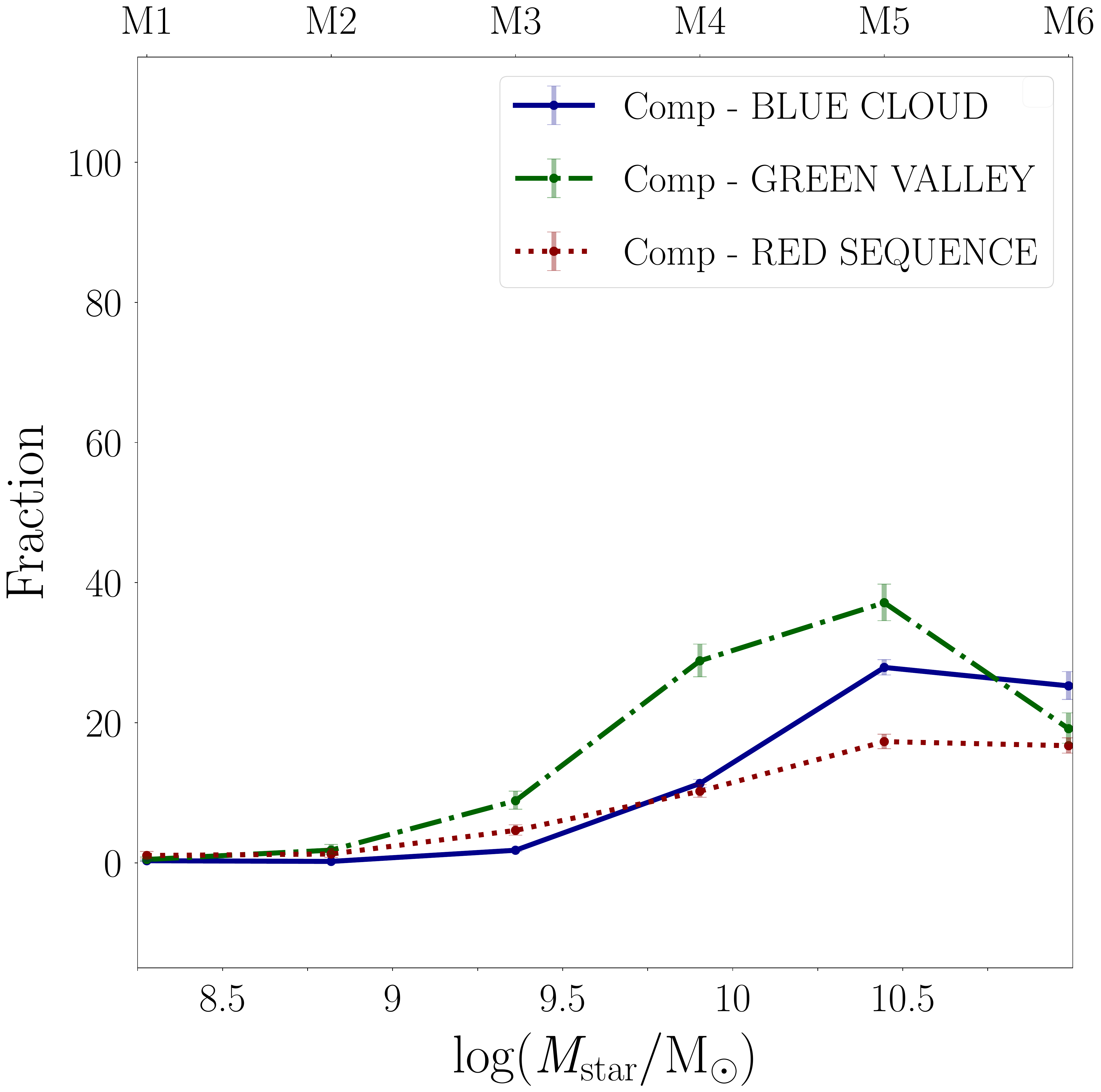}
\includegraphics[width=0.24\textwidth]{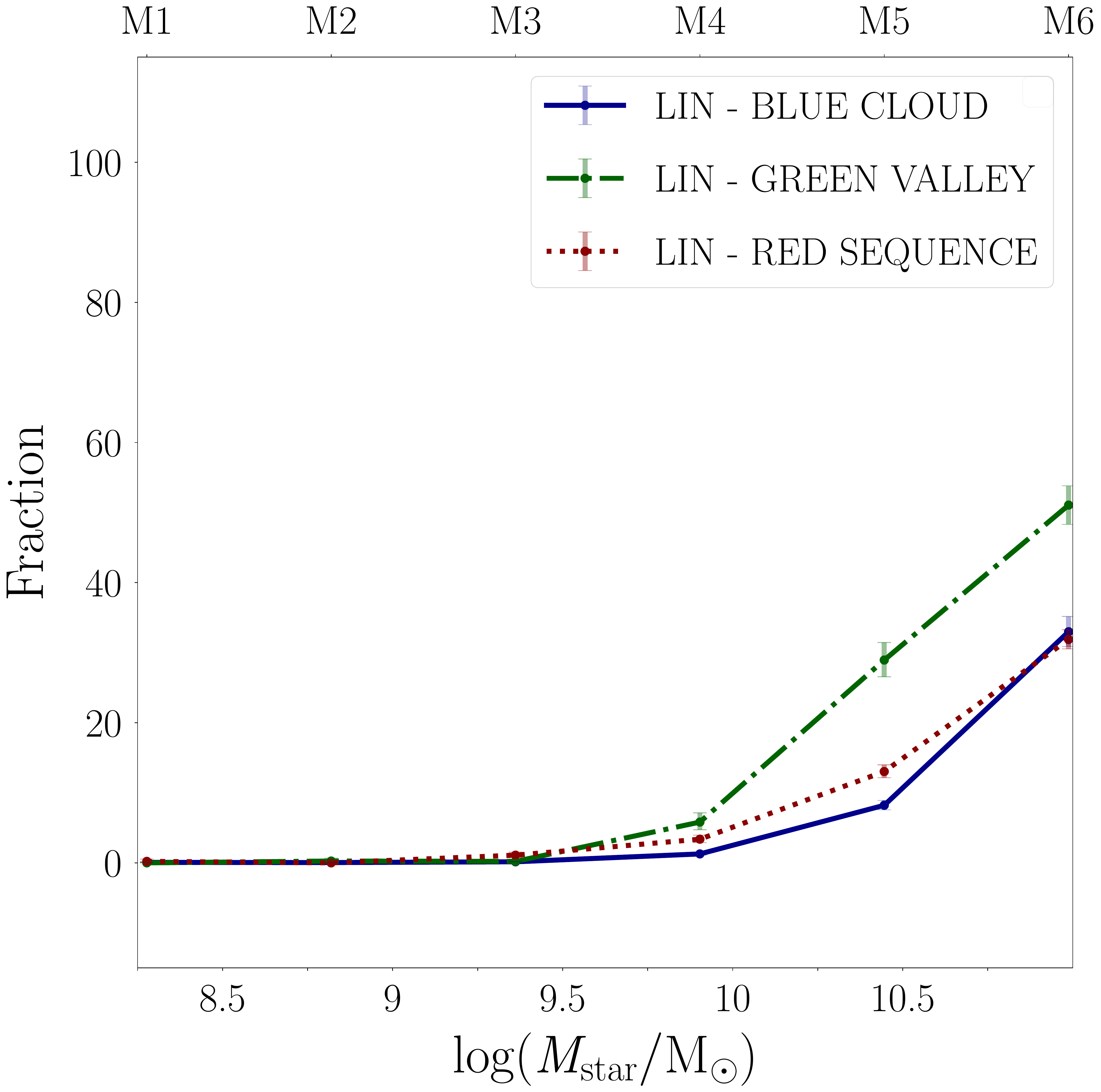}
\includegraphics[width=0.24\textwidth]{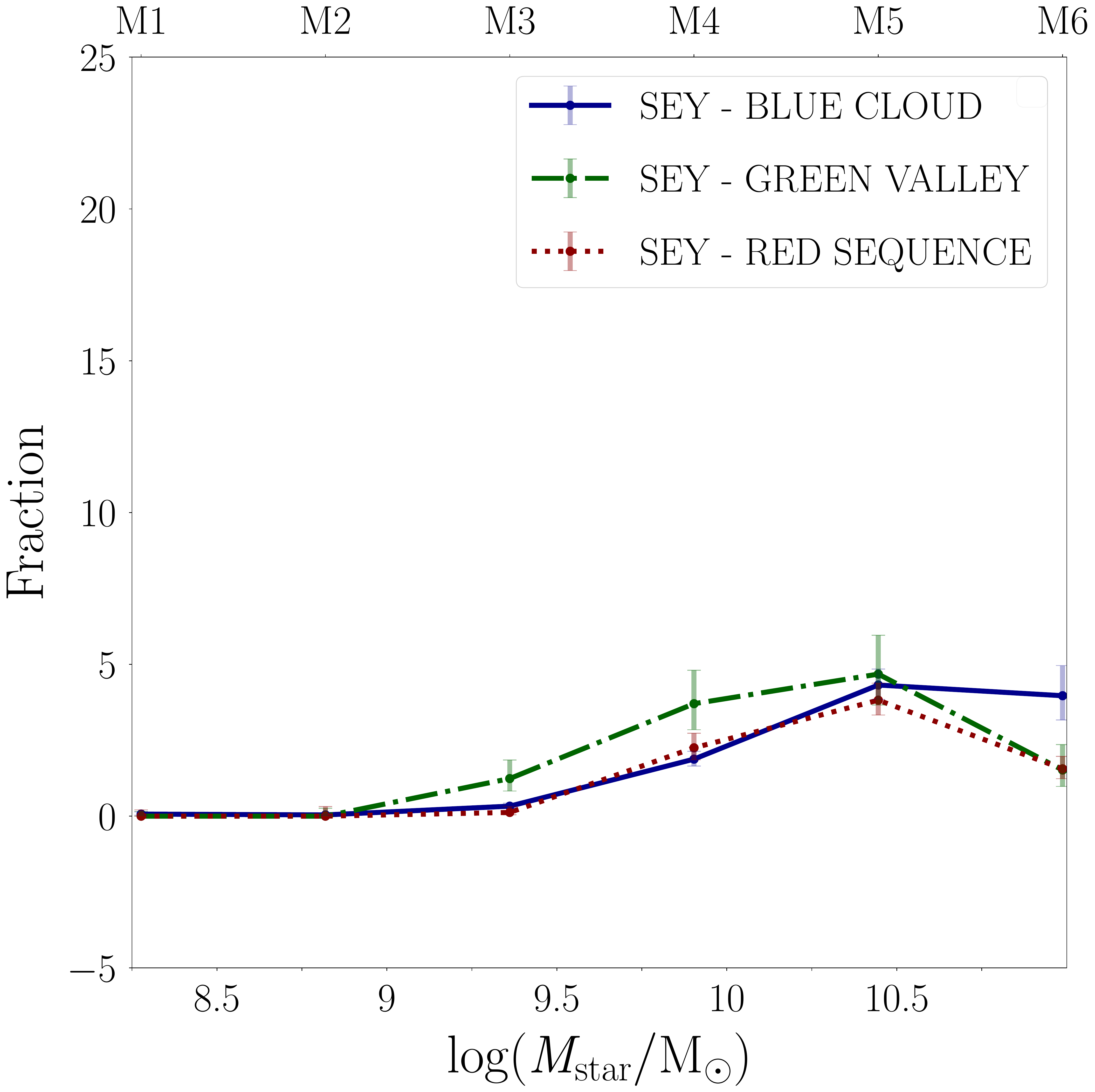}\\
\end{center}
\caption{The BPT-computed frequencies of star-forming (SF), AGN (SEY), LINER (LIN) and composite (Comp) as a function of $\log ( M_{\mathrm{star}}/\mathrm{M}_{\odot})$ (represented by the six bins M1, ..., M6) in the three chromatic regimes (blue cloud, green valley, red sequence, see Sec. $\S$ \ref{sec:photometry}). The $y$-axis scale on the right panel (SEY) is magnified to highlight the differences between the three chromatic classes.}
\label{fig:nuclear_bins1}
\end{figure*}

\begin{figure*}
\begin{center} 
\includegraphics[width=0.24\textwidth]{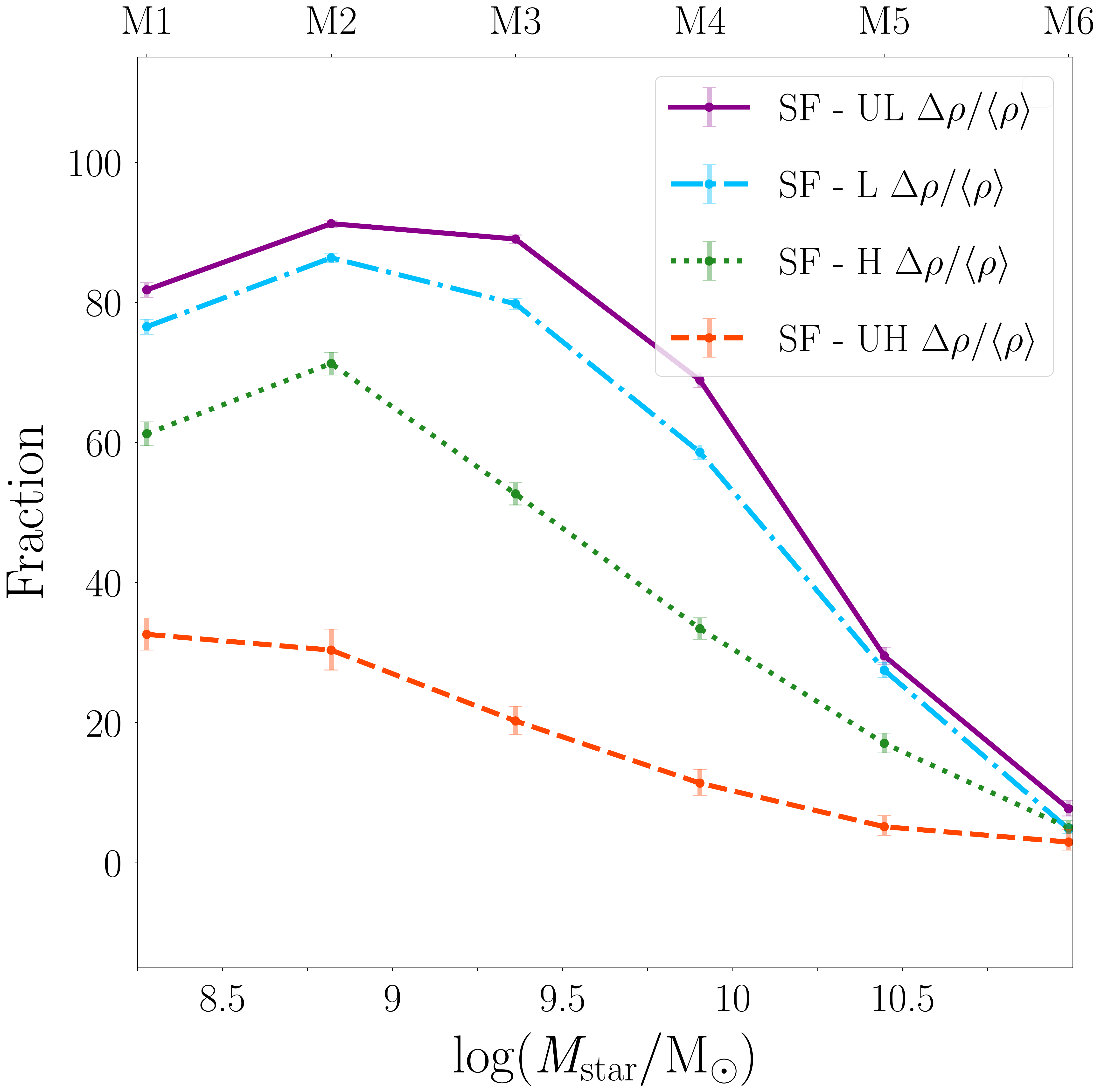}
\includegraphics[width=0.24\textwidth]{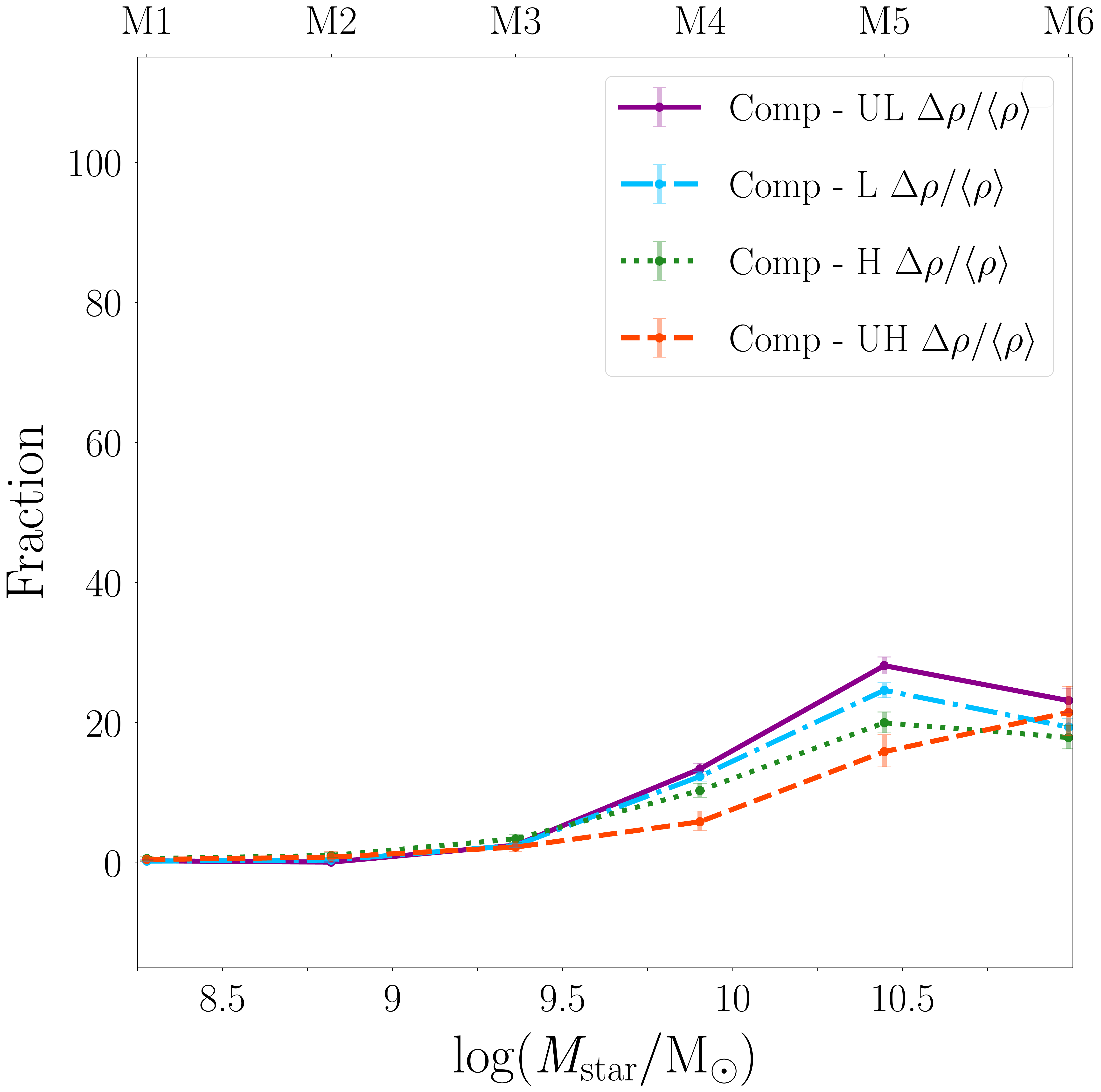}
\includegraphics[width=0.24\textwidth]{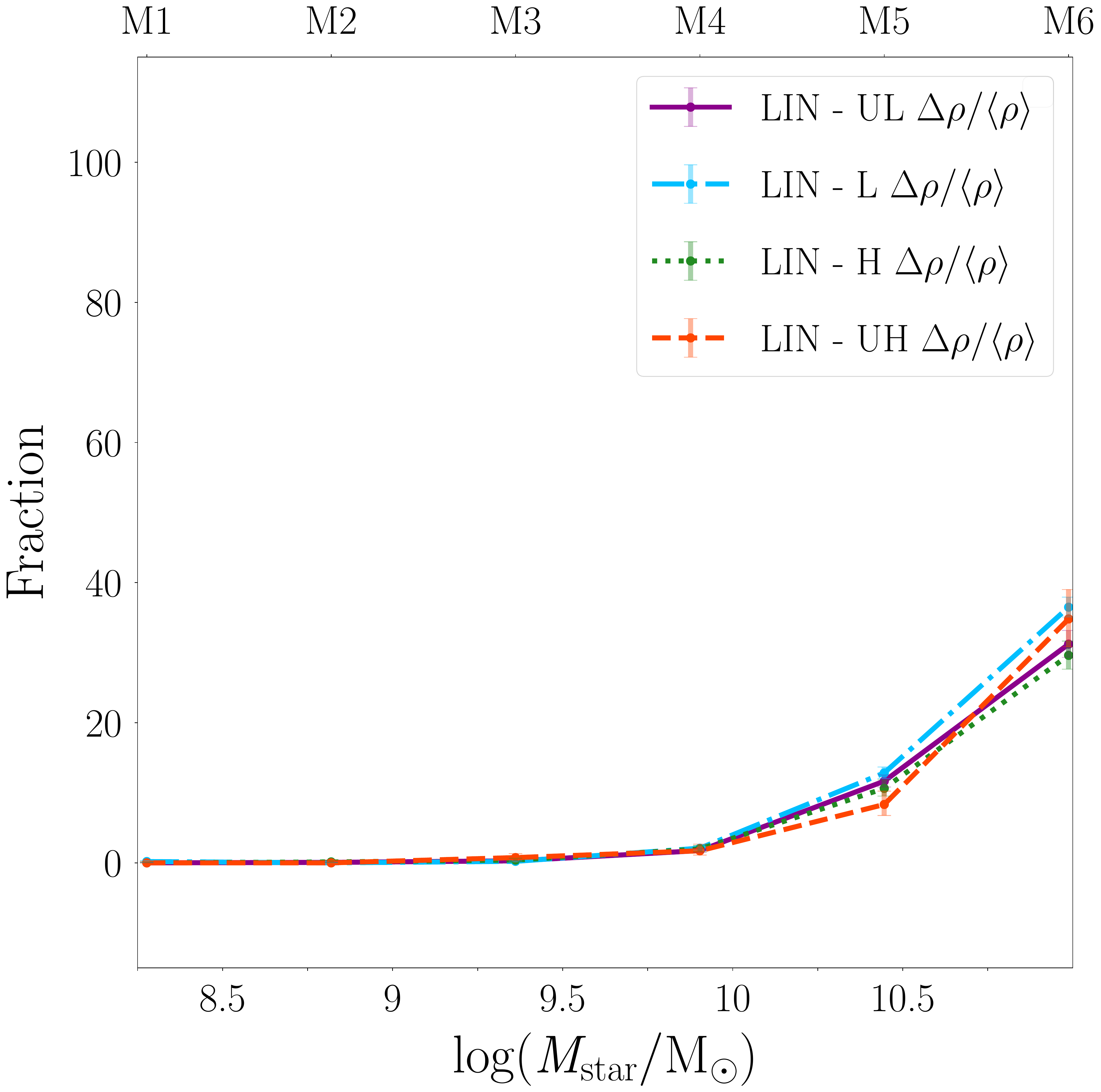}
\includegraphics[width=0.24\textwidth]{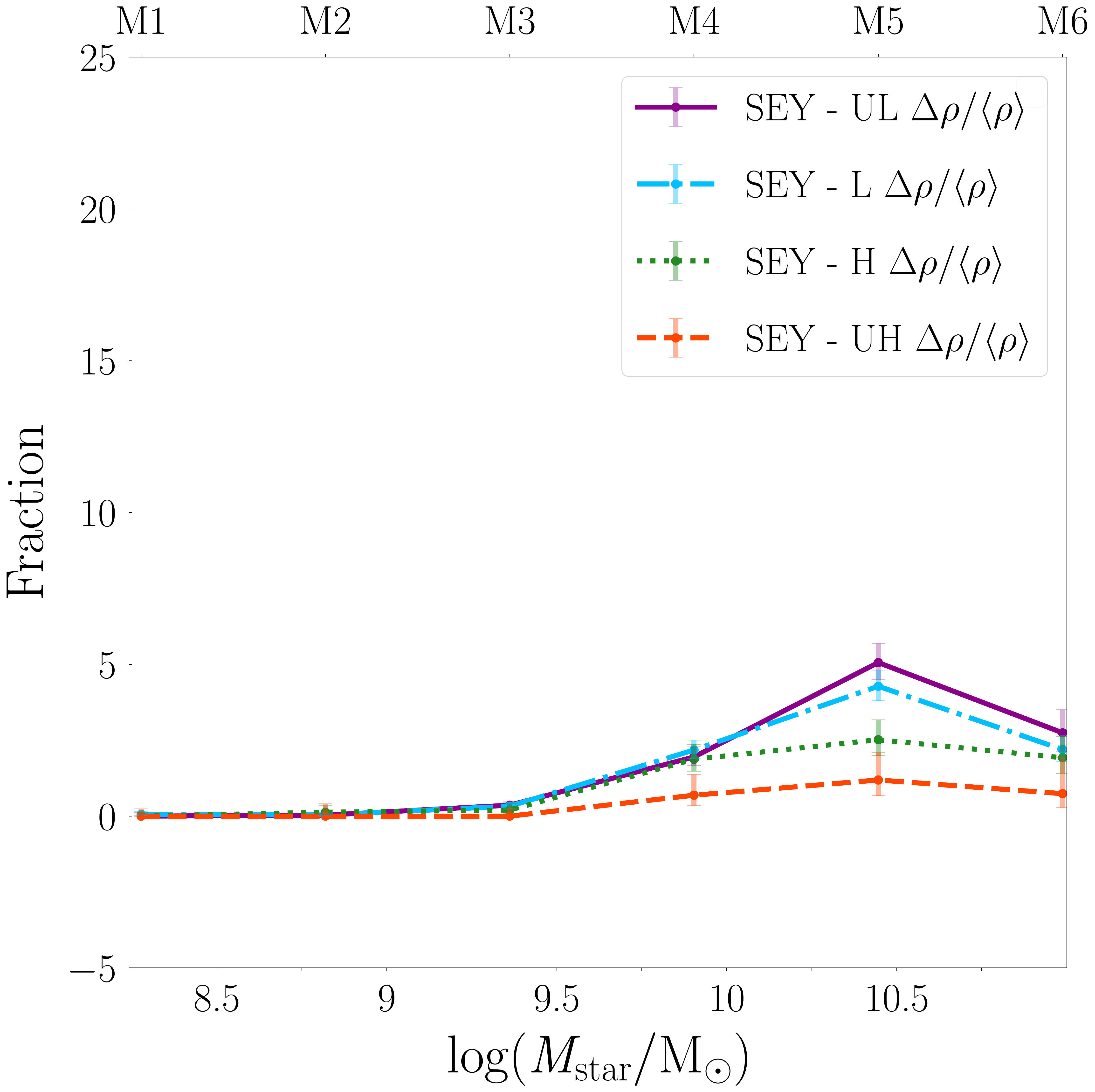}\\
\end{center}
\caption{The BPT-computed frequencies of star-forming (SF), AGN (SEY), LINER (LIN) and composite (Comp) as a function of $\log ( M_{\mathrm{star}}/\mathrm{M}_{\odot})$ (represented by the six bins M1, ..., M6) in the four overdensity bins defined in Sec. $\S$ \ref{subsec:overdensity}). The $y$-axis scale on the right panel (SEY) is magnified to highlight the differences between the four overdensity classes.}
\label{fig:nuclear_bins2}
\end{figure*}

\begin{figure*}
\begin{center} 
\includegraphics[width=0.24\textwidth]{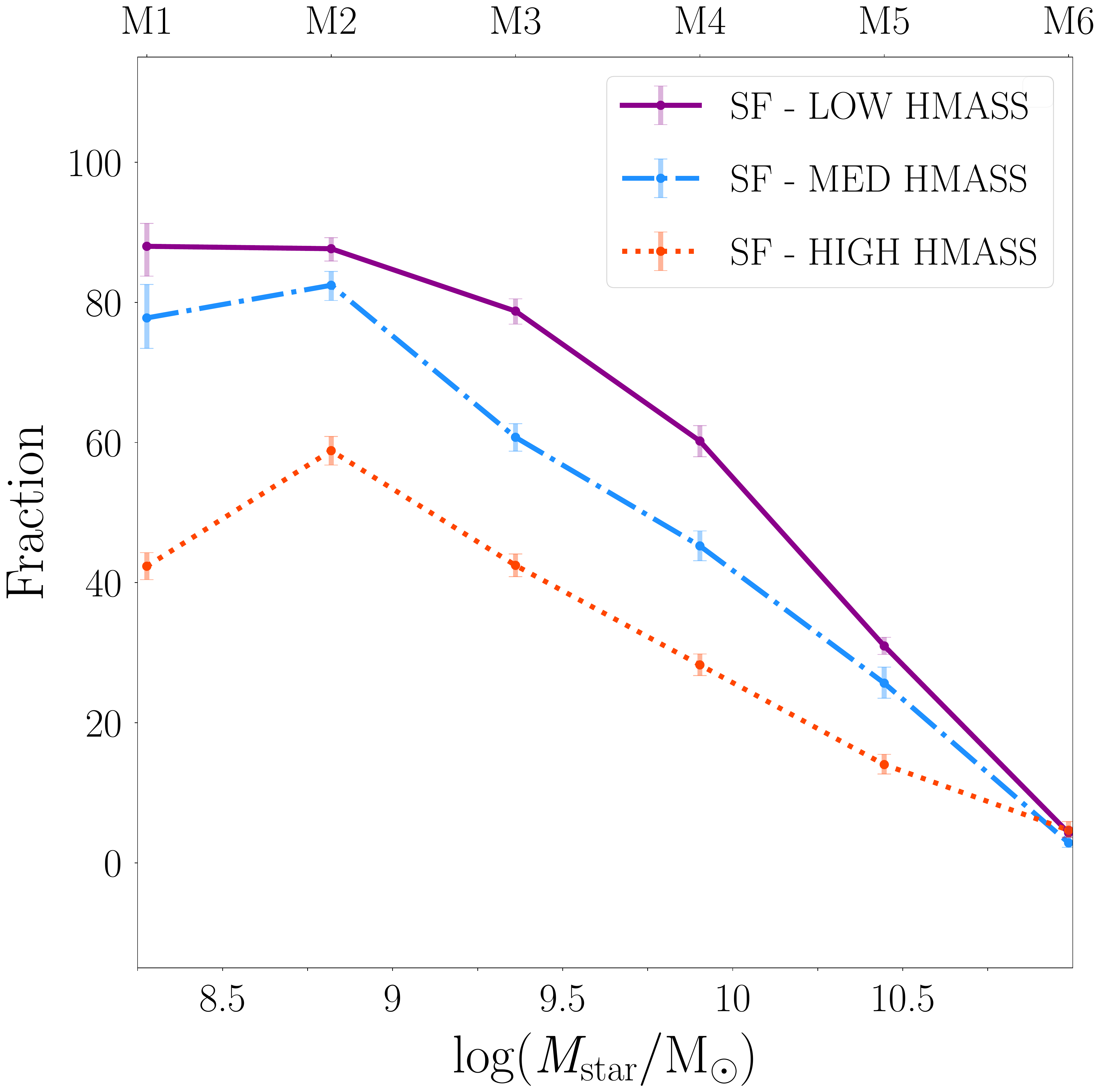}
\includegraphics[width=0.24\textwidth]{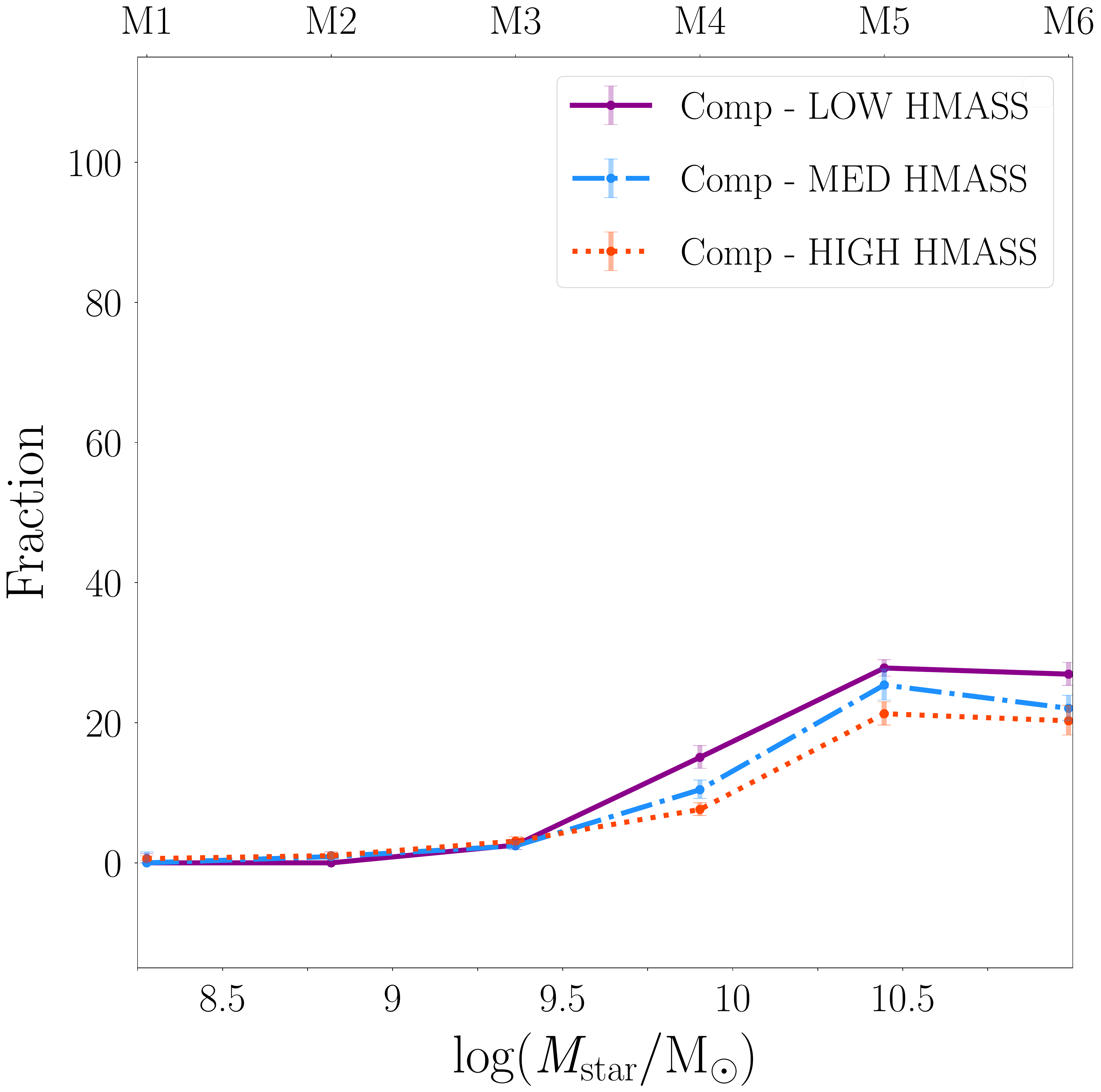}
\includegraphics[width=0.24\textwidth]{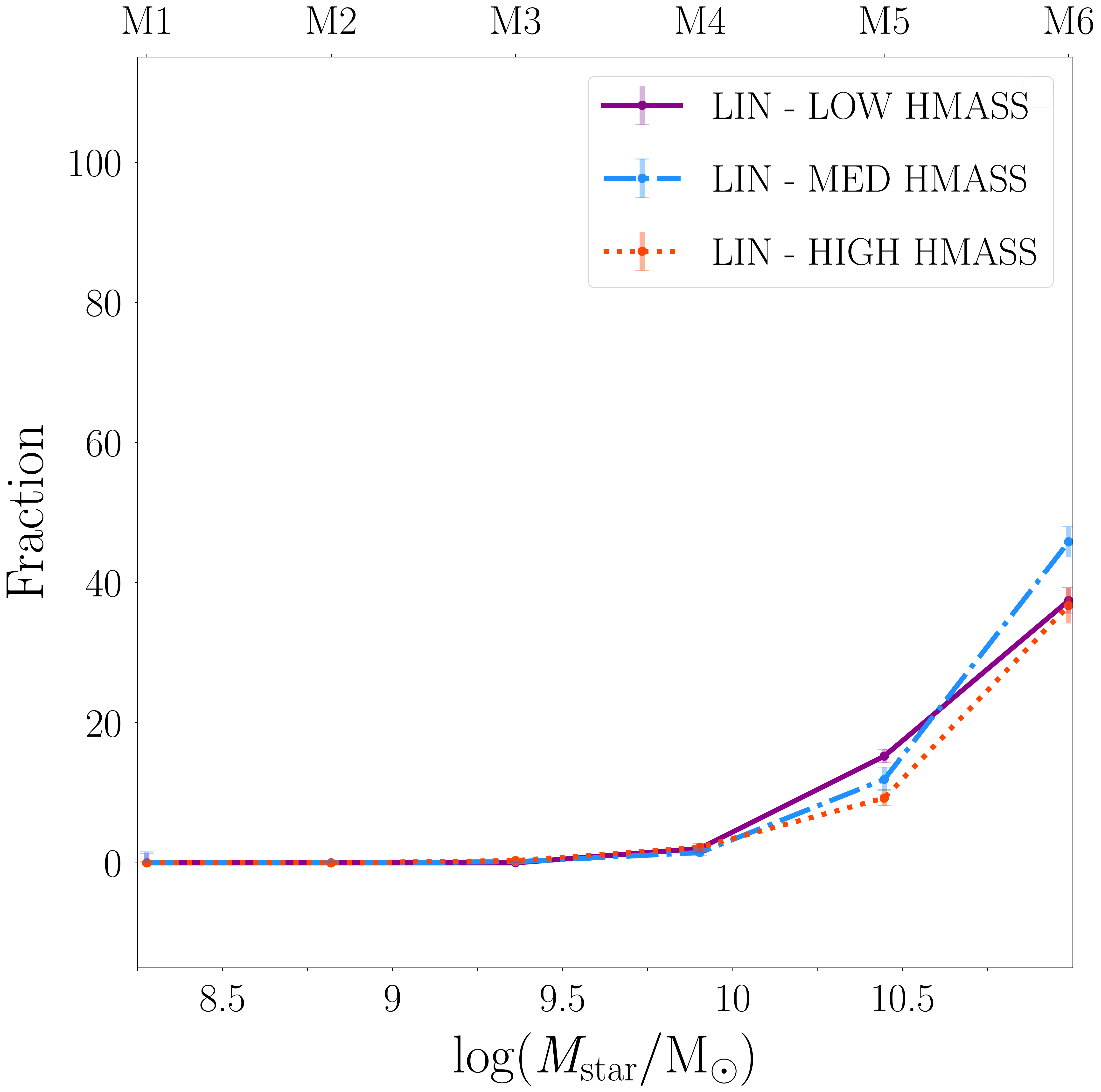}
\includegraphics[width=0.24\textwidth]{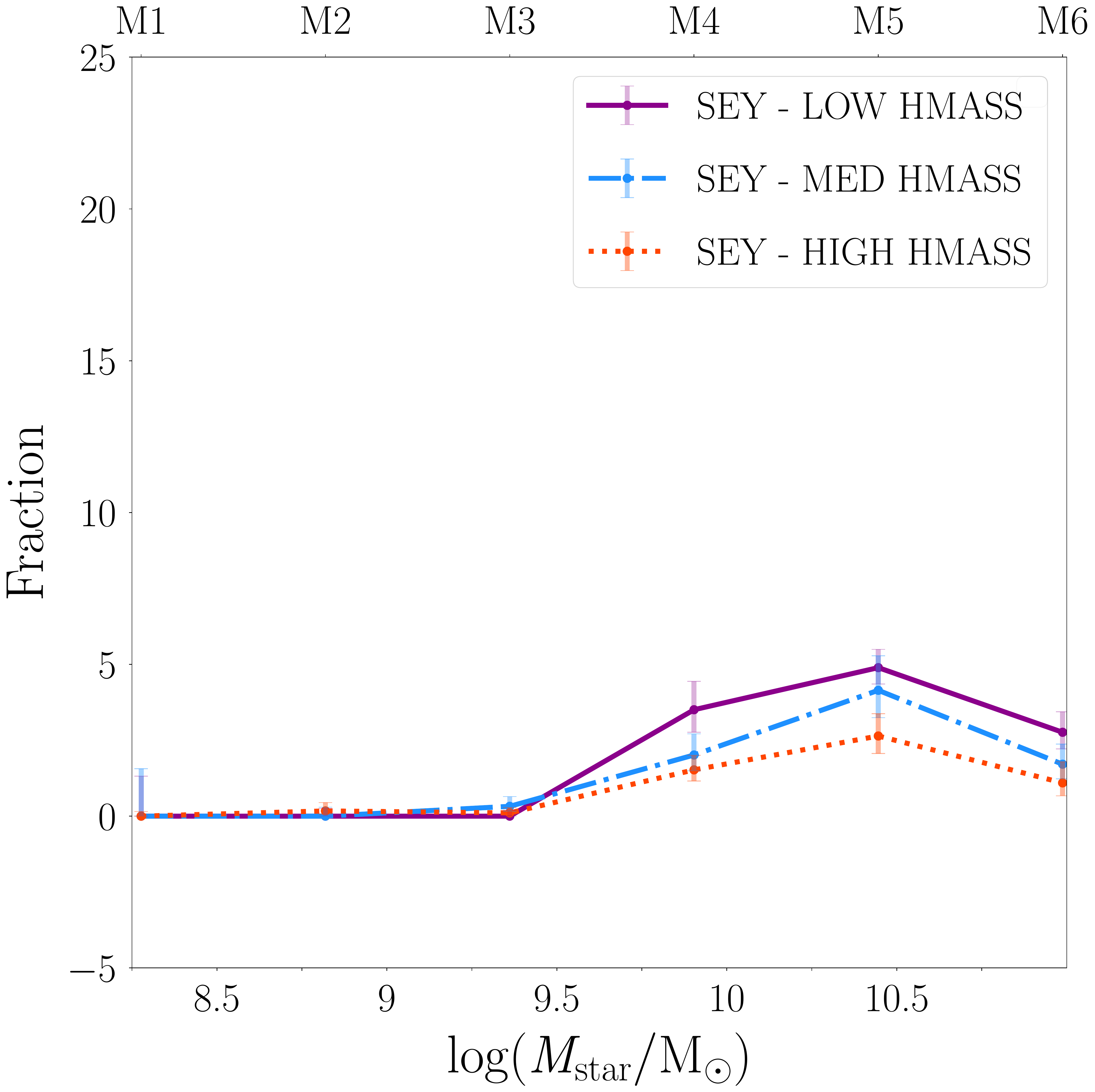}\\
\end{center}
\caption{The BPT-computed frequencies of star-forming (SF), AGN (SEY), LINER (LIN) and composite (Comp) as a function of $\log ( M_{\mathrm{star}}/\mathrm{M}_{\odot})$ (represented by the six bins M1, ..., M6) in the three halo-mass bins defined in Sec. $\S$ \ref{subsec:halomass}). The $y$-axis scale on the right panel (SEY) is magnified to highlight the differences between the three halo-mass classes.}
\label{fig:nuclear_bins3}
\end{figure*}

\begin{figure*}
\begin{center} 
\includegraphics[width=0.24\textwidth]{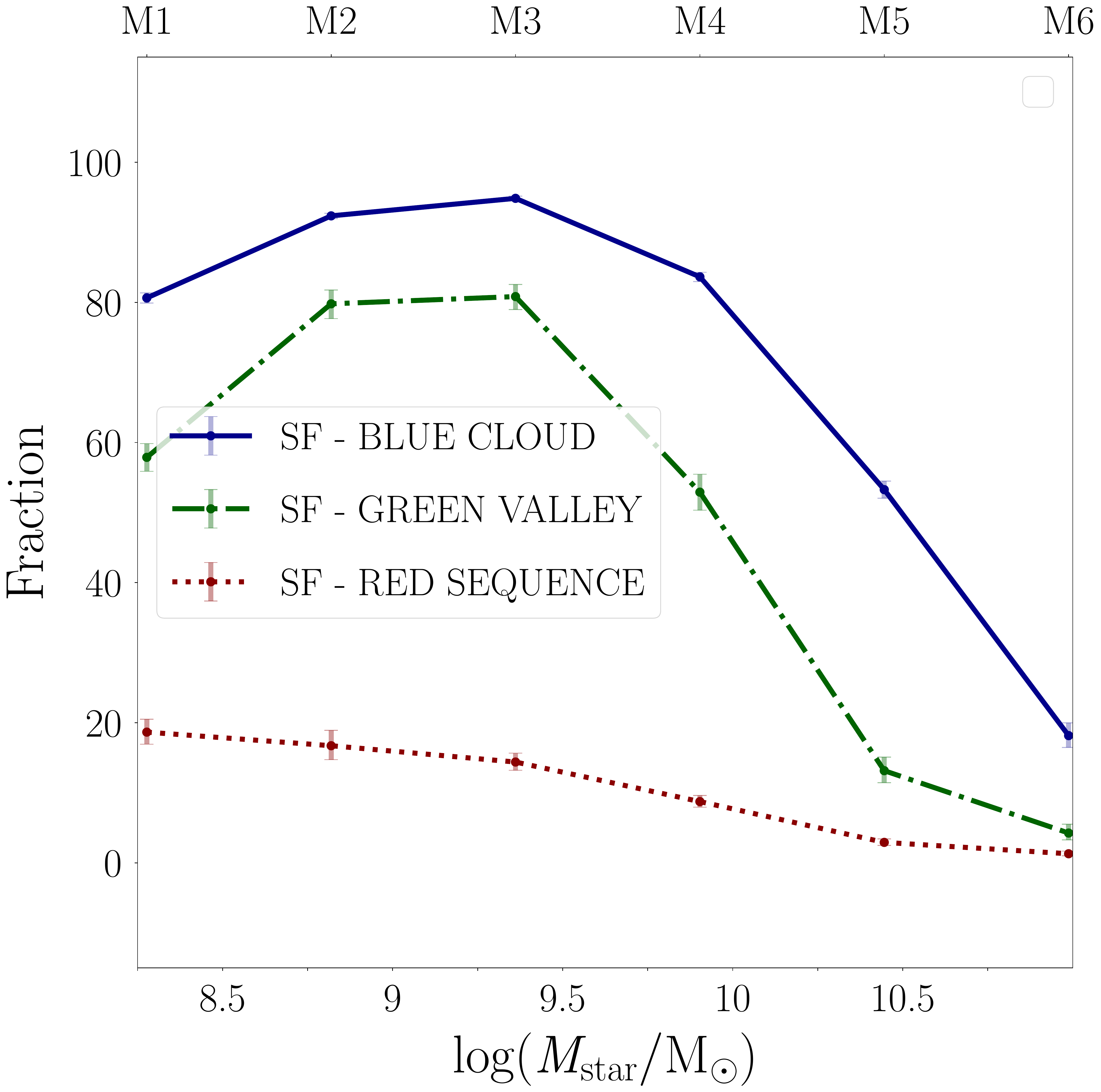}
\includegraphics[width=0.24\textwidth]{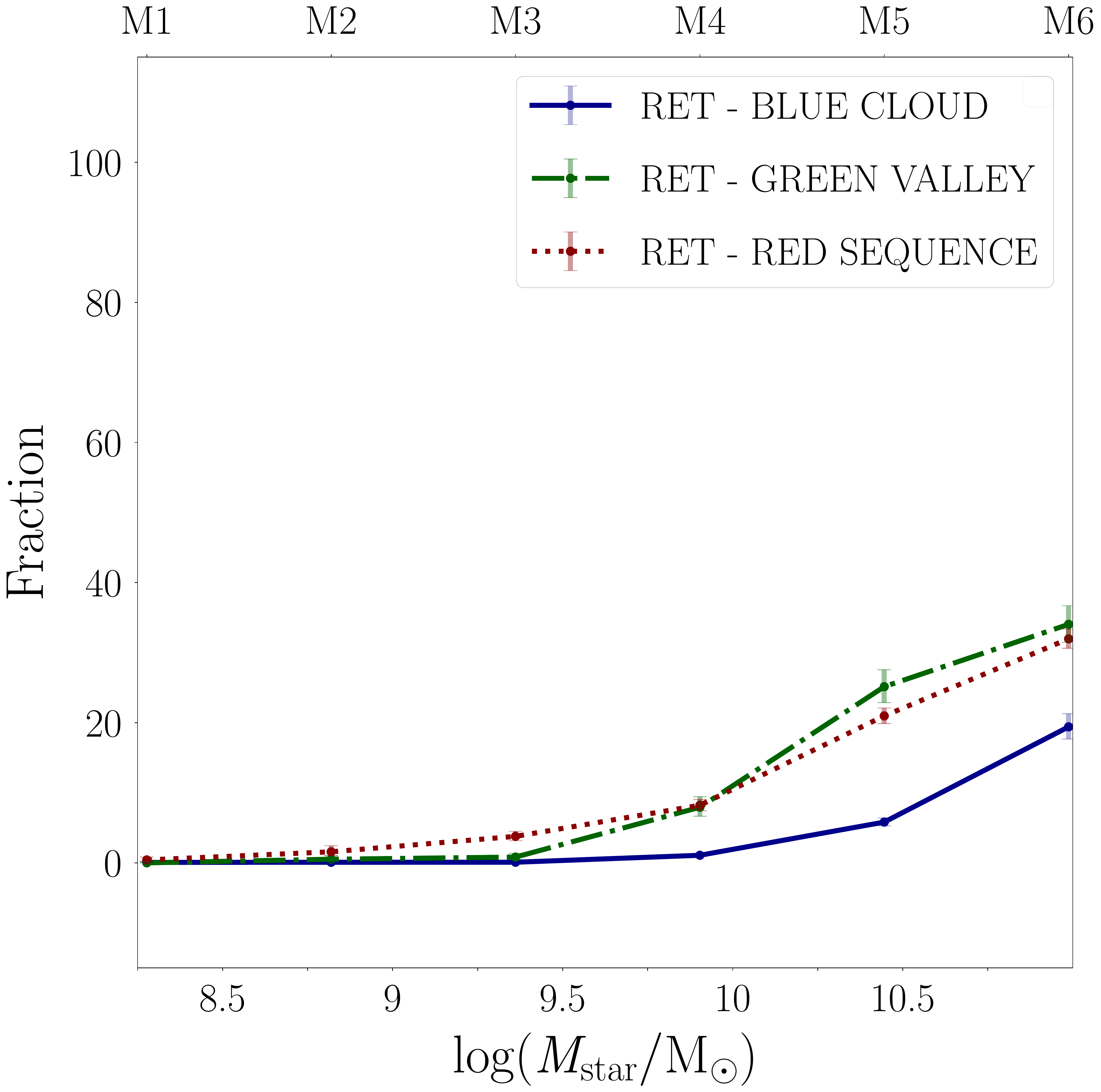}
\includegraphics[width=0.24\textwidth]{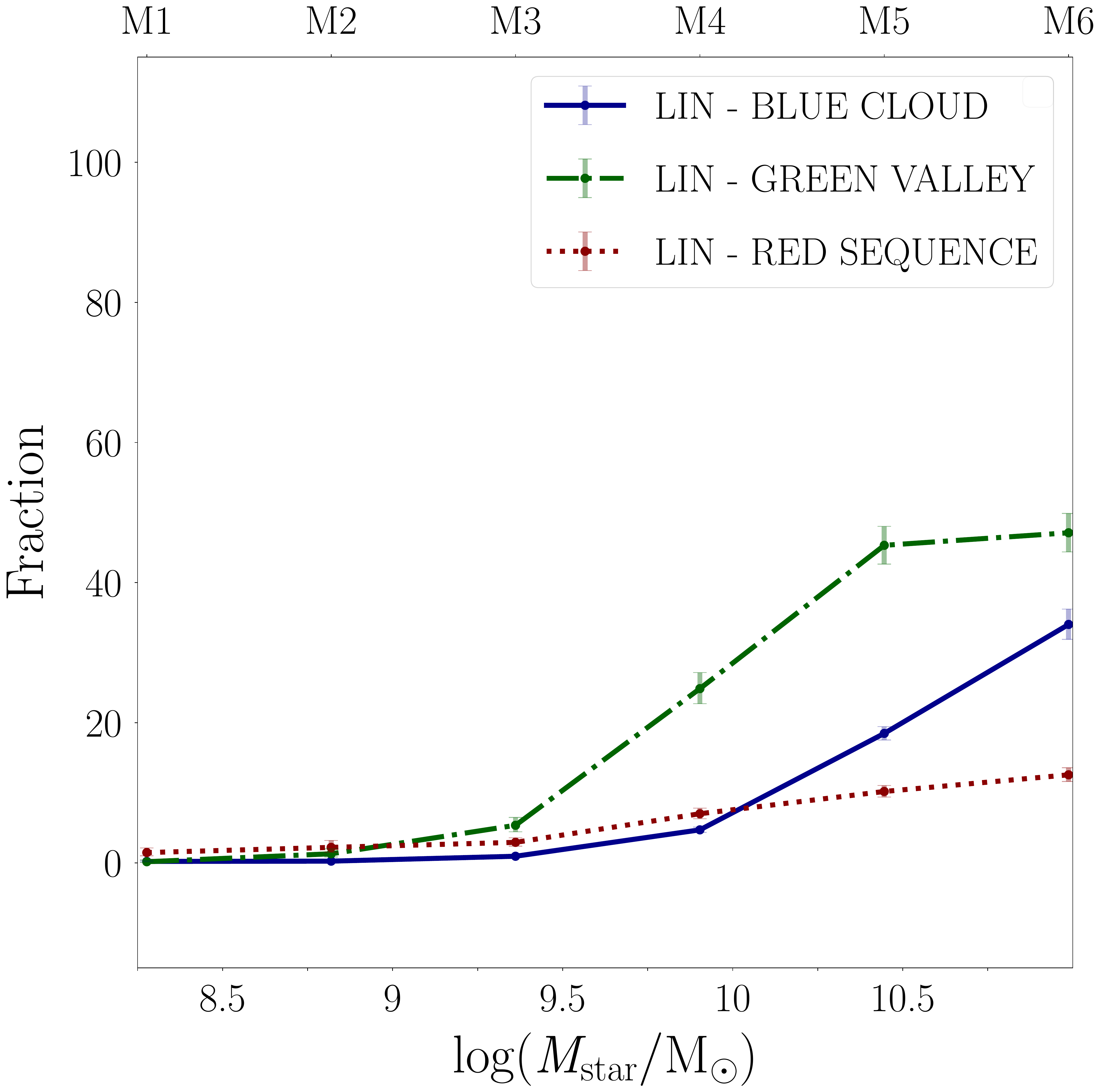}
\includegraphics[width=0.24\textwidth]{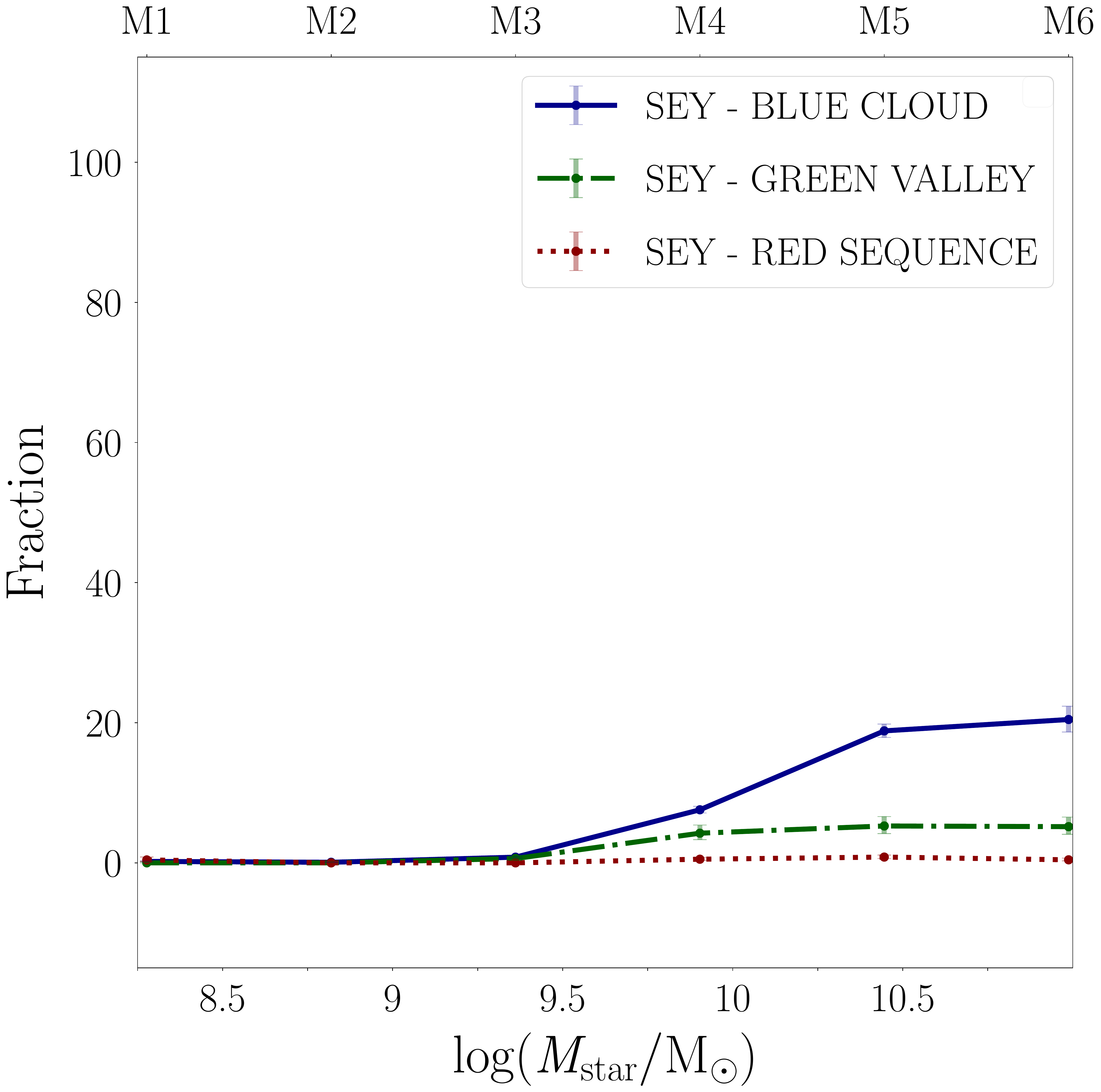}\\
\end{center}
\caption{The WHAN-computed frequencies (criteria of \cite{Gavazzi-2011}) of star-forming (SF), AGN (SEY), LINER (LIN) and retired (RET) as a function of $\log ( M_{\mathrm{star}}/\mathrm{M}_{\odot})$ (represented by the six bins M1, ..., M6) in the three chromatic regimes (blue cloud, green valley, red sequence, see Sec. $\S$ \ref{sec:photometry}).}
\label{fig:whan_nuclear_bins1}
\end{figure*}

\begin{figure*}
\begin{center} 
\includegraphics[width=0.24\textwidth]{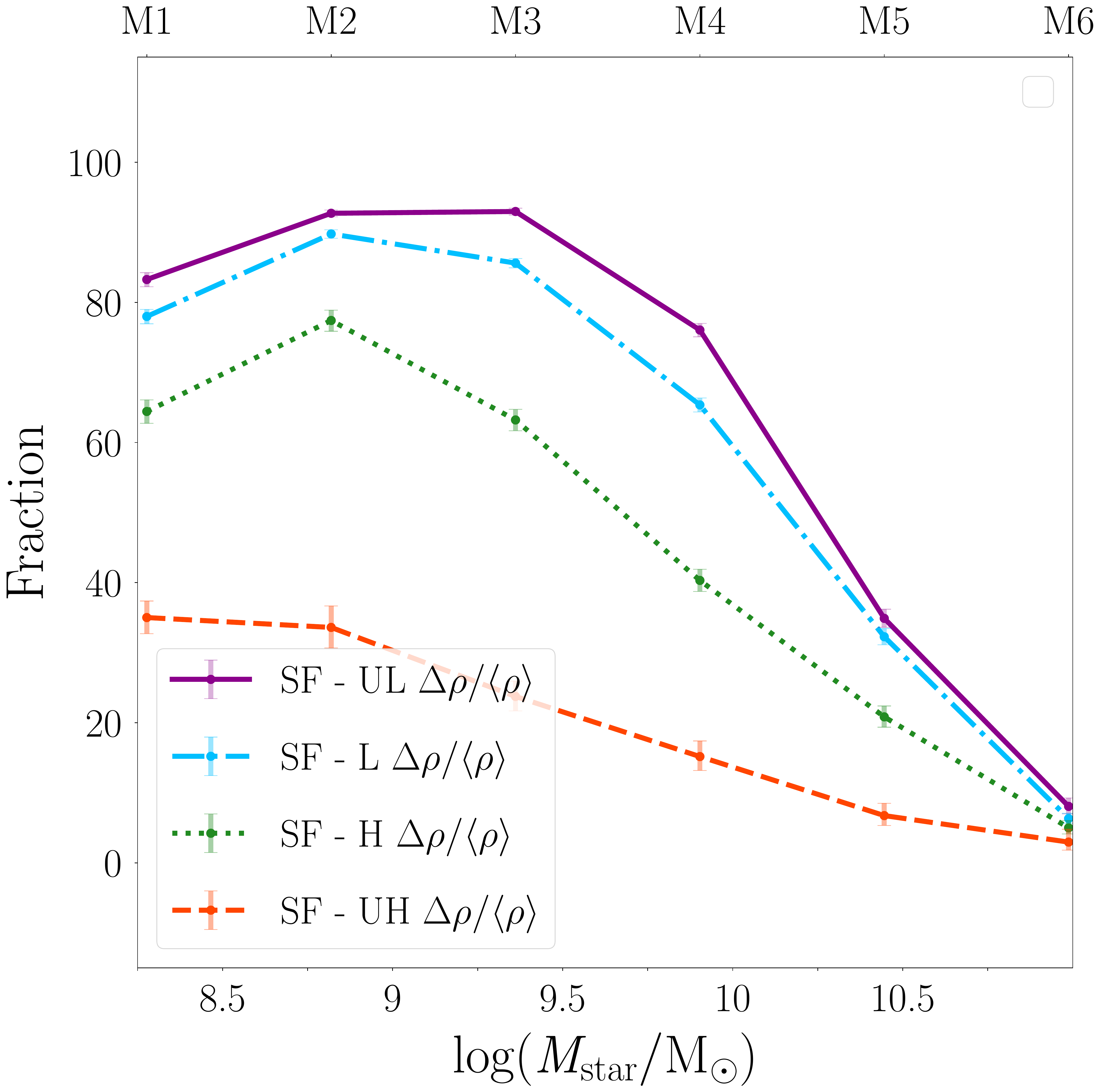}
\includegraphics[width=0.24\textwidth]{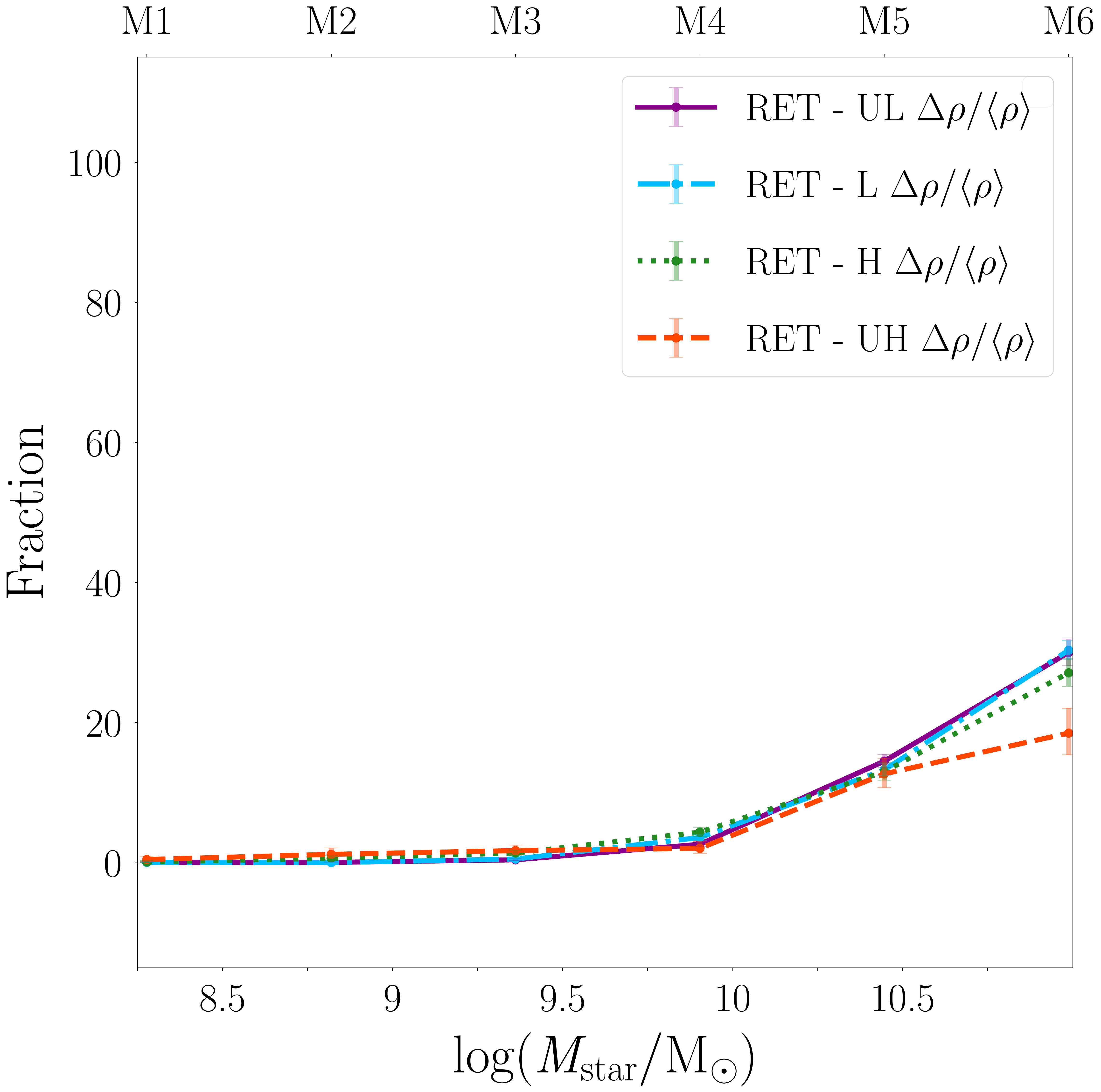}
\includegraphics[width=0.24\textwidth]{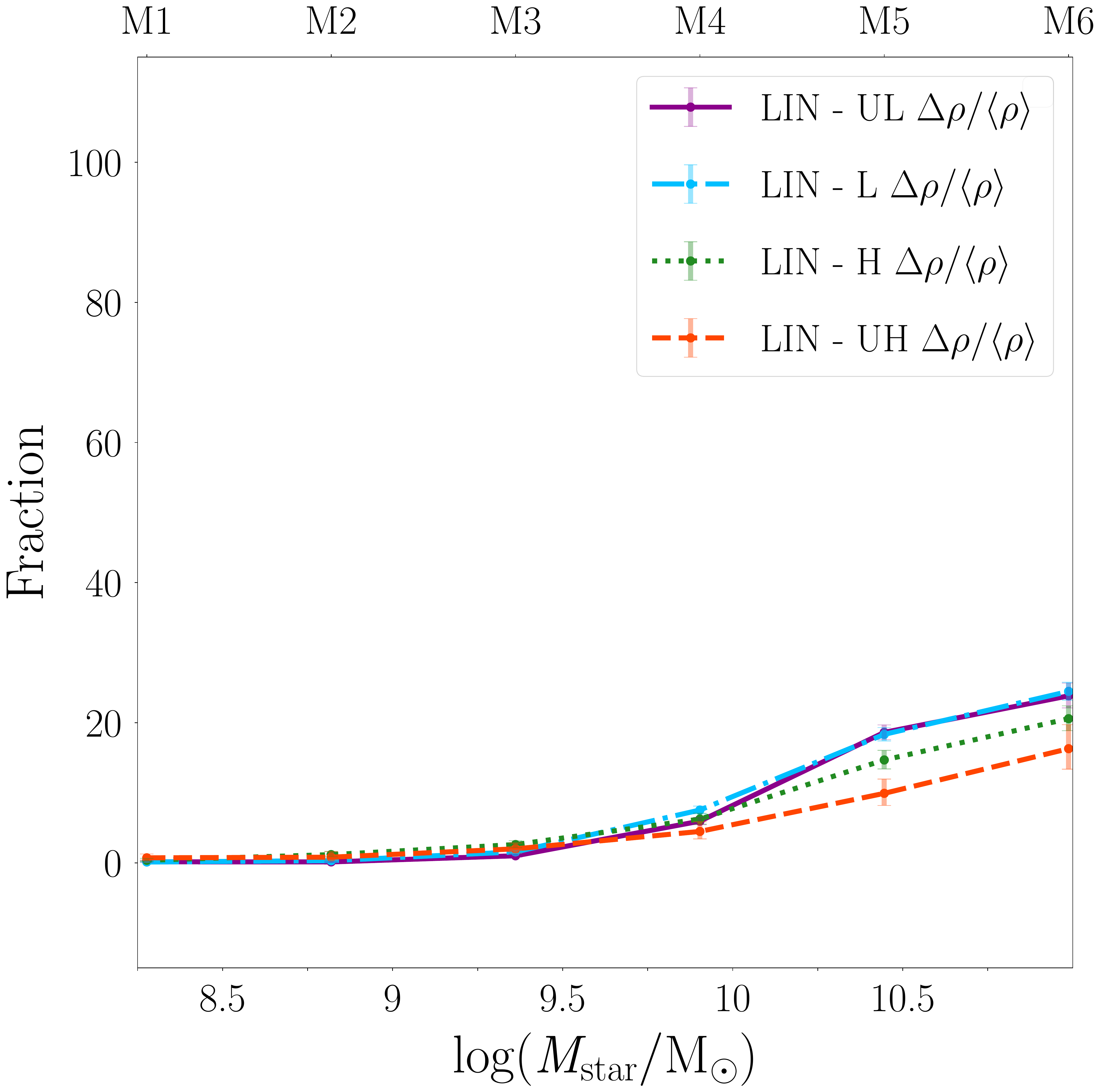}
\includegraphics[width=0.24\textwidth]{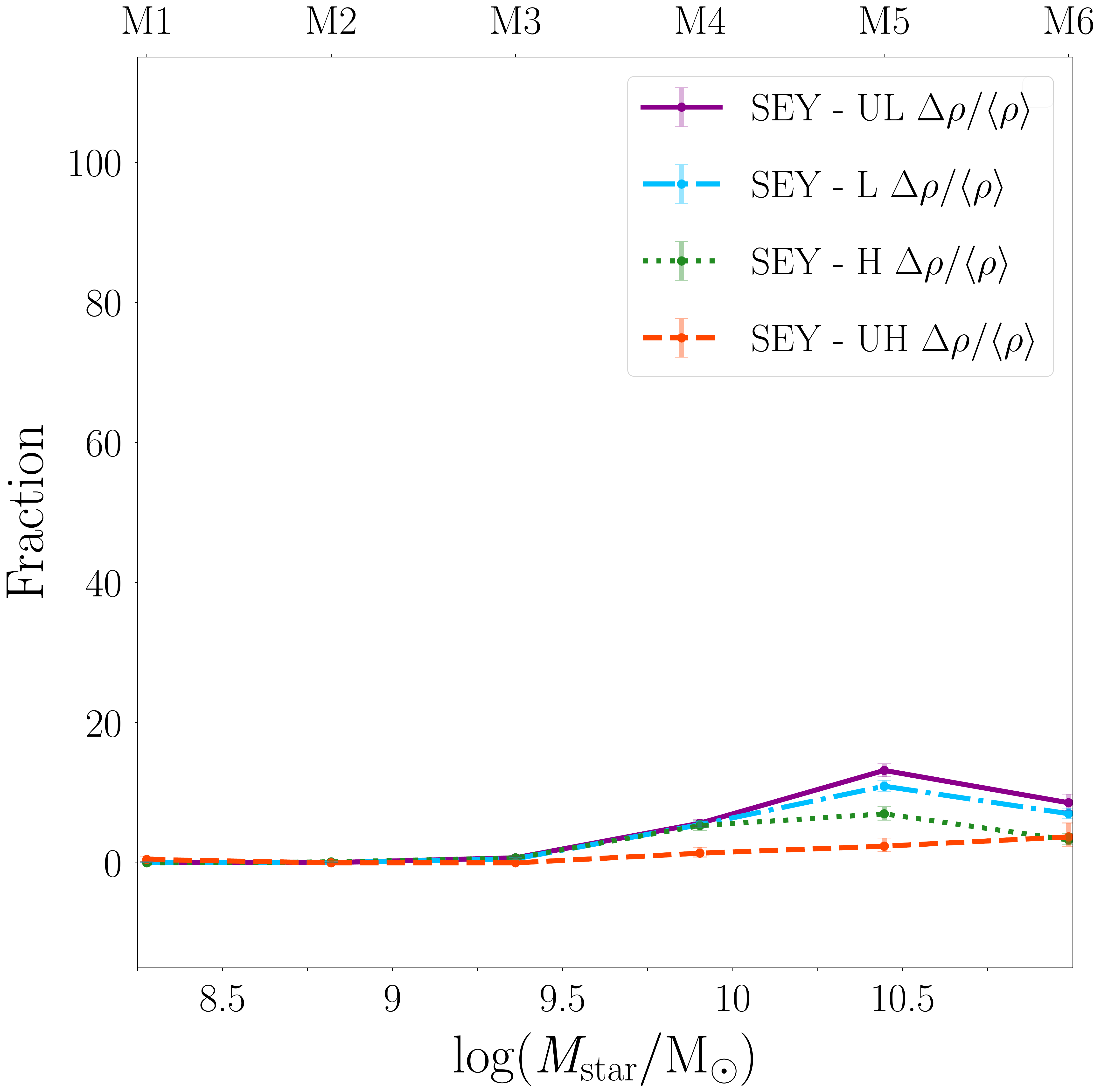}\\
\end{center}
\caption{The WHAN-computed frequencies (criteria of \cite{Gavazzi-2011}) of star-forming (SF), AGN (SEY), LINER (LIN) and retired (RET) as a function of $\log ( M_{\mathrm{star}}/\mathrm{M}_{\odot})$ (represented by the six bins M1, ..., M6) in the four overdensity bins defined in Sec. $\S$ \ref{subsec:overdensity}).}
\label{fig:whan_nuclear_bins2}
\end{figure*}

\begin{figure*}
\begin{center} 
\includegraphics[width=0.24\textwidth]{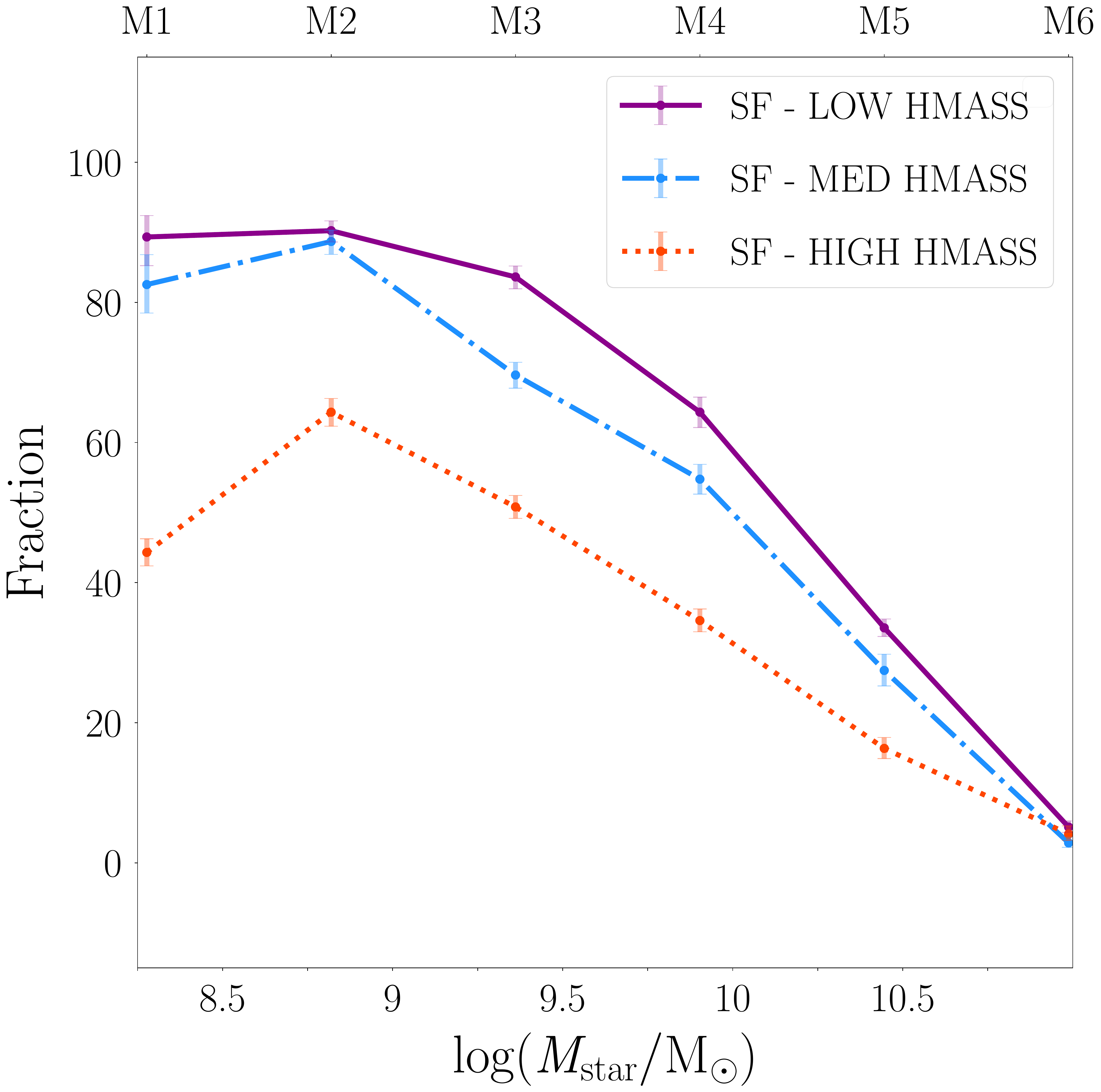}
\includegraphics[width=0.24\textwidth]{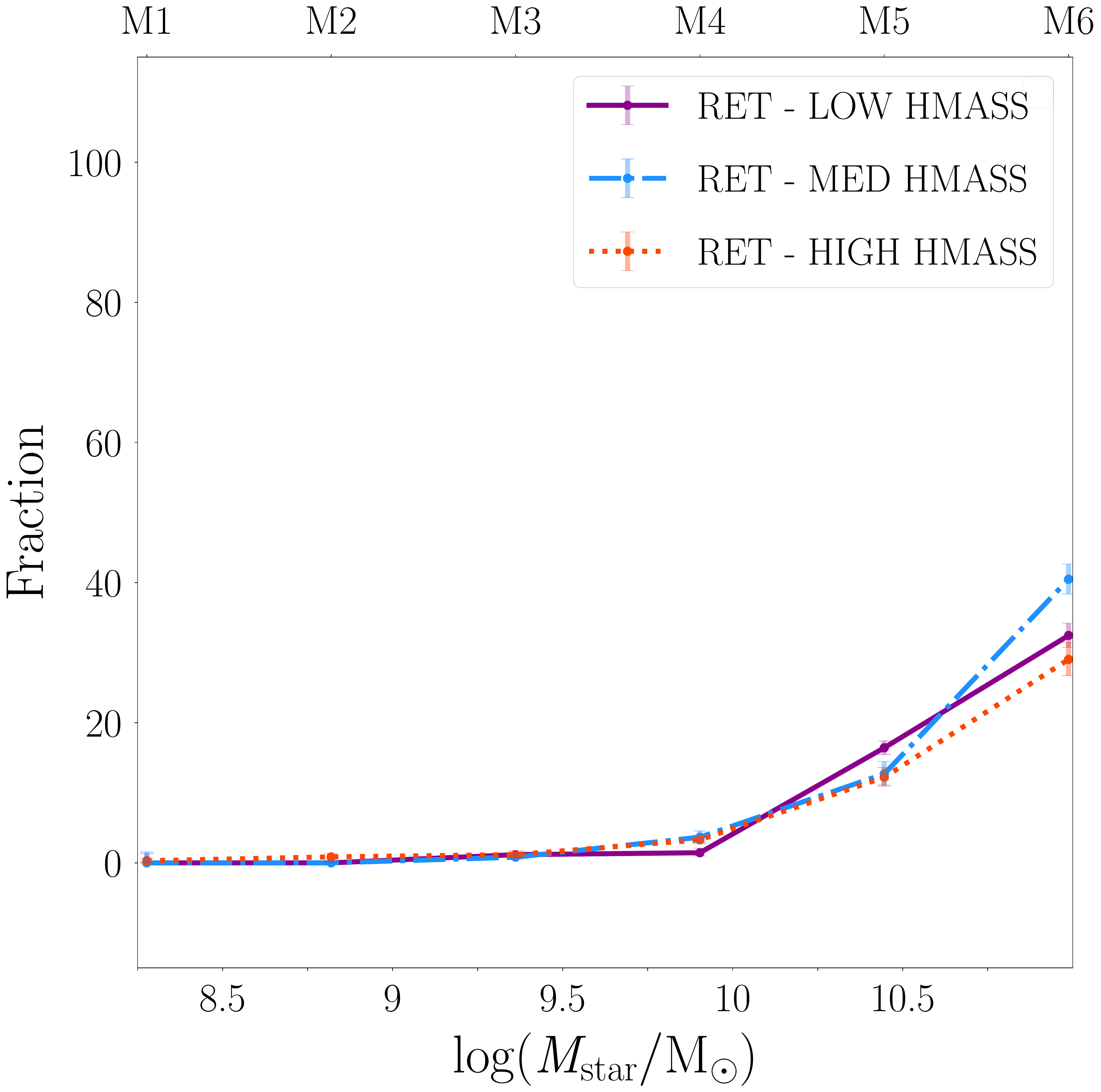}
\includegraphics[width=0.24\textwidth]{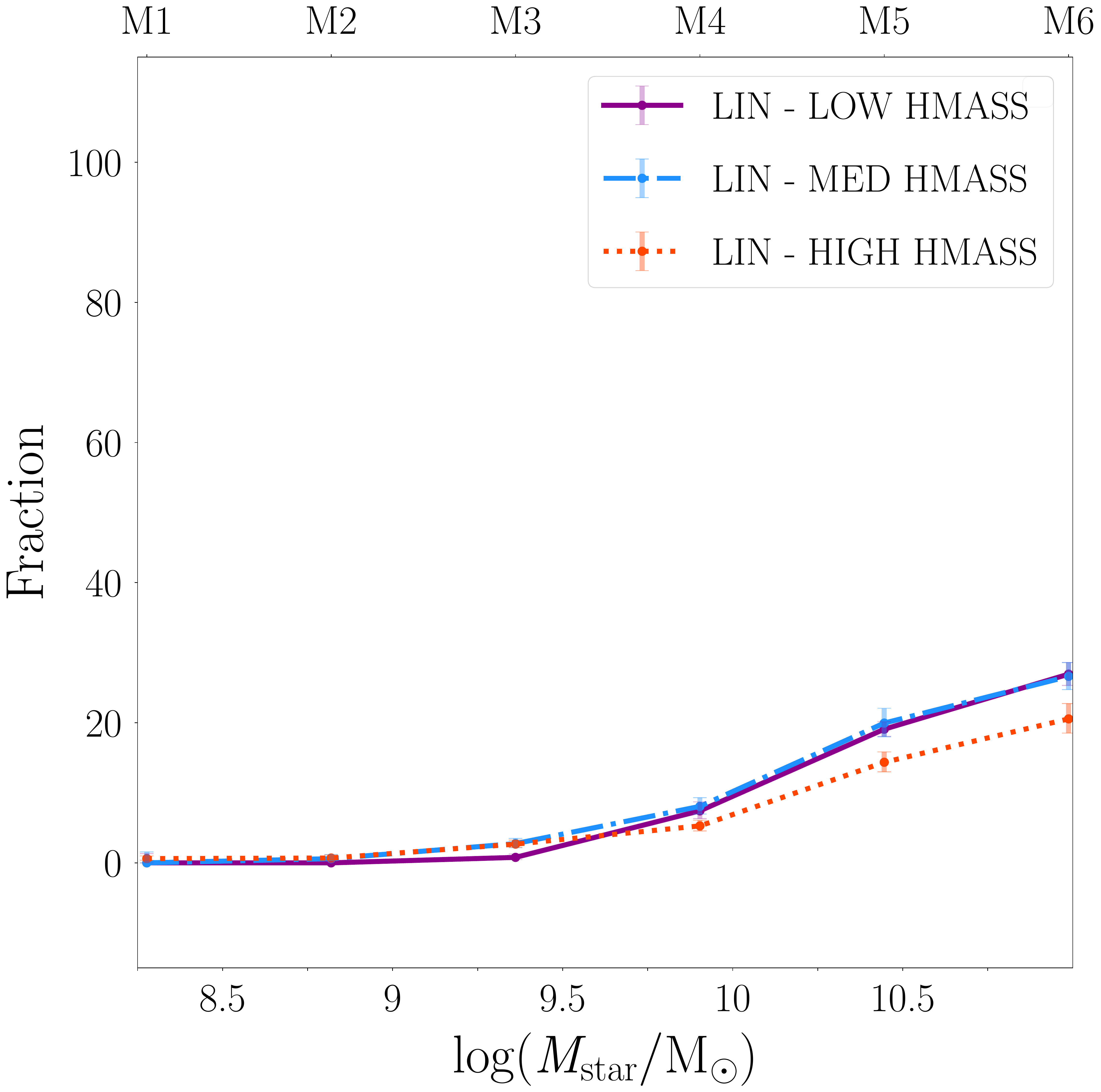}
\includegraphics[width=0.24\textwidth]{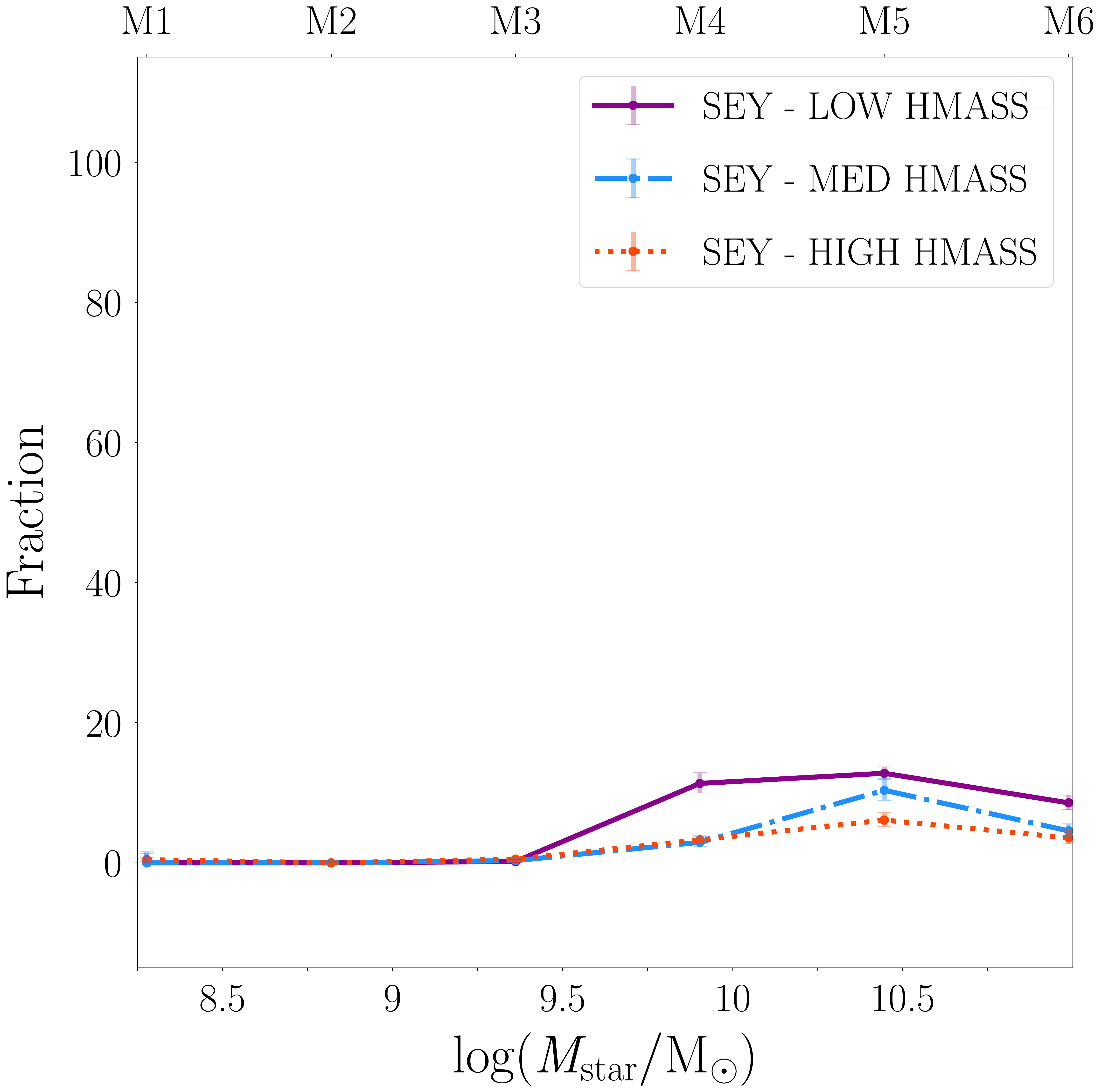}\\
\end{center}
\caption{The WHAN-computed frequencies (criteria of \cite{Gavazzi-2011}) of star-forming (SF), AGN (SEY), LINER (LIN) and retired (RET) as a function of $\log ( M_{\mathrm{star}}/\mathrm{M}_{\odot})$ (represented by the six bins M1, ..., M6) in the three halo-mass bins defined in Sec. $\S$ \ref{subsec:halomass}).}
\label{fig:whan_nuclear_bins3}
\end{figure*}

To verify if the incidence of the nuclear excitation properties is a general feature of high/low-mass galaxies or, conversely, depends on other environmental features, we replicate Fig. \ref{fig:FIG5ALL} dividing the sample according to $(i)$ the galaxy chromatic class (RS, BC, and GV, Sec. $\S$\ref{sec:photometry}), $(ii)$ the overdensity parameter (Sec. $\S$\ref{subsec:overdensity}) and $(iii)$ the halo-mass (Sec. $\S$\ref{subsec:halomass}). Our results are reported in Figures \ref{fig:nuclear_bins1}-\ref{fig:whan_nuclear_bins3} and summarised in the following:
\begin{itemize}
    \item (Star-forming): In general, the frequency of SF nuclei is highest in the lower-mass bins M1 - M3 ($M_{\mathrm{star}}\leq10^{9.5} \ \mathrm{M}_{\odot}$), and decreases as the stellar-mass increases. For a fixed bin of stellar-mass, it diminishes migrating from isolated galaxies to galaxies in clusters,
    indicating that star-forming systems live in relatively lower-density environments \cite[see also][]{Constantin-2006}.
    A similar decrease is observed as we increase the halo-mass or we move from the blue cloud (BC) to the red sequence (RS). For example, in the M5 bin we find that the BPT-computed SF nuclei are $\sim$45\% in the BC, $\sim$10\% in the GV, and $\sim$5\% in the RS (see Fig. \ref{fig:whan_nuclear_bins1}, left panel).\\
    The BPT and WHAN classifications for SF nuclei are in good agreement.
    \item (LINER): The frequency of LINERs computed with the BPT diagram results little sensitive to the galaxy overdensity and halo-mass (Figs. \ref{fig:nuclear_bins2}-\ref{fig:nuclear_bins3} and \ref{fig:whan_nuclear_bins2}-\ref{fig:whan_nuclear_bins3}). It exhibits the same strong dependence on the stellar-mass in each overdensity and halo-mass bin, as it is consistent with $\sim$0\% for $M_{\mathrm{star}} < 10^{10} \ \mathrm{M}_{\odot}$, and increases up to $\sim$30\%-40\% for $M_{\mathrm{star}} > 10^{10.5} \ \mathrm{M}_{\odot}$. We find a dependence of the LIN fraction on the chromatic class in the mass range $M_{\mathrm{star}} > 10^{10} \ \mathrm{M}_{\odot}$, where it is $\sim$20\% higher in the GV than in the BC and RS (Fig. \ref{fig:nuclear_bins1}, third panel), whereas LIN fractions in the BC and RS are consistent with each other across the whole mass range. \\
    Conversely, the LIN fraction computed with the WHAN scheme exhibits mild overdensity and halo-mass dependence for $M_{\mathrm{star}} > 10^{9.5} \ \mathrm{M}_{\odot}$, increasing by $\sim$5\%-10\% as galaxies move from high to low overdensity/halo-mass. Consistent with the BPT-computed fraction, we find a similar chromatic dependence of the LIN frequency on the chromatic class in the mass range $M_{\mathrm{star}} > 10^{9.5} \ \mathrm{M}_{\odot}$, where galaxies in the GV show a steep increase up to $\sim$50\% in the M5 and M6 bins; the fractions of LIN galaxies in the BC and RS are consistent with each other up to $M_{\mathrm{star}} \sim 10^{9.5} \ \mathrm{M}_{\odot}$; in the M5-M6 bins, the LIN fraction in the BC increases to $\sim$30\%, whereas the RS fraction settles to less than 10\% (Fig. \ref{fig:whan_nuclear_bins1}, third panel).
    \item (Seyfert): Unlike the frequency of LINERs, which is found to be increasing with stellar-mass up to 40-50\%, the BPT-computed fraction of Seyfert galaxies shows a much milder increase with the stellar-mass and, in general, reaches its maximum ($\sim$5\%) in the M5 bin ($10^{10} < M_{\mathrm{star}}/\mathrm{M}_{\odot} < 10^{10.5}$). We find a similar dependence of the SEY fraction on the overdensity and halo-mass indicators (\textit{i.e.}, it decreases moving from high to low overdensity/halo-mass, see Figs. \ref{fig:nuclear_bins2}-\ref{fig:nuclear_bins3}, right panels). Conversely, it does not show considerable dependence on the chromatic class. \\
    The WHAN-computed SEY fraction displays significant variations with respect to the BPT counterpart. In general, it is consistent with $\sim$0\% for $M_{\mathrm{star}} < 10^{9.5} \ \mathrm{M}_{\odot}$, and increases for higher masses, reaching a peak in the M5 bin which rises up to $\sim$15\% as we move from high to low overdensity/halo-mass galaxies. These results are consistent with \cite{Constantin-2006}, who observed that Seyfert galaxies reside in less massive halos than those that host LINERs (see Figs. \ref{fig:nuclear_bins3} and \ref{fig:whan_nuclear_bins3}).
    The fraction of SEY among galaxies in the BC furtherly increases for masses $M_{\mathrm{star}} > 10^{10.5} \ \mathrm{M}_{\odot}$ reaching $\sim$18\% and $\sim$25\% in bins M5 and M6, respectively.
    \item (Composite): As the LINERs frequency, the frequency of composite (Comp) galaxies reaches its maximum in the highest-stellar-mass bins M5 or M6 and shows a mild dependence on the galaxy overdensity and the halo-mass; \textit{i.e.}, the fraction of composite nuclei in the M4-M6 bins moderately increases migrating from galaxies in clusters (UH) to isolated (UL) galaxies; similarly, the Comp fraction in the higher stellar-mass range is $\sim5\%$ higher in lower halo-mass galaxies (Fig. \ref{fig:nuclear_bins3}, second panel). A higher fraction of Comp galaxies is found in galaxies belonging to the GV for all but one (M6) mass bins (Fig. \ref{fig:nuclear_bins1}, second panel). The Comp frequency in the GV reaches its maximum ($\sim$40\%) in the M5 bin.
    \item (Retired): The fractions of RET in the three overdensity and four halo-mass bins display similar trends, steadily increasing for $M_{\mathrm{star}} > 10^{9.5} \ \mathrm{M}_{\odot}$ up to $\sim$30\% in the M5 and M6 bins and showing little dependence on the overdensity and the halo-mass;  For $10^9 <M_{\mathrm{star}}\mathrm{M}_{\odot} < 10^{9.5}$, it is is few percent higher in the RS than in the BC/GV. For masses $M_{\mathrm{star}} > 10^{9.5} \ \mathrm{M}_{\odot}$, it follows similar trends for RS and GV galaxies and is about $\sim$15\% lower for galaxies in the BC.
\end{itemize}
Again, the disparities in the features of non-star-forming nuclei may be attributed to deviations among the BPT and WHAN classifications (Tables \ref{tab:BPT_WHAN}-\ref{tab:WHAN_BPT}). In general, our results show that one of the most (if not the most) important factor that affects the triggering of nuclear activity is the galaxy stellar-mass \cite[][]{Kauffmann-2003}. The dependence of the prevalence of nuclear activity on the local galaxy density and halo-mass appears to indicate that the fraction of Seyfert nuclei decreases towards the highest-density regions and highest halo-masses \cite[see also][]{Sabater-2013, Sabater-2015, Li-2019, Lopes-2017}, as well as the fraction of star-forming nuclei. The fraction of LINERs displays minor dependencies on the galaxy density and halo-mass, but appears to be related to the galaxy colour in the mass range $M_{\mathrm{star}} > 10^{10} \ \mathrm{M}_{\odot}$, where it is $\sim$20\% higher in the green valley.
\section{Summary}\label{conclusion}
In this work we present the SPRING catalogue, a complete photometric database ($ r<17.7$ mag) of local galaxies in the Northern spring sky ($10^h<\mathrm{RA}<16^h$; $0^{\circ}<\mathrm{Dec}<65^{\circ}$, $cz<10000 \ \mathrm{km \ s}^{-1}$) by merging the Portsmouth emission-line table by \cite{Thomas-2013} extracted from the SDSS and the NASA-Sloan atlas \cite[][]{Blanton-2011}. To further reduce the bright-end incompleteness, we complement our sample with the updated Zwicky catalogue (UZC) \cite{Zwicky-1961, Falco-1999}, and with the UGC catalogue \cite{Nilson-1973}.
We get the available nuclear spectroscopy in the SDSS, and integrate it with optical spectra taken by other surveys \cite[][]{Zwicky-1961, Nilson-1973, Ho-1993, Ho-1995, Ho-1997, Falco-1999, Gavazzi-2011, Gavazzi-2013}, and with the spectroscopy obtained by us in 2014-2020 using the Cassini 1.5m telescope of the Loiano Observatory. The SPRING catalogue includes 891 galaxies with optical spectra observed independently from SDSS.
We claim that the resulting sample of 30597 targets constitutes the cleanest and most complete catalogue of galaxies available so far for the Northern spring sky limited to $ r<17.7$ mag and $cz<10000 \ \mathrm{km \ s}^{-1}$.
The construction of such a sample constitutes the first goal of the present work (Sec. $\S$ \ref{sample}).

We study the extensive properties (colours, stellar-mass, H\thinspace{\scriptsize I} content) of galaxies in such a sample, confirming that galaxy colour ($\mathrm{NUV}-i$) is a steep function of stellar-mass, with the red sequence well separated from the blue cloud by the green valley \cite[][Sec. $\S$ \ref{sec:photometry}]{Boselli-2014}. 

We perform a recalibration of the optical diameter-based H\thinspace{\scriptsize I}-deficiency parameter employing a reference sample of 3416 isolated galaxies extracted from our catalogue, and find that the slope of the $\log(M(\mathrm{H} \thinspace \scriptsize{\text{I}}))$ vs $\log(D^2)$ relation is strongly dependent on the galaxy stellar-mass (Sec. $\S$ \ref{subsec:HIdef}).
We observe that most green-valley objects are consistent with galaxies with deficient H\thinspace{\scriptsize I} content, stressing the role of gas in fuelling (and quenching) star-formation, as claimed by \cite{Gavazzi-2013Ha3_b}.

Remarkably, we find that a filamentary structure of galaxies extending in the spring sky at intermediate distances between Virgo and Coma contains $\sim$30.7\% of H\thinspace{\scriptsize I}-deficient galaxies, which significantly exceeds the average H\thinspace{\scriptsize I}-deficiency fraction ($\sim$12\%) of field galaxies. To probe the relevance of the deficiency of H\thinspace{\scriptsize I} in filaments, we construct ten thousand overdensity-matched samples and investigate their H\thinspace{\scriptsize I} content. We find that the average fraction of H\thinspace{\scriptsize I}-deficient galaxies in the matched-samples is (20.2 $\pm$ 1.6)\%. This could point out that galaxies in filaments are more H\thinspace{\scriptsize I}-deficient than field galaxies.

The nuclear classification of galaxies in our sample (Sec. $\S$ \ref{sec:results}) is based on the four-lines BPT and the two-lines WHAN diagnostic diagrams.
We confirm that the fraction of AGN (especially LINERs) over SF nuclei is a steep function of stellar-mass \cite[][]{Kauffmann-2003}, \textit{i.e.}, the fraction of AGN (Seyfert and LINERs) is consistent with zero up to $M_{\mathrm{star}}< 10^{9.5}$ M$_{\odot}$, and becomes $\sim$ 40 \% for $M_{\mathrm{star}} > 10^{10.5}$ M$_{\odot}$.
A general agreement between the BPT and WHAN classifications is observed. We investigate the mass-dependency of the excitation properties determined with both diagnostics, and find that star-forming nuclei, Seyfert and LINERs display similar trends (Sec. $\S$ \ref{sec:bpt}-\ref{sec:whan}). 
In addition, we investigate whether the nuclear-excitation fractions depend predominantly on the stellar-mass or on other environmental related indicators (Sec. $\S$ \ref{subsec:stmass-nuclear}). To this end, we examine the connection between the mass-dependence of the galaxy excitation properties and their environment, which we determine using three different indicators, \textit{i.e.}, \textit{(i)} a chromatic criterion, \textit{(ii)} the local galaxy density, and \textit{(iii)} the halo-mass of the group to which galaxies belong.
In general, we observe that the mass-dependency of the fraction of Seyfert nuclei is little sensitive to the galaxy environment (with noticeable variations in the SEY percentage depending on the adopted diagnostic scheme), while the fraction of star-forming nuclei is a steeper function of stellar-mass in lower-density environments and in blue-cloud galaxies. The fraction of LINERs exhibits a dependence on the galaxy colour and, for $M_{\rm star} \gtrsim 10^{9.5-10}$ M$_{\odot}$, is significantly larger in galaxies belonging to the green valley.

\section{Discussion}

The analysis presented in this work \textendash based on a statistically-significant sample\textendash \ points out that the main parameter driving the level of nuclear activity is the galaxy stellar-mass \cite[][]{Kauffmann-2003, Decarli-2007, Juneau-2011, RodriguezDP-2017} and shows that the fraction of AGN does not vary significantly with the galaxy density, the mass of the hosting halo, or the colour of galaxies (see Sec. \S \ref{subsec:stmass-nuclear}). This result overall agrees with the results of \cite{Kauffmann-2004} based on a larger SDSS optically-selected sample of galaxies, and  with \cite{Constantin-2006}. They also agree with the results gathered by other teams which analyse optically-  and X-rays-selected clusters \cite[][]{Vollmer-2001, Vollmer-2004, Vollmer-2008, Steyrleithner-2020}.

Interesting is the fact that the highest density regions in our sample are those within well known clusters of galaxies such as Coma, A1367, and Virgo. Detailed multi-frequency studies of these clusters clearly indicate that the dominant perturbing mechanism affecting the evolution of the infalling, gas-rich spiral population is ram-pressure. This mechanism has been identified without ambiguity by the presence of long cometary tails in H\thinspace{\scriptsize I} \cite[][]{Gavazzi-1989, Dickey-1991, Bravo-Alfaro-2001, Scott-2010, Scott-2012, Chung-2007}, radio continuum \cite[][]{Gavazzi-1978, Gavazzi-Jaffe-1985, Gavazzi-Jaffe-1987, Gavazzi-1995, Miller-2009, Chen-2020, Lal-2020, Vollmer-2004, Vollmer-2010, Crowl-2005, Kantharia-2008}, and H$\alpha$ \cite[][]{Yoshida-2002, Boselli-2018a, Fossati-2018, Yagi-2007, Yagi-2010, Yagi-2017}, by detailed multifrequency studies of large statistical samples \cite[\textit{e.g.},][]{Vollmer-2001, Boselli-2014, Gavazzi-2012Ha3}, and of individual objects \cite[][]{Boselli-2016, Kenney-1999, Kenney-2004, Kenney-2008, Vollmer-2003, Vollmer-2004, Vollmer-2008}.

A relation between the incidence rate of galaxies hosting an AGN and the  density of the environment in which they reside has been looked for in  several statistical studies, since it is expected in different evolutionary scenarios \cite[\textit{e.g.},][]{Kauffmann-2004}.
In a starvation scenario \cite[\textit{e.g.},][]{Larson-1980}, in which the hot halo surrounding galaxies is removed once galaxies become satellites of larger structures, the feedback mechanism induced by a strong nuclear activity easily removes most of the cold-gas component host on the galaxy disc as indicated by the most recent hydrodynamical  simulations \cite[][]{Stevens-2021, Tonnesen-Bryan-2009, Boselli-2021}. For this reason, the activity of star formation of the host galaxies is strongly reduced, producing quiescent systems.
Recently, it has also been claimed that the loss of  angular momentum which follows a ram-pressure-stripping event might induce  gas infall in the galaxy centre, feeding nuclear activity \cite[\textit{e.g.},][]{Schulz-Struck-2001, Tonnesen-Bryan-2009, Poggianti-2017, Martinez-2018}.

Our analysis of the nuclear properties of galaxies in different environments and stellar-mass ranges shows that the incidence of Seyfert nuclei in galaxies within the core of a cluster (UH overdensity) is minimum (Figs. \ref{fig:nuclear_bins2}-\ref{fig:whan_nuclear_bins2}). This result agrees with a number of recent works aimed at studying the incidence of strong AGN in different
environments \cite[][]{Sabater-2013, Sabater-2015, Li-2019, Lopes-2017}. Conversely, the incidence of nuclei with BPT-classified LINER-like emission-line ratios displays no dependence on the local galaxy overdensity (Fig. \ref{fig:nuclear_bins2}, third panel), but is strongly connected with the galaxy stellar-mass and reaches $\sim40\%$ for $M_{\mathrm{star}} > 10^{10.5} \ \mathrm{M}_{\odot}$ (similar results yield for LINERs classified by the WHAN diagnostic).
This evidence is at odds with the results of \cite{Poggianti-2017}, who found an enhanced AGN (Seyfert) fraction in a sample of ram-pressure-stripped (jellyfish) objects (5/7). However, such results were not confirmed by the follow-up analysis of \cite{Peluso-2022} on a larger selection of jellyfish galaxies. 
The recent analysis by \cite{Roman-Oliveira-2019} of a sample of ram-pressure-stripped galaxies did not find a high incidence of AGN (5/70). The disparities between these studies may be attributed to the different nuclear-activity classification schemes which they adopt. \cite{Poggianti-2017} and \cite{Peluso-2022} use the BPT diagnostic, while the classification of the sample analysed in \cite{Roman-Oliveira-2019} employs the WHAN (see also \cite{RDP-2017}). As we point out in Sec. \S \ref{sec:whan}, the two diagrams display noticeable deviations in the classification of non-star-forming nuclei (\textit{e.g.}, among the galaxies classified as Seyfert by the BPT, only $\sim16\%$ are classified as SEY by the WHAN). 

Another potential element of discrepancy among different analyses of the nuclear activity vs environment relation is the characterization of AGN. \cite{Poggianti-2017} denotes as AGN those galaxies which are classified as Seyfert by the BPT diagram; \cite{Peluso-2022} classifies a galaxy as an AGN if its nuclear region has a Seyfert or LINER BPT-classification; similarly, \cite{RDP-2017} labels as AGN those galaxies which are classified as Seyfert or LINERs by the WHAN.
As we see in Figs. \ref{fig:nuclear_bins1}-\ref{fig:nuclear_bins3}, active nuclei classified with the BPT diagnostic diagram are dominated by the LINERs, suggesting that more lenient AGN definitions are biased toward LINER-like behavior. LINERs differ from Seyferts in many of their characteristics; also, the ionisation source of LINER-like objects is still uncertain. Thus, caution should be exercised when investigating the interplay between the galaxy environmental properties (\textit{e.g.}, being subject to ram-pressure-stripping) and the triggering of nuclear activity.

As extensively discussed in \cite{Boselli-2022-REV}, the results of \cite{Poggianti-2017} may be due to the fact that their sample of 7 galaxies was composed of very massive objects ($M_{\mathrm{star}} > 4 \times 10^{10} \ \mathrm{M}_{\odot}$) which are more likely to host an active nucleus.
The sample examined by \cite{Poggianti-2017} has been further expanded in \cite{Peluso-2022}, who considered a total of 51 galaxies observed in the context of the GASP survey \cite{Poggianti-2017-GASP} along with additional 82 galaxies retrieved from the literature and found that the fraction of galaxies which host a AGN (Seyfert or LINER) is $\sim27\%$ at masses $M_{\mathrm{star}} > 10^9 \ \mathrm{M}_{\odot}$, and $\sim 51\%$ at masses $M_{\mathrm{star}} > 10^{10} \ \mathrm{M}_{\odot}$. The majority of objects considered by \cite{Peluso-2022} are high-mass galaxies (all galaxies hosting a AGN in GASP have $M_{\mathrm{star}} > 10^{10.5} \ \mathrm{M}_{\odot}$). 
In contrast to \cite{Poggianti-2017}, the AGN fractions considered by \cite{Peluso-2022} include both Seyfert and LINERs. Also, the AGN fractions are calculated with respect to a sample of emission-line galaxies only, introducing possible bias. For comparison, the BPT-computed SEY+LIN fraction $f_{\mathrm{(SEY+LIN)}}/f_{\mathrm{ELG}}$ of SPRING galaxies in the M6 bin (\textit{i.e.}, with stellar-mass  $M_{\mathrm{star}} > 10^{10.5} \ \mathrm{M}_{\odot}$) calculated over all emission-line galaxies in the same stellar-mass bin is $\sim58.2\%$, whereas the fraction $f_{\mathrm{(SEY+LIN)}}/f_{\mathrm{TOT}}$ in the same mass range (where $f_{\mathrm{TOT}}$ accounts for both emission-line and passive galaxies) is $\sim37\%$. In general, our analysis suggests that the investigation of the occurrence of AGN in different galaxy environments requires care, as one needs to take into account possible biases that may arise due to \textit{(i)} the particular diagnostic adopted (BPT vs WHAN vs other diagnostics), \textit{(ii)} the characterization of AGN (are LINERs true AGN?), and \textit{(iii)} the sample which defines the underlying galaxy population.

Finally, our analysis suggests that the incidence of LINERs is higher in galaxies belonging to the green-valley (Figs. \ref{fig:nuclear_bins1} and \ref{fig:whan_nuclear_bins1}).
In a future companion paper, we aim to further investigate the nuclear activity vs environment relation of a sample of SPRING galaxies which are subject to ram-pressure-stripping in the cluster environment.

\begin{acknowledgements}
We are grateful to Paolo Franzetti and Alessandro Donati for their contribution to GoldMine, the Galaxy On Line Database \cite[][]{Gavazzi-2003} extensively used in this work (\url{http://goldmine.mib.infn.it}). We thank Andrea Macciò for useful discussion.
We also thank the coordinator of the TAC, Giovannna Stirpe for the generous time allocation at Loiano.
This research has made use of the NASA/IPAC Extragalactic Database (NED), which is operated by the Jet Propulsion Laboratory, California Institute of Technology, under contract with the National Aeronautics and Space Administration. 
The present study made extensive use of SDSS. Funding for the Sloan Digital Sky Survey (SDSS) and SDSS-II has been provided by the Alfred P. Sloan Foundation, the Participating Institutions, the National Science Foundation, the U.S. Department of Energy, the National Aeronautics and Space Administration, the Japanese Monbukagakusho, and the Max Planck Society, and the Higher Education Funding Council for England. 
The SDSS Web site is http://www.sdss.org/.
The SDSS is managed by the Astrophysical Research Consortium (ARC) for the Participating Institutions. 
The Participating Institutions are the American Museum of Natural History, Astrophysical Institute Potsdam,  University of Basel, University of Cambridge, Case Western Reserve University, The University of Chicago, Drexel University, Fermilab, the Institute for Advanced Study, the Japan Participation Group, the Johns Hopkins University, the Joint Institute for Nuclear Astrophysics, the Kavli Institute for Particle Astrophysics and Cosmology, the Korean Scientist Group, the Chinese Academy of Sciences (LAMOST), the Los Alamos National Laboratory, the Max-Planck-Institute for Astronomy (MPIA), the Max-Planck-Institute for Astrophysics (MPA), New Mexico State University, Ohio State University, University of Pittsburgh, the University of Portsmouth, Princeton University, the United States Naval Observatory, and the University of Washington.
\end{acknowledgements}

\bibliography{biblio-spring} 

\begin{thebibliography}{158}
\expandafter\ifx\csname natexlab\endcsname\relax\def\natexlab#1{#1}\fi

\bibitem[{Agostino \& Salim(2019)}]{Agostino-2019}
Agostino, C.~J. \& Salim, S. 2019, ApJ, 876, 12

\bibitem[{{Baldwin} {et~al.}(1981){Baldwin}, {Phillips}, \&
  {Terlevich}}]{Baldwin-1981}
{Baldwin}, J.~A., {Phillips}, M.~M., \& {Terlevich}, R. 1981, \pasp, 93, 5

\bibitem[{{Ballo} {et~al.}(2012){Ballo}, {Heras}, {Barcons}, \&
  {Carrera}}]{Ballo-2012}
{Ballo}, L., {Heras}, F.~J.~H., {Barcons}, X., \& {Carrera}, F.~J. 2012, \aap,
  545, A66

\bibitem[{Bianchi {et~al.}(2017)Bianchi, Shiao, \& Thilker}]{Bianchi-2017}
Bianchi, L., Shiao, B., \& Thilker, D. 2017, The Astrophysical Journal
  Supplement Series, 230, 24

\bibitem[{{Binette} {et~al.}(1994){Binette}, {Magris}, {Stasi{\'n}ska}, \&
  {Bruzual}}]{Binette-1994}
{Binette}, L., {Magris}, C.~G., {Stasi{\'n}ska}, G., \& {Bruzual}, A.~G. 1994,
  \aap, 292, 13

\bibitem[{{Binggeli} {et~al.}(1985){Binggeli}, {Sandage}, \&
  {Tammann}}]{Binggeli-1985}
{Binggeli}, B., {Sandage}, A., \& {Tammann}, G.~A. 1985, \aj, 90, 1681

\bibitem[{Blanton {et~al.}(2001)Blanton, Dalcanton, Eisenstein, Loveday,
  Strauss, SubbaRao, Weinberg, John E.~Anderson, Annis, Bahcall, Bernardi,
  Brinkmann, Brunner, Burles, Carey, Castander, Connolly, Csabai, Doi,
  Finkbeiner, Friedman, Frieman, Fukugita, Gunn, Hennessy, Hindsley, Hogg,
  Ichikawa, Ivezi{\'{c}}, Kent, Knapp, Lamb, Leger, Long, Lupton, McKay,
  Meiksin, Merelli, Munn, Narayanan, Newcomb, Nichol, Okamura, Owen, Pier,
  Pope, Postman, Quinn, Rockosi, Schlegel, Schneider, Shimasaku, Siegmund,
  Smee, Snir, Stoughton, Stubbs, Szalay, Szokoly, Thakar, Tremonti, Tucker,
  Uomoto, Berk, Vogeley, Waddell, Yanny, Yasuda, \& York}]{Blanton-2001}
Blanton, M.~R., Dalcanton, J., Eisenstein, D., {et~al.} 2001, The Astronomical
  Journal, 121, 2358

\bibitem[{{Blanton} {et~al.}(2005{\natexlab{a}}){Blanton}, {Eisenstein},
  {Hogg}, {Schlegel}, \& {Brinkmann}}]{Blanton-2005a}
{Blanton}, M.~R., {Eisenstein}, D., {Hogg}, D.~W., {Schlegel}, D.~J., \&
  {Brinkmann}, J. 2005{\natexlab{a}}, \apj, 629, 143

\bibitem[{{Blanton} {et~al.}(2011){Blanton}, {Kazin}, {Muna}, {Weaver}, \&
  {Price-Whelan}}]{Blanton-2011}
{Blanton}, M.~R., {Kazin}, E., {Muna}, D., {Weaver}, B.~A., \& {Price-Whelan},
  A. 2011, \aj, 142, 31

\bibitem[{{Blanton} {et~al.}(2005{\natexlab{b}}){Blanton}, {Lupton},
  {Schlegel}, {Strauss}, {Brinkmann}, {Fukugita}, \& {Loveday}}]{Blanton-2005b}
{Blanton}, M.~R., {Lupton}, R.~H., {Schlegel}, D.~J., {et~al.}
  2005{\natexlab{b}}, \apj, 631, 208

\bibitem[{{Blanton} {et~al.}(2005{\natexlab{c}}){Blanton}, {Schlegel},
  {Strauss}, {Brinkmann}, {Finkbeiner}, {Fukugita}, {Gunn}, {Hogg},
  {Ivezi{\'c}}, {Knapp}, {Lupton}, {Munn}, {Schneider}, {Tegmark}, \&
  {Zehavi}}]{Blanton-2005c}
{Blanton}, M.~R., {Schlegel}, D.~J., {Strauss}, M.~A., {et~al.}
  2005{\natexlab{c}}, \aj, 129, 2562

\bibitem[{{Blanton} {et~al.}(2017)}]{Blanton-2017}
{Blanton}, M.~R. {et~al.} 2017, \aj, 154, 28

\bibitem[{Boselli {et~al.}(2009)Boselli, Boissier, Cortese, Buat, Hughes, \&
  Gavazzi}]{Boselli-2009}
Boselli, A., Boissier, S., Cortese, L., {et~al.} 2009, \apj, 706, 1527

\bibitem[{{Boselli} {et~al.}(2008){Boselli}, {Boissier}, {Cortese}, \&
  {Gavazzi}}]{Boselli-2008a}
{Boselli}, A., {Boissier}, S., {Cortese}, L., \& {Gavazzi}, G. 2008, \apj, 674,
  742

\bibitem[{{Boselli} {et~al.}(2016){Boselli}, {Cuillandre}, {Fossati},
  {Boissier}, {Bomans}, {Consolandi}, {Anselmi}, {Cortese}, {C{\^o}t{\'e}},
  {Durrell}, {Ferrarese}, {Fumagalli}, {Gavazzi}, {Gwyn}, {Hensler}, {Sun}, \&
  {Toloba}}]{Boselli-2016}
{Boselli}, A., {Cuillandre}, J.~C., {Fossati}, M., {et~al.} 2016, \aap, 587,
  A68

\bibitem[{{Boselli} {et~al.}(2018){Boselli}, {Fossati}, {Consolandi}, {Amram},
  {Ge}, {Sun}, {Anderson}, {Boissier}, {Boquien}, {Buat}, {Burgarella},
  {Cortese}, {C{\^o}t{\'e}}, {Cuillandre}, {Durrell}, {Epinat}, {Ferrarese},
  {Fumagalli}, {Galbany}, {Gavazzi}, {G{\'o}mez-L{\'o}pez}, {Gwyn}, {Hensler},
  {Kuncarayakti}, {Marcelin}, {Mendes de Oliveira}, {Quint}, {Roediger},
  {Roehlly}, {Sanchez}, {Sanchez-Janssen}, {Toloba}, {Trinchieri}, \&
  {Vollmer}}]{Boselli-2018a}
{Boselli}, A., {Fossati}, M., {Consolandi}, G., {et~al.} 2018, \aap, 620, A164

\bibitem[{Boselli {et~al.}(2022)Boselli, Fossati, Roediger, Boquien, Fumagalli,
  Balogh, Boissier, Braine, Ciesla, Côté, Cuillandre, Ferrarese, Gavazzi,
  Gwyn, {Junais}, Hensler, Longobardi, \& Sun}]{Boselli-2022a}
Boselli, A., Fossati, M., Roediger, J., {et~al.} 2022

\bibitem[{{Boselli} {et~al.}(2022){Boselli}, {Fossati}, \&
  {Sun}}]{Boselli-2022-REV}
{Boselli}, A., {Fossati}, M., \& {Sun}, M. 2022, \aapr, 30, 3

\bibitem[{{Boselli} \& {Gavazzi}(2006)}]{Boselli-2006}
{Boselli}, A. \& {Gavazzi}, G. 2006, \pasp, 118, 517

\bibitem[{{Boselli} {et~al.}(2021){Boselli}, {Lupi}, {Epinat}, {Amram},
  {Fossati}, {Anderson}, {Boissier}, {Boquien}, {Consolandi}, {C{\^o}t{\'e}},
  {Cuillandre}, {Ferrarese}, {Galbany}, {Gavazzi}, {G{\'o}mez-L{\'o}pez},
  {Gwyn}, {Hensler}, {Hutchings}, {Kuncarayakti}, {Longobardi}, {Peng},
  {Plana}, {Postma}, {Roediger}, {Roehlly}, {Schimd}, {Trinchieri}, \&
  {Vollmer}}]{Boselli-2021}
{Boselli}, A., {Lupi}, A., {Epinat}, B., {et~al.} 2021, \aap, 646, A139

\bibitem[{{Boselli} {et~al.}(2014){Boselli}, {Voyer}, {Boissier}, {Cucciati},
  {Consolandi}, {Cortese}, {Fumagalli}, {Gavazzi}, {Heinis}, {Roehlly}, \&
  {Toloba}}]{Boselli-2014}
{Boselli}, A., {Voyer}, E., {Boissier}, S., {et~al.} 2014, \aap, 570, A69

\bibitem[{{Bravo-Alfaro} {et~al.}(2001){Bravo-Alfaro}, {Cayatte}, {van Gorkom},
  \& {Balkowski}}]{Bravo-Alfaro-2001}
{Bravo-Alfaro}, H., {Cayatte}, V., {van Gorkom}, J.~H., \& {Balkowski}, C.
  2001, \aap, 379, 347

\bibitem[{{Bregman}(1990)}]{Bregman-1990}
{Bregman}, J.~N. 1990, \aapr, 2, 125

\bibitem[{{Capetti} \& {Baldi}(2011)}]{Capetti-2011}
{Capetti}, A. \& {Baldi}, R.~D. 2011, A\&A, 529, A126

\bibitem[{{Cardelli} {et~al.}(1989){Cardelli}, {Clayton}, \&
  {Mathis}}]{Cardelli-1989}
{Cardelli}, J.~A., {Clayton}, G.~C., \& {Mathis}, J.~S. 1989, \apj, 345, 245

\bibitem[{{Castignani} {et~al.}(2022{\natexlab{a}}){Castignani}, {Combes},
  {Jablonka}, {Finn}, {Rudnick}, {Vulcani}, {Desai}, {Zaritsky}, \&
  {Salom{\'e}}}]{Castignani-2022a}
{Castignani}, G., {Combes}, F., {Jablonka}, P., {et~al.} 2022{\natexlab{a}},
  \aap, 657, A9

\bibitem[{{Castignani} {et~al.}(2022{\natexlab{b}}){Castignani}, {Vulcani},
  {Finn}, {Combes}, {Jablonka}, {Rudnick}, {Zaritsky}, {Whalen}, {Conger}, {De
  Lucia}, {Desai}, {Koopmann}, {Moustakas}, {Norman}, \&
  {Townsend}}]{Castignani-2022b}
{Castignani}, G., {Vulcani}, B., {Finn}, R.~A., {et~al.} 2022{\natexlab{b}},
  \apjs, 259, 43

\bibitem[{{Chabrier}(2003)}]{Chabrier-2003}
{Chabrier}, G. 2003, \pasp, 115, 763

\bibitem[{{Chen} {et~al.}(2020){Chen}, {Sun}, {Yagi}, {Bravo-Alfaro}, {Brinks},
  {Kenney}, {Combes}, {Sivanandam}, {Jachym}, {Fossati}, {Gavazzi}, {Boselli},
  {Nulsen}, {Sarazin}, {Ge}, {Yoshida}, \& {Roediger}}]{Chen-2020}
{Chen}, H., {Sun}, M., {Yagi}, M., {et~al.} 2020, \mnras, 496, 4654

\bibitem[{{Chung} {et~al.}(2007){Chung}, {van Gorkom}, {Kenney}, \&
  {Vollmer}}]{Chung-2007}
{Chung}, A., {van Gorkom}, J.~H., {Kenney}, J. D.~P., \& {Vollmer}, B. 2007,
  \apjl, 659, L115

\bibitem[{{Cid Fernandes} {et~al.}(2004){Cid Fernandes}, {Gonz{\'a}lez
  Delgado}, {Schmitt}, {Storchi-Bergmann}, {Martins}, {P{\'e}rez}, {Heckman},
  {Leitherer}, \& {Schaerer}}]{Cid-Fernandes-2004}
{Cid Fernandes}, R., {Gonz{\'a}lez Delgado}, R.~M., {Schmitt}, H., {et~al.}
  2004, \apj, 605, 105

\bibitem[{Cid~Fernandes {et~al.}(2011)Cid~Fernandes, Stasi{\'n}ska, Mateus, \&
  Asari}]{Cid_Fernandez-2011}
Cid~Fernandes, R., Stasi{\'n}ska, G., Mateus, A., \& Asari, N.~V. 2011, \mnras,
  413, 1687

\bibitem[{Cid~Fernandes {et~al.}(2010)Cid~Fernandes, Stasi{\'n}ska,
  Schlickmann, Mateus, Asari, Schoenell, Sodr\'e, \& (the
  SEAGal~collaboration)}]{Cid_Fernandez-2010}
Cid~Fernandes, R., Stasi{\'n}ska, G., Schlickmann, M.~S., {et~al.} 2010,
  \mnras, 403, 1036

\bibitem[{{Consolandi} {et~al.}(2017){Consolandi}, {Gavazzi}, {Fossati},
  {Fumagalli}, {Boselli}, {Yagi}, \& {Yoshida}}]{Consolandi-2017}
{Consolandi}, G., {Gavazzi}, G., {Fossati}, M., {et~al.} 2017, \aap, 606, A83

\bibitem[{Constantin \& Vogeley(2006)}]{Constantin-2006}
Constantin, A. \& Vogeley, M.~S. 2006, \apj, 650, 727

\bibitem[{Cortese {et~al.}(2021)Cortese, Catinella, \& Smith}]{Cortese-2021}
Cortese, L., Catinella, B., \& Smith, R. 2021, Publications of the Astronomical
  Society of Australia, 38, e035

\bibitem[{{Cortese} \& {Hughes}(2009)}]{Cortese-2009}
{Cortese}, L. \& {Hughes}, T.~M. 2009, \mnras, 400, 1225

\bibitem[{{Corwin} {et~al.}(1994){Corwin}, {Buta}, \& {de Vaucouleurs}}]{RC3}
{Corwin}, Harold~G., J., {Buta}, R.~J., \& {de Vaucouleurs}, G. 1994, \aj, 108,
  2128

\bibitem[{{Crone Odekon} {et~al.}(2018){Crone Odekon}, {Hallenbeck}, {Haynes},
  {Koopmann}, {Phi}, \& {Wolfe}}]{Odekon-2018}
{Crone Odekon}, M., {Hallenbeck}, G., {Haynes}, M.~P., {et~al.} 2018, \apj,
  852, 142

\bibitem[{Crowl {et~al.}(2005)Crowl, Kenney, van Gorkom, \&
  Vollmer}]{Crowl-2005}
Crowl, H.~H., Kenney, J. D.~P., van Gorkom, J.~H., \& Vollmer, B. 2005, \aj,
  130, 65

\bibitem[{Cybulski {et~al.}(2014)Cybulski, Yun, Fazio, \&
  Gutermuth}]{Cybulski-2014}
Cybulski, R., Yun, M.~S., Fazio, G.~G., \& Gutermuth, R.~A. 2014, MNRAS, 439,
  3564

\bibitem[{Decarli {et~al.}(2007)Decarli, Gavazzi, Arosio, Cortese, Boselli,
  Bonfanti, \& Colpi}]{Decarli-2007}
Decarli, R., Gavazzi, G., Arosio, I., {et~al.} 2007, MNRAS, 381, 136

\bibitem[{{Dewangan} {et~al.}(2008){Dewangan}, {Mathur}, {Griffiths}, \&
  {Rao}}]{Dewangan-2008}
{Dewangan}, G.~C., {Mathur}, S., {Griffiths}, R.~E., \& {Rao}, A.~R. 2008,
  \apj, 689, 762

\bibitem[{{Dickey} \& {Gavazzi}(1991)}]{Dickey-1991}
{Dickey}, J.~M. \& {Gavazzi}, G. 1991, \apj, 373, 347

\bibitem[{{Dressler}(1980)}]{Dressler-1980}
{Dressler}, A. 1980, \apjs, 42, 565

\bibitem[{{Durbala} {et~al.}(2020){Durbala}, {Finn}, {Crone Odekon}, {Haynes},
  {Koopmann}, \& {O'Donoghue}}]{Durbala-2020}
{Durbala}, A., {Finn}, R.~A., {Crone Odekon}, M., {et~al.} 2020, \aj, 160, 271

\bibitem[{{Einasto} {et~al.}(1983){Einasto}, {Corwin}, {Huchra}, {Miller}, \&
  {Tarenghi}}]{Einasto-1983}
{Einasto}, J., {Corwin}, H.~G., J., {Huchra}, J., {Miller}, R.~H., \&
  {Tarenghi}, M. 1983, Highlights of Astronomy, 6, 757

\bibitem[{{Falco} {et~al.}(1999){Falco}, {Kurtz}, {Geller}, {Huchra}, {Peters},
  {Berlind}, {Mink}, {Tokarz}, \& {Elwell}}]{Falco-1999}
{Falco}, E.~E., {Kurtz}, M.~J., {Geller}, M.~J., {et~al.} 1999, \pasp, 111, 438

\bibitem[{{Ferland} \& {Netzer}(1983)}]{Ferland-1983}
{Ferland}, G.~J. \& {Netzer}, H. 1983, \apj, 264, 105

\bibitem[{{Fossati} {et~al.}(2019){Fossati}, {Fumagalli}, {Gavazzi},
  {Consolandi}, {Boselli}, {Yagi}, {Sun}, \& {Wilman}}]{Fossati-2019}
{Fossati}, M., {Fumagalli}, M., {Gavazzi}, G., {et~al.} 2019, \mnras, 484, 2212

\bibitem[{{Fossati} {et~al.}(2013){Fossati}, {Gavazzi}, {Savorgnan},
  {Fumagalli}, {Boselli}, {Guti{\'e}rrez}, {Hern{\'a}ndez Toledo},
  {Giovanelli}, \& {Haynes}}]{Gavazzi-2013Ha3_c}
{Fossati}, M., {Gavazzi}, G., {Savorgnan}, G., {et~al.} 2013, \aap, 553, A91

\bibitem[{{Fossati} {et~al.}(2018){Fossati}, {Mendel}, {Boselli}, {Cuilland
  re}, {Vollmer}, {Boissier}, {Consolandi}, {Ferrarese}, {Gwyn}, {Amram},
  {Boquien}, {Buat}, {Burgarella}, {Cortese}, {C{\^o}t{\'e}}, {C{\^o}t{\'e}},
  {Durrell}, {Fumagalli}, {Gavazzi}, {Gomez-Lopez}, {Hensler}, {Koribalski},
  {Longobardi}, {Peng}, {Roediger}, {Sun}, \& {Toloba}}]{Fossati-2018}
{Fossati}, M., {Mendel}, J.~T., {Boselli}, A., {et~al.} 2018, \aap, 614, A57

\bibitem[{{Fossati} {et~al.}(2015){Fossati}, {Wilman}, {Fontanot}, {De Lucia},
  {Monaco}, {Hirschmann}, {Mendel}, {Beifiori}, \& {Contini}}]{Fossati-2015}
{Fossati}, M., {Wilman}, D.~J., {Fontanot}, F., {et~al.} 2015, \mnras, 446,
  2582

\bibitem[{{Gavazzi}(1978)}]{Gavazzi-1978}
{Gavazzi}, G. 1978, \aap, 69, 355

\bibitem[{{Gavazzi}(1989)}]{Gavazzi-1989}
{Gavazzi}, G. 1989, \apj, 346, 59

\bibitem[{{Gavazzi} {et~al.}(2002){Gavazzi}, {Bonfanti}, {Sanvito}, {Boselli},
  \& {Scodeggio}}]{Gavazzi-2002}
{Gavazzi}, G., {Bonfanti}, C., {Sanvito}, G., {Boselli}, A., \& {Scodeggio}, M.
  2002, \apj, 576, 135

\bibitem[{{Gavazzi} {et~al.}(2003){Gavazzi}, {Boselli}, {Donati}, {Franzetti},
  \& {Scodeggio}}]{Gavazzi-2003}
{Gavazzi}, G., {Boselli}, A., {Donati}, A., {Franzetti}, P., \& {Scodeggio}, M.
  2003, \aap, 400, 451

\bibitem[{{Gavazzi} {et~al.}(2018){Gavazzi}, {Consolandi}, {Belladitta},
  {Boselli}, \& {Fossati}}]{Gavazzi-2018a}
{Gavazzi}, G., {Consolandi}, G., {Belladitta}, S., {Boselli}, A., \& {Fossati},
  M. 2018, \aap, 615, A104

\bibitem[{{Gavazzi} {et~al.}(2015{\natexlab{a}}){Gavazzi}, {Consolandi},
  {Dotti}, {Fanali}, {Fossati}, {Fumagalli}, {Viscardi}, {Savorgnan},
  {Boselli}, {Guti{\'e}rrez}, {Hern{\'a}ndez Toledo}, {Giovanelli}, \&
  {Haynes}}]{Gavazzi-2015Ha3_b}
{Gavazzi}, G., {Consolandi}, G., {Dotti}, M., {et~al.} 2015{\natexlab{a}},
  \aap, 580, A116

\bibitem[{{Gavazzi} {et~al.}(2013{\natexlab{a}}){Gavazzi}, {Consolandi},
  {Dotti}, {Fossati}, {Savorgnan}, {Gualandi, R.}, \& {Bruni,
  I.}}]{Gavazzi-2013}
{Gavazzi}, G., {Consolandi}, G., {Dotti}, M., {et~al.} 2013{\natexlab{a}},
  A\&A, 558, A68

\bibitem[{{Gavazzi} {et~al.}(2015{\natexlab{b}}){Gavazzi}, {Consolandi},
  {Viscardi}, {Fossati}, {Savorgnan}, {Fumagalli}, {Gutierrez}, {Hernandez
  Toledo}, {Boselli}, {Giovanelli}, \& {Haynes}}]{Gavazzi-2015Ha3_a}
{Gavazzi}, G., {Consolandi}, G., {Viscardi}, E., {et~al.} 2015{\natexlab{b}},
  \aap, 576, A16

\bibitem[{{Gavazzi} {et~al.}(1995){Gavazzi}, {Contursi}, {Carrasco}, {Boselli},
  {Kennicutt}, {Scodeggio}, \& {Jaffe}}]{Gavazzi-1995}
{Gavazzi}, G., {Contursi}, A., {Carrasco}, L., {et~al.} 1995, \aap, 304, 325

\bibitem[{{Gavazzi} {et~al.}(2010){Gavazzi}, {Fumagalli}, {Cucciati}, \&
  {Boselli}}]{Gavazzi-2010}
{Gavazzi}, G., {Fumagalli}, M., {Cucciati}, O., \& {Boselli}, A. 2010, \aap,
  517, A73

\bibitem[{{Gavazzi} {et~al.}(2013{\natexlab{b}}){Gavazzi}, {Fumagalli},
  {Fossati}, {Galardo}, {Grossetti}, {Boselli}, {Giovanelli}, \&
  {Haynes}}]{Gavazzi-2013Ha3_a}
{Gavazzi}, G., {Fumagalli}, M., {Fossati}, M., {et~al.} 2013{\natexlab{b}},
  \aap, 553, A89

\bibitem[{{Gavazzi} {et~al.}(2012){Gavazzi}, {Fumagalli}, {Galardo},
  {Grossetti}, {Boselli}, {Giovanelli}, {Haynes}, \&
  {Fabello}}]{Gavazzi-2012Ha3}
{Gavazzi}, G., {Fumagalli}, M., {Galardo}, V., {et~al.} 2012, \aap, 545, A16

\bibitem[{{Gavazzi} \& {Jaffe}(1985)}]{Gavazzi-Jaffe-1985}
{Gavazzi}, G. \& {Jaffe}, W. 1985, \apjl, 294, L89

\bibitem[{{Gavazzi} \& {Jaffe}(1987)}]{Gavazzi-Jaffe-1987}
{Gavazzi}, G. \& {Jaffe}, W. 1987, \aap, 186, L1

\bibitem[{{Gavazzi} {et~al.}(1996){Gavazzi}, {Pierini}, \&
  {Boselli}}]{Gavazzi-1996}
{Gavazzi}, G., {Pierini}, D., \& {Boselli}, A. 1996, \aap, 312, 397

\bibitem[{{Gavazzi} {et~al.}(2013{\natexlab{c}}){Gavazzi}, {Savorgnan},
  {Fossati}, {Dotti}, {Fumagalli}, {Boselli}, {Guti{\'e}rrez}, {Hern{\'a}ndez
  Toledo}, {Giovanelli}, \& {Haynes}}]{Gavazzi-2013Ha3_b}
{Gavazzi}, G., {Savorgnan}, G., {Fossati}, M., {et~al.} 2013{\natexlab{c}},
  \aap, 553, A90

\bibitem[{{Gavazzi} {et~al.}(2011){Gavazzi}, {Savorgnan, G.}, \& {Fumagalli,
  Mattia}}]{Gavazzi-2011}
{Gavazzi}, G., {Savorgnan, G.}, \& {Fumagalli, Mattia}. 2011, A\&A, 534, A31

\bibitem[{{Gil de Paz} {et~al.}(2007){Gil de Paz}, {Boissier}, {Madore},
  {Seibert}, {Joe}, {Boselli}, {Wyder}, {Thilker}, {Bianchi}, {Rey}, {Rich},
  {Barlow}, {Conrow}, {Forster}, {Friedman}, {Martin}, {Morrissey}, {Neff},
  {Schiminovich}, {Small}, {Donas}, {Heckman}, {Lee}, {Milliard}, {Szalay}, \&
  {Yi}}]{Gil-de-Paz-2007}
{Gil de Paz}, A., {Boissier}, S., {Madore}, B.~F., {et~al.} 2007, \apjs, 173,
  185

\bibitem[{{Giovanelli} \& {Haynes}(1985)}]{Giovanelli-1985}
{Giovanelli}, R. \& {Haynes}, M.~P. 1985, \apj, 292, 404

\bibitem[{{G{\'o}mez} {et~al.}(2003)}]{Gomez-2003}
{G{\'o}mez}, P.~L. {et~al.} 2003, \apj, 584, 210

\bibitem[{{Gualandi} \& {Merighi}(2001)}]{BFOSC}
{Gualandi}, R. \& {Merighi}, R. 2001, INAF Report

\bibitem[{{Halpern} \& {Steiner}(1983)}]{Halpern-1983}
{Halpern}, J.~P. \& {Steiner}, J.~E. 1983, \apjl, 269, L37

\bibitem[{{Hao} {et~al.}(2011){Hao}, {Kennicutt}, {Johnson}, {Calzetti},
  {Dale}, \& {Moustakas}}]{Hao-2011}
{Hao}, C.-N., {Kennicutt}, R.~C., {Johnson}, B.~D., {et~al.} 2011, \apj, 741,
  124

\bibitem[{{Haynes} \& {Giovanelli}(1984)}]{Haynes-1984}
{Haynes}, M.~P. \& {Giovanelli}, R. 1984, \aj, 89, 758

\bibitem[{{Haynes} {et~al.}(2011){Haynes}, {Giovanelli}, {Martin}, {Hess},
  {Saintonge}, {Adams}, {Hallenbeck}, {Hoffman}, {Huang}, {Kent}, {Koopmann},
  {Papastergis}, {Stierwalt}, {Balonek}, {Craig}, {Higdon}, {Kornreich},
  {Miller}, {O'Donoghue}, {Olowin}, {Rosenberg}, {Spekkens}, {Troischt}, \&
  {Wilcots}}]{Haynes-2011}
{Haynes}, M.~P., {Giovanelli}, R., {Martin}, A.~M., {et~al.} 2011, \aj, 142,
  170

\bibitem[{{Heckman}(1980)}]{Heckman-1980}
{Heckman}, T.~M. 1980, \aap, 87, 152

\bibitem[{{Ho} {et~al.}(1993){Ho}, {Filippenko}, \& {Sargent}}]{Ho-1993}
{Ho}, L.~C., {Filippenko}, A.~V., \& {Sargent}, W. L.~W. 1993, in American
  Astronomical Society Meeting Abstracts, Vol. 182, , 17.05

\bibitem[{{Ho} {et~al.}(1995){Ho}, {Filippenko}, \& {Sargent}}]{Ho-1995}
{Ho}, L.~C., {Filippenko}, A.~V., \& {Sargent}, W. L.~W. 1995, \apjs, 98, 477

\bibitem[{{Ho} {et~al.}(1997){Ho}, {Filippenko}, \& {Sargent}}]{Ho-1997}
{Ho}, L.~C., {Filippenko}, A.~V., \& {Sargent}, W. L.~W. 1997, \apjs, 112, 315

\bibitem[{{Hughes} \& {Cortese}(2009)}]{Hughes-2009}
{Hughes}, T.~M. \& {Cortese}, L. 2009, \mnras, 396, L41

\bibitem[{Juneau {et~al.}(2011)Juneau, Dickinson, Alexander, \&
  Salim}]{Juneau-2011}
Juneau, S., Dickinson, M., Alexander, D.~M., \& Salim, S. 2011, \apj, 736, 104

\bibitem[{{Kantharia} {et~al.}(2008){Kantharia}, {Rao}, \&
  {Sirothia}}]{Kantharia-2008}
{Kantharia}, N.~G., {Rao}, A.~P., \& {Sirothia}, S.~K. 2008, \mnras, 383, 173

\bibitem[{{Kauffmann} \& {Haehnelt}(2000)}]{Kauffmann-2000}
{Kauffmann}, G. \& {Haehnelt}, M. 2000, \mnras

\bibitem[{{Kauffmann} {et~al.}(2003{\natexlab{a}}){Kauffmann}, {Heckman},
  {Tremonti}, {Brinchmann}, {Charlot}, {White}, {Ridgway}, {Brinkmann},
  {Fukugita}, {Hall}, {Ivezi{\'c}}, {Richards}, \&
  {Schneider}}]{Kauffmann-2003}
{Kauffmann}, G., {Heckman}, T.~M., {Tremonti}, C., {et~al.} 2003{\natexlab{a}},
  \mnras, 346, 1055

\bibitem[{Kauffmann {et~al.}(2004)Kauffmann, White, Heckman, MÃ©nard,
  Brinchmann, Charlot, Tremonti, \& Brinkmann}]{Kauffmann-2004}
Kauffmann, G., White, S. D.~M., Heckman, T.~M., {et~al.} 2004, MNRAS, 353, 713

\bibitem[{{Kauffmann} {et~al.}(2003{\natexlab{b}})}]{Kauffmann-2003b}
{Kauffmann}, G. {et~al.} 2003{\natexlab{b}}, \mnras, 341, 33

\bibitem[{Kenney \& Koopmann(1999)}]{Kenney-1999}
Kenney, J. D.~P. \& Koopmann, R.~A. 1999, \aj, 117, 181

\bibitem[{{Kenney} {et~al.}(2008){Kenney}, {Tal}, {Crowl}, {Feldmeier}, \&
  {Jacoby}}]{Kenney-2008}
{Kenney}, J. D.~P., {Tal}, T., {Crowl}, H.~H., {Feldmeier}, J., \& {Jacoby},
  G.~H. 2008, \apjl, 687, L69

\bibitem[{Kenney {et~al.}(2004)Kenney, van Gorkom, \& Vollmer}]{Kenney-2004}
Kenney, J. D.~P., van Gorkom, J.~H., \& Vollmer, B. 2004, \aj, 127, 3361

\bibitem[{{Kennicutt}(1998)}]{Kennicutt-1998}
{Kennicutt}, Robert~C., J. 1998, \araa, 36, 189

\bibitem[{{Kewley} {et~al.}(2001){Kewley}, {Dopita}, {Sutherland}, {Heisler},
  \& {Trevena}}]{Kewley-2001}
{Kewley}, L.~J., {Dopita}, M.~A., {Sutherland}, R.~S., {Heisler}, C.~A., \&
  {Trevena}, J. 2001, \apj, 556, 121

\bibitem[{{Kewley} {et~al.}(2006){Kewley}, {Groves}, {Kauffmann}, \&
  {Heckman}}]{Kewley-2006}
{Kewley}, L.~J., {Groves}, B., {Kauffmann}, G., \& {Heckman}, T. 2006, \mnras,
  372, 961

\bibitem[{Lal(2020)}]{Lal-2020}
Lal, D.~V. 2020, The Astrophysical Journal Supplement Series, 250, 22

\bibitem[{{Larson} {et~al.}(1980){Larson}, {Tinsley}, \&
  {Caldwell}}]{Larson-1980}
{Larson}, R.~B., {Tinsley}, B.~M., \& {Caldwell}, C.~N. 1980, \apj, 237, 692

\bibitem[{Lee {et~al.}(2021)Lee, Kim, Rey, \& Chung}]{Lee-2021}
Lee, Y., Kim, S., Rey, S.-C., \& Chung, J. 2021, \apj, 906, 68

\bibitem[{Li {et~al.}(2019)Li, Wuyts, Lei, Lin, Lam, Boquien, Andrews, \&
  Schneider}]{Li-2019}
Li, H., Wuyts, S., Lei, H., {et~al.} 2019, \apj, 872, 63

\bibitem[{{Lopes} {et~al.}(2017){Lopes}, {Ribeiro}, \& {Rembold}}]{Lopes-2017}
{Lopes}, P.~A.~A., {Ribeiro}, A.~L.~B., \& {Rembold}, S.~B. 2017, \mnras, 472,
  409

\bibitem[{{Macchetto}(1996)}]{Macchetto-1996}
{Macchetto}, F.~D. 1996, in IAU Symposium, Vol. 175, Extragalactic Radio
  Sources, ed. R.~D. {Ekers}, C.~{Fanti}, \& L.~{Padrielli}

\bibitem[{{Makarov} {et~al.}(2014){Makarov}, {Prugniel}, {Terekhova},
  {Courtois}, \& {Vauglin}}]{Makarov-2014}
{Makarov}, D., {Prugniel}, P., {Terekhova}, N., {Courtois}, H., \& {Vauglin},
  I. 2014, \aap, 570, A13

\bibitem[{{Martin} {et~al.}(2007){Martin}, {Wyder}, {Schiminovich}, {Barlow},
  {Forster}, {Friedman}, {Morrissey}, {Neff}, {Seibert}, {Small}, {Welsh},
  {Bianchi}, {Donas}, {Heckman}, {Lee}, {Madore}, {Milliard}, {Rich}, {Szalay},
  \& {Yi}}]{Martin-2007}
{Martin}, D.~C., {Wyder}, T.~K., {Schiminovich}, D., {et~al.} 2007, \apjs, 173,
  342

\bibitem[{Martin {et~al.}(2005)}]{GALEX}
Martin, D.~C. {et~al.} 2005, \apj, 619, L1

\bibitem[{{Meyer} {et~al.}(2017){Meyer}, {Robotham}, {Obreschkow}, {Westmeier},
  {Duffy}, \& {Staveley-Smith}}]{Meyer-2017}
{Meyer}, M., {Robotham}, A., {Obreschkow}, D., {et~al.} 2017, \pasa, 34, 52

\bibitem[{Miller {et~al.}(2009)Miller, Hornschemeier, Mobasher, Bridges,
  Hudson, Marzke, \& Smith}]{Miller-2009}
Miller, N.~A., Hornschemeier, A.~E., Mobasher, B., {et~al.} 2009, \aj, 137,
  4450

\bibitem[{{Moln{\'a}r} {et~al.}(2022){Moln{\'a}r}, {Serra}, {van der Hulst},
  {Jarrett}, {Boselli}, {Cortese}, {Healy}, {de Blok}, {Cappellari}, {Hess},
  {J{\'o}zsa}, {McDermid}, {Oosterloo}, \& {Verheijen}}]{Molnar-2022}
{Moln{\'a}r}, D.~C., {Serra}, P., {van der Hulst}, T., {et~al.} 2022, \aap,
  659, A94

\bibitem[{{Morrissey} {et~al.}(2007){Morrissey}, {Conrow}, {Barlow}, {Small},
  {Seibert}, {Wyder}, {Budav{\'a}ri}, {Arnouts}, {Friedman}, {Forster},
  {Martin}, {Neff}, {Schiminovich}, {Bianchi}, {Donas}, {Heckman}, {Lee},
  {Madore}, {Milliard}, {Rich}, {Szalay}, {Welsh}, \& {Yi}}]{GALEX07}
{Morrissey}, P., {Conrow}, T., {Barlow}, T.~A., {et~al.} 2007, \apjs, 173, 682

\bibitem[{{Muldrew} {et~al.}(2012){Muldrew}, {Croton}, {Skibba}, {Pearce},
  {Ann}, {Baldry}, {Brough}, {Choi}, {Conselice}, {Cowan}, {Gallazzi}, {Gray},
  {Gr{\"u}tzbauch}, {Li}, {Park}, {Pilipenko}, {Podgorzec}, {Robotham},
  {Wilman}, {Yang}, {Zhang}, \& {Zibetti}}]{Muldrew-2012}
{Muldrew}, S.~I., {Croton}, D.~J., {Skibba}, R.~A., {et~al.} 2012, \mnras, 419,
  2670

\bibitem[{{Nilson}(1973)}]{Nilson-1973}
{Nilson}, P. 1973, Acta Universitatis Upsaliensis. Nova Acta Regiae Societatis
  Scientiarum Upsaliensis

\bibitem[{{Oliveira} {et~al.}(2022){Oliveira}, {Krabbe}, {Hernandez-Jimenez},
  {Dors}, {Zinchenko}, {H{\"a}gele}, {Cardaci}, \& {Monteiro}}]{Oliveira-2022}
{Oliveira}, C.~B., {Krabbe}, A.~C., {Hernandez-Jimenez}, J.~A., {et~al.} 2022,
  \mnras, 515, 6093

\bibitem[{{Pedrini} {et~al.}(2022){Pedrini}, {Fossati}, {Gavazzi}, {Fumagalli},
  {Boselli}, {Consolandi}, {Sun}, {Yagi}, \& {Yoshida}}]{Pedrini-2022}
{Pedrini}, A., {Fossati}, M., {Gavazzi}, G., {et~al.} 2022, \mnras, 511, 5180

\bibitem[{{Peluso} {et~al.}(2022){Peluso}, {Vulcani}, {Poggianti}, {Moretti},
  {Radovich}, {Smith}, {Jaff{\'e}}, {Crossett}, {Gullieuszik}, {Fritz}, \&
  {Ignesti}}]{Peluso-2022}
{Peluso}, G., {Vulcani}, B., {Poggianti}, B.~M., {et~al.} 2022, \apj, 927, 130

\bibitem[{{Peng} {et~al.}(2010){Peng}, {Lilly}, {Kova{\v{c}}}, {Bolzonella},
  {Pozzetti}, {Renzini}, {Zamorani}, {Ilbert}, {Knobel}, {Iovino}, {Maier},
  {Cucciati}, {Tasca}, {Carollo}, {Silverman}, {Kampczyk}, {de Ravel},
  {Sanders}, {Scoville}, {Contini}, {Mainieri}, {Scodeggio}, {Kneib}, {Le
  F{\`e}vre}, {Bardelli}, {Bongiorno}, {Caputi}, {Coppa}, {de la Torre},
  {Franzetti}, {Garilli}, {Lamareille}, {Le Borgne}, {Le Brun}, {Mignoli},
  {Perez Montero}, {Pello}, {Ricciardelli}, {Tanaka}, {Tresse}, {Vergani},
  {Welikala}, {Zucca}, {Oesch}, {Abbas}, {Barnes}, {Bordoloi}, {Bottini},
  {Cappi}, {Cassata}, {Cimatti}, {Fumana}, {Hasinger}, {Koekemoer},
  {Leauthaud}, {Maccagni}, {Marinoni}, {McCracken}, {Memeo}, {Meneux}, {Nair},
  {Porciani}, {Presotto}, \& {Scaramella}}]{Peng-2010}
{Peng}, Y.-j., {Lilly}, S.~J., {Kova{\v{c}}}, K., {et~al.} 2010, \apj, 721, 193

\bibitem[{Poggianti {et~al.}(2017{\natexlab{a}})Poggianti, Moretti,
  Gullieuszik, Fritz, Jaff{\'{e}}, Bettoni, Fasano, Bellhouse, Hau, Vulcani,
  Biviano, Omizzolo, Paccagnella, D'Onofrio, Cava, Sheen, Couch, \&
  Owers}]{Poggianti-2017-GASP}
Poggianti, B.~M., Moretti, A., Gullieuszik, M., {et~al.} 2017{\natexlab{a}},
  \apj, 844, 48

\bibitem[{Poggianti {et~al.}(2017{\natexlab{b}})}]{Poggianti-2017}
Poggianti, B.~M. {et~al.} 2017{\natexlab{b}}, Nature, 548, 304

\bibitem[{Ramos-Mart\'{i}nez {et~al.}(2018)Ramos-Mart\'{i}nez, G\'{o}mez, \&
  P\'{e}rez-Villegas}]{Martinez-2018}
Ramos-Mart\'{i}nez, M., G\'{o}mez, G.~C., \& P\'{e}rez-Villegas, A. 2018,
  MNRAS, 476, 3781

\bibitem[{{Reines} {et~al.}(2013){Reines}, {Greene}, \& {Geha}}]{Reines-2013}
{Reines}, A.~E., {Greene}, J.~E., \& {Geha}, M. 2013, \apj, 775, 116

\bibitem[{{Rodr{\'\i}guez del Pino} {et~al.}(2017{\natexlab{a}}){Rodr{\'\i}guez
  del Pino}, {Arag{\'o}n-Salamanca}, {Chies-Santos}, {Weinzirl}, {Bamford},
  {Gray}, {B{\"o}hm}, {Wolf}, \& {Maltby}}]{RodriguezDP-2017}
{Rodr{\'\i}guez del Pino}, B., {Arag{\'o}n-Salamanca}, A., {Chies-Santos},
  A.~L., {et~al.} 2017{\natexlab{a}}, \mnras, 467, 4200

\bibitem[{{Rodr{\'\i}guez del Pino} {et~al.}(2017{\natexlab{b}}){Rodr{\'\i}guez
  del Pino}, {Arag{\'o}n-Salamanca}, {Chies-Santos}, {Weinzirl}, {Bamford},
  {Gray}, {B{\"o}hm}, {Wolf}, \& {Maltby}}]{RDP-2017}
{Rodr{\'\i}guez del Pino}, B., {Arag{\'o}n-Salamanca}, A., {Chies-Santos},
  A.~L., {et~al.} 2017{\natexlab{b}}, \mnras, 467, 4200

\bibitem[{{Roman-Oliveira} {et~al.}(2019){Roman-Oliveira}, {Chies-Santos},
  {Rodr{\'\i}guez del Pino}, {Arag{\'o}n-Salamanca}, {Gray}, \&
  {Bamford}}]{Roman-Oliveira-2019}
{Roman-Oliveira}, F.~V., {Chies-Santos}, A.~L., {Rodr{\'\i}guez del Pino}, B.,
  {et~al.} 2019, \mnras, 484, 892

\bibitem[{{Sabater} {et~al.}(2013){Sabater}, {Best}, \&
  {Argudo-Fern{\'a}ndez}}]{Sabater-2013}
{Sabater}, J., {Best}, P.~N., \& {Argudo-Fern{\'a}ndez}, M. 2013, \mnras, 430,
  638

\bibitem[{{Sabater} {et~al.}(2015){Sabater}, {Best}, \&
  {Heckman}}]{Sabater-2015}
{Sabater}, J., {Best}, P.~N., \& {Heckman}, T.~M. 2015, \mnras, 447, 110

\bibitem[{Sarzi {et~al.}(2010)Sarzi, Shields, Schawinski,
  {et~al.}}]{Sarzi-2010}
Sarzi, M., Shields, J.~C., Schawinski, K., {et~al.} 2010, MNRAS, 402, 2187

\bibitem[{Schawinski {et~al.}(2007)Schawinski, Thomas, Sarzi, Maraston,
  Kaviraj, Joo, Yi, \& Silk}]{Schawinski-2007}
Schawinski, K., Thomas, D., Sarzi, M., {et~al.} 2007, MNRAS, 382, 1415

\bibitem[{{Schechter}(1976)}]{Schechter-1976}
{Schechter}, P. 1976, \apj, 203, 297

\bibitem[{Schulz \& Struck(2001)}]{Schulz-Struck-2001}
Schulz, S. \& Struck, C. 2001, MNRAS, 328, 185

\bibitem[{{Scott} {et~al.}(2010){Scott}, {Bravo-Alfaro}, {Brinks}, {Caretta},
  {Cortese}, {Boselli}, {Hardcastle}, {Croston}, \& {Plauchu}}]{Scott-2010}
{Scott}, T.~C., {Bravo-Alfaro}, H., {Brinks}, E., {et~al.} 2010, \mnras, 403,
  1175

\bibitem[{{Scott} {et~al.}(2012){Scott}, {Cortese}, {Brinks}, {Bravo-Alfaro},
  {Auld}, \& {Minchin}}]{Scott-2012}
{Scott}, T.~C., {Cortese}, L., {Brinks}, E., {et~al.} 2012, \mnras, 419, L19

\bibitem[{{Shields}(1992)}]{Shields-1992}
{Shields}, J.~C. 1992, \apjl, 399, L27

\bibitem[{{Stasi{\'n}ska} {et~al.}(2008){Stasi{\'n}ska}, {Vale Asari}, {Cid
  Fernandes}, {Gomes}, {Schlickmann}, {Mateus}, {Schoenell}, {Sodr{\'e}}, \&
  {Seagal Collaboration}}]{Stasinska-2008}
{Stasi{\'n}ska}, G., {Vale Asari}, N., {Cid Fernandes}, R., {et~al.} 2008,
  \mnras, 391, L29

\bibitem[{{Stevens} {et~al.}(2021){Stevens}, {Lagos}, {Cortese}, {Catinella},
  {Diemer}, {Nelson}, {Pillepich}, {Hernquist}, {Marinacci}, \&
  {Vogelsberger}}]{Stevens-2021}
{Stevens}, A. R.~H., {Lagos}, C. d.~P., {Cortese}, L., {et~al.} 2021, \mnras,
  502, 3158

\bibitem[{Steyrleithner {et~al.}(2020)Steyrleithner, Hensler, \&
  Boselli}]{Steyrleithner-2020}
Steyrleithner, P., Hensler, G., \& Boselli, A. 2020, Monthly Notices of the
  Royal Astronomical Society, 494, 1114

\bibitem[{{Terlevich} \& {Melnick}(1985)}]{Terlevich-1985}
{Terlevich}, R. \& {Melnick}, J. 1985, \mnras, 213, 841

\bibitem[{Thomas {et~al.}(2019)Thomas, Kewley, Dopita, Groves, Hopkins, \&
  Sutherland}]{Thomas-2019}
Thomas, A.~D., Kewley, L.~J., Dopita, M.~A., {et~al.} 2019, \apj, 874, 100

\bibitem[{{Thomas} {et~al.}(2013){Thomas}, {Steele}, {Maraston}, {Johansson},
  {Beifiori}, {Pforr}, {Str{\"o}mb{\"a}ck}, {Tremonti}, {Wake}, {Bizyaev},
  {Bolton}, {Brewington}, {Brownstein}, {Comparat}, {Kneib}, {Malanushenko},
  {Malanushenko}, {Oravetz}, {Pan}, {Parejko}, {Schneider}, {Shelden},
  {Simmons}, {Snedden}, {Tanaka}, {Weaver}, \& {Yan}}]{Thomas-2013}
{Thomas}, D., {Steele}, O., {Maraston}, C., {et~al.} 2013, \mnras, 431, 1383

\bibitem[{Tonnesen \& Bryan(2009)}]{Tonnesen-Bryan-2009}
Tonnesen, S. \& Bryan, G.~L. 2009, \apj, 694, 789

\bibitem[{{Trinchieri} \& {di Serego Alighieri}(1991)}]{Trinchieri-1991}
{Trinchieri}, G. \& {di Serego Alighieri}, S. 1991, \aj, 101, 1647

\bibitem[{{Tully}(1982)}]{Tully-1982}
{Tully}, R.~B. 1982, \apj, 257, 389

\bibitem[{{Vollmer}(2003)}]{Vollmer-2003}
{Vollmer}, B. 2003, \aap, 398, 525

\bibitem[{{Vollmer} {et~al.}(2004){Vollmer}, {Balkowski}, {Cayatte}, {van
  Driel}, \& {Huchtmeier}}]{Vollmer-2004}
{Vollmer}, B., {Balkowski}, C., {Cayatte}, V., {van Driel}, W., \&
  {Huchtmeier}, W. 2004, \aap, 419, 35

\bibitem[{{Vollmer} {et~al.}(2001){Vollmer}, {Cayatte}, {Balkowski}, \&
  {Duschl}}]{Vollmer-2001}
{Vollmer}, B., {Cayatte}, V., {Balkowski}, C., \& {Duschl}, W.~J. 2001, \apj,
  561, 708

\bibitem[{{Vollmer} {et~al.}(2010){Vollmer}, {Soida}, {Chung}, {Beck},
  {Urbanik}, {Chy{\.z}y}, {Otmianowska-Mazur}, \& {van Gorkom}}]{Vollmer-2010}
{Vollmer}, B., {Soida}, M., {Chung}, A., {et~al.} 2010, \aap, 512, A36

\bibitem[{{Vollmer, B.} {et~al.}(2008){Vollmer, B.}, {Braine, J.}, {Pappalardo,
  C.}, \& {Hily-Blant, P.}}]{Vollmer-2008}
{Vollmer, B.}, {Braine, J.}, {Pappalardo, C.}, \& {Hily-Blant, P.} 2008, A\&A,
  491, 455

\bibitem[{{Voyer} {et~al.}(2014){Voyer}, {Boselli}, {Boissier}, {Heinis},
  {Cortese}, {Ferrarese}, {Cote}, {Cuillandre}, {Gwyn}, {Peng}, {Zhang}, \&
  {Liu}}]{Voyer-2014}
{Voyer}, E.~N., {Boselli}, A., {Boissier}, S., {et~al.} 2014, \aap, 569, A124

\bibitem[{Wilson(1927)}]{Wilson-1927}
Wilson, E.~B. 1927, Journal of the American Statistical Association, 22, 209

\bibitem[{{Wright} {et~al.}(2010)}]{WISE}
{Wright}, E.~L. {et~al.} 2010, \aj, 140, 1868

\bibitem[{Yagi {et~al.}(2007)Yagi, Komiyama, Yoshida, Furusawa, Kashikawa,
  Koyama, \& Okamura}]{Yagi-2007}
Yagi, M., Komiyama, Y., Yoshida, M., {et~al.} 2007, \apj, 660, 1209

\bibitem[{{Yagi} {et~al.}(2017){Yagi}, {Yoshida}, {Gavazzi}, {Komiyama},
  {Kashikawa}, \& {Okamura}}]{Yagi-2017}
{Yagi}, M., {Yoshida}, M., {Gavazzi}, G., {et~al.} 2017, \apj, 839, 65

\bibitem[{{Yagi} {et~al.}(2010){Yagi}, {Yoshida}, {Komiyama}, {Kashikawa},
  {Furusawa}, {Okamura}, {Graham}, {Miller}, {Carter}, {Mobasher}, \&
  {Jogee}}]{Yagi-2010}
{Yagi}, M., {Yoshida}, M., {Komiyama}, Y., {et~al.} 2010, \aj, 140, 1814

\bibitem[{Yang {et~al.}(2008)Yang, Mo, \& van~den Bosch}]{Yang-2008}
Yang, X., Mo, H.~J., \& van~den Bosch, F.~C. 2008, ApJ, 676, 248

\bibitem[{Yang {et~al.}(2009)Yang, Mo, \& van~den Bosch}]{Yang-2009}
Yang, X., Mo, H.~J., \& van~den Bosch, F.~C. 2009, ApJ, 695, 900

\bibitem[{Yang {et~al.}(2007)Yang, Mo, van~den Bosch, Pasquali, Li, \&
  Barden}]{Yang-2007}
Yang, X., Mo, H.~J., van~den Bosch, F.~C., {et~al.} 2007, \apj, 671, 153

\bibitem[{{Yates} {et~al.}(2012){Yates}, {Kauffmann}, \& {Guo}}]{Yates-2012}
{Yates}, R.~M., {Kauffmann}, G., \& {Guo}, Q. 2012, \mnras, 422, 215

\bibitem[{{York} {et~al.}(2000)}]{York-2000}
{York}, D.~G. {et~al.} 2000, \aj, 120, 1579

\bibitem[{Yoshida {et~al.}(2002)Yoshida, Yagi, Okamura, Aoki, Ohyama, Komiyama,
  Yasuda, Iye, Kashikawa, Doi, Furusawa, Hamabe, Kimura, Miyazaki, Miyazaki,
  Nakata, Ouchi, Sekiguchi, Shimasaku, \& Ohtani}]{Yoshida-2002}
Yoshida, M., Yagi, M., Okamura, S., {et~al.} 2002, \apj, 567, 118

\bibitem[{Zibetti {et~al.}(2009)Zibetti, Charlot, \& Rix}]{Zibetti-2009}
Zibetti, S., Charlot, S., \& Rix, H.-W. 2009, MNRAS, 400, 1181

\bibitem[{{Zwicky} {et~al.}(1961){Zwicky}, {Herzog}, {Wild}, {Karpowicz}, \&
  {Kowal}}]{Zwicky-1961}
{Zwicky}, F., {Herzog}, E., {Wild}, P., {Karpowicz}, M., \& {Kowal}, C.~T.
  1961, Catalogue of galaxies and of clusters of galaxies, Vol. I (Pasadena:
  California Institute of Technology)

\end{thebibliography}

\appendix 
\begin{onecolumn}
\section{Loiano observations}\label{appendix-Loiano}
We used the Bologna Faint Object Spectrograph and Camera
\cite[BFOSC,][]{BFOSC} attached to the 152cm F/8 Cassini Telescope locatedin Loiano, which belongs to the Observatory of Bologna, to obtain optical spectra of the nuclei of 376 galaxies\footnote{The long-slit optical nuclear spectra taken at Loiano are available in machine-readable format via the Strasbourg Astronomical Data Center (CDS) and NED.}. The observations took place from 2014 to 2020.  The long-slit spectra were taken through a slit of 2 or 2.5 arcsec width (depending on the seeing conditions), with a intermediate-resolution red-channel grism (R $\sim$ 2200) that covers the 6100 - 8200 \text{\normalfont\AA} portion of the spectrum, which contains H$\alpha$, [NII], and [SII] lines (three galaxies were observed also using a  blue-channel grism covering H$\beta$ and [OIII]).

\vspace{.2cm}
The detector used by BFOSC is an EEV LN/1300-EB/1 CCD of 1300x1340 pixels, with 90\% quantum efficiency near 5500 \text{\normalfont\AA}.  Its spatial scale of 0.58 arcsec/pixel results in a field of view of $12.6'\times13'$. The  dispersion of the red-channel grism is 8.8 nm/mm and results in spectra with 1.6 \ \text{\normalfont\AA}/pix. The instrumental broadening is typically   6 \text{\normalfont\AA} full-width-half-maximum (FWHM), as checked on the 6300.3 \text{\normalfont\AA} sky line.
We obtained exposures of 3-5 minutes, repeated typically three-six times per run (to remove the cosmic rays), 
but several galaxies were re-observed in more than one run (see Table \ref{data2}). The seeing at Loiano is typically 1.5 - 2.5 arcsec.
The slit was mostly set in the E-W direction, except when it was positioned along the galaxy major axis
or along the direction connecting two adjacent objects to accommodate both in one exposure. 

\vspace{.2cm}
The wavelength was calibrated using
frequent exposures of a He-Ar hollow-cathode lamp. We used several sky lines to check \textit{a posteriori} the wavelength calibration.
The spectrograph response was obtained by daily exposures of the star Feige-34.
\vspace{.3cm}
\noindent

Table \ref{data2} summarises the spectra obtained at Loiano (2014-2020) as follows:

\vspace{-.1cm}
\noindent
\begin{enumerate}
    \item[] Column 1: Galaxy name;\\
    \vspace{-0.2cm}
    \item[] Column 2,3: Right Ascension; Declination (J2000);\\
    \vspace{-0.2cm}    
    \item[] Column 4:  redshift $cz$ (km/s);\\
    \vspace{-0.2cm}    
    \item[] Column 5: $r$-band magnitude;\\
    \vspace{-0.2cm}    
    \item[] Column 6: duration of the individual exposures;\\
    \vspace{-0.2cm}    
    \item[] Column 7: number of individual exposures;\\
    \vspace{-0.2cm}    
    \item[] Column 8: observing date;\\
    \vspace{-0.2cm}    
    \item[] Column 9: slit orientation (counterclockwise from N) 270 corresponds to East-West;\\
    \vspace{-0.2cm}    
    \item[] Column 10: measured EW of the H$\alpha$ line (negative EW values represent emission);\\
    \vspace{-0.2cm}    
    \item[] Column 11: measured EW  of [NII]$\mathrm{\lambdaup} \ 6584$ line;\\
    \vspace{-0.2cm}    
    \item[] Column 12: measured EW  of H$\beta$ line;\\
    \vspace{-0.2cm}    
    \item[] Column 13: measured EW  of [OIII]$\mathrm{\lambdaup} \ 5007$ line.
\end{enumerate}

 \tiny
\begin{longtable}{lllcccccccccc}
  \caption{{\bf List of 302 galaxies whose spectra have been taken at Loiano between 2014 and 2020.}}\\
 \hline
 \noalign{\smallskip}
  ID                &     RA       &    DEC   &   $cz$           &   $r$    &   Tint & N   &  Obs date  &  PA slit & EW $H\alpha$  & EW NII   & EW $H\beta$    & EW [OIII] \\                          
                    &   Deg        & Deg      &  $\rm km s^{-1}$ &   mag    &    sec &     &  Y-M-D     & Deg      &  \text{\normalfont\AA}          &    \text{\normalfont\AA}   &   \text{\normalfont\AA} & \text{\normalfont\AA}                  \\
   (1)              &   (2)        & (3)      &   (4)            &    (5)   & (6)    & (7) &   (8)      &  (9)     &  (10)         &   (11)   &  (12) &  (13)                 \\
 \hline                           \endfirsthead
 \hline
 \noalign{\smallskip}    
  ID                &     RA       &    DEC   &   $cz$           &   $r$    &   Tint & N   &  Obs date  &  PA slit & EW $H\alpha$  & EW NII   & EW $H\beta$    & EW [OIII] \\                          
                    &   Deg        & Deg      &  $\rm km s^{-1}$ &   mag    &    sec &     &  Y-M-D     & Deg      &  \text{\normalfont\AA}          &    \text{\normalfont\AA}   &   \text{\normalfont\AA} & \text{\normalfont\AA}                  \\
   (1)              &   (2)        & (3)      &   (4)            &    (5)   & (6)    & (7) &   (8)      &  (9)     &  (10)         &   (11)   &  (12) &  (13)                 \\
 \hline                           \endhead
   J073637+3175303    &   114.15417  & 17.88417 &  5037 & 11.65  &  300  & 3 & 2014-02-27   &  270  &   0     & -0.97    & -  & -  \\  
   CGCG 119-029       &   124.35751  & 21.68547 &  3566 & 13.55  &  600  & 3 & 2014-02-27   &  290  &  -20.59 & -7.62    & -  & -  \\  
   CGCG 119-057       &   124.95133  & 22.03145 &  3584 & 13.55  &  600  & 3 & 2014-02-27   &  345  &   0     & -1.18    & -  & -  \\  
   NGC-3065           &   150.47922  & 72.17047 &  2000 &  -     &  300  & 3 & 2014-03-27   &  336  &    0    & -2.16    & -  & -  \\  
   NGC-3066           &   150.54463  & 72.12587 &  2049 &  -     &  300  & 3 & 2014-03-27   &  336  &  -29.33 & -12.84   & -  & -  \\  
   NGC-3094           &   150.35809  & 15.77007 &  2404 & 14.41  &  300  & 3 & 2014-03-29   &  270  &  -20.79 & -16.15   & -  & -  \\  
   NGC-3098           &   150.56956  & 24.71110 &  1387 & 12.04  &  300  & 3 & 2014-02-27   &  268  &  -0.39  & 0	 & -  & -  \\  
   NGC-3119S          &   151.71613  & 14.37298 &  8825 & 12.47  &  300  & 3 & 2014-02-27   &  209  &	0     &  0	 & -  & -  \\
\end{longtable}
   \normalsize
   \end{onecolumn}
\clearpage
\section{the Local Spring database}\label{appendix-springSample}

A sample of the SPRING database containing the first 30 of 30597 galaxies is given in Table \ref{tab:springsample} (full table available at the CDS) as follows:
\begin{itemize}  
 \item Column 1: Right Ascension (in degrees);\\
    \vspace{-0.2cm}    
 \item Column 2: Declination  (in degrees);\\
    \vspace{-0.2cm}    
 \item Column 3-6: corrected magnitudes FUV, NUV, $g, i$;\\
    \vspace{-0.2cm}    
 \item Column 7: recessional velocity (in $\mathrm{km \ s}^{-1}$);\\
    \vspace{-0.2cm}    
 \item Column 8: adopted distance in $h^{-1}$ Mpc;\\
    \vspace{-0.2cm}    
 \item Column 9: origin of the line measurement (SL=SDSS, HO=\cite{Ho-1995, Ho-1997}, ZW = \cite{Falco-1999}, LOI=Loiano \cite[][this work]{Gavazzi-2011, Gavazzi-2013},
 NO=no nuclear spectrum available);\\
    \vspace{-0.2cm}    
 \item Column 10-13: Equivalent width (\text{\normalfont\AA}) of H$\alpha$, N[II], H$\beta$ , O[III];\\
    \vspace{-0.2cm}    
 \item Column 14-15: adopted nuclear classification according to the BPT and WHAN diagnostic diagrams;\\
    \vspace{-0.2cm}    
 \item Column 16: adopted stellar-mass $\log (M_{\mathrm{star}}/\mathrm{M}_{\odot})$ calculated following \cite{Zibetti-2009};\\
    \vspace{-0.2cm}    
 \item Column 17: local galaxy overdensity $(\rho-\langle\rho\rangle/\langle\rho\rangle$;\\
    \vspace{-0.2cm} 
 \item Column 18: halo-mass $M_{\mathrm{Halo}}$;\\
     \vspace{-0.2cm} 
 \item Column 19: H\thinspace{\scriptsize I}-deficiency parameter.
 \end{itemize}  
 \begin{onecolumn}
 \tiny
 \begin{landscape}
 \begin{longtable}{ccccccccccccccccccc}
 \caption{The first 30 among 30597 galaxies analysed in this work, The full table is available at CDS}\label{tab:springsample}\\
 \hline
 \noalign{\smallskip}
RA	  & Dec    &	FUV$_{\mathrm{corr}}$  &   NUV$_{\mathrm{corr}}$  &   $g_{\mathrm{corr}}$    &  $i_{\mathrm{corr}}$  &    $cz$   &   Dist &  Dbase  &     $H\alpha$EW  &  [NII]EW &  $H\beta$EW &  [OIII]EW &	 BPT  &   WHAN &  $M_{\rm star}$ &	 $\Delta \rho/\langle\rho\rangle$  & $M_{\mathrm{Halo}}$  &   $\mathrm{H} \thinspace \scriptsize{\text{I}}_{\mathrm{def}}$  \\
Deg       & Deg     &  mag   &  mag   & mag    &  mag   &   $\rm km s^{-1}$ &  $h^{-1}$ Mpc &   & \text{\normalfont\AA} & \text{\normalfont\AA}	 & \text{\normalfont\AA}   & \text{\normalfont\AA}   &  &  & $\log M_{\odot}$ &   &  $\log M_{\odot}$  &  \\
(1)       & (2)     &  (3)   &  (4)   & (5)    &  (6)	&  (7)  &   (8)	    & (9) & (10)  & (11) & (12) & (13)  & (14) & (15) & (16)	 & (17)  & (18)  & (19)  \\
 \hline
 \hline
 \endhead
150.00089 & 59.79984 & 18.63 & 18.14 & 16.51 & 15.95 & 9457.6 & 128.67 &  SL  & -19.52  & -5.11  &  -5.80	&  -2.96  & SF   &   SF   &   9.28  &   -0.72   &   -       &   -	  \\	 
150.00138 & 12.93467 & - & - & 18.09 & 17.05 & 7190.1 &  97.83 &  SL  & -0.06   & -0.27  &  0.29	&  -0.19  & -    &    -   &   9.09  &   0.42    &   12.61   &   -	  \\	 
150.00229 & 35.35190 & 18.34 & 17.99 & 16.60 & 16.11 & 5161.9 &  70.23 &  SL  & -17.11  & -3.10  &  -5.99	&  -4.07  & SF   &   SF   &   8.61  &   -0.43   &   11.88   &   -	  \\	 
150.01284 & 17.58867 &  18.89     & 18.57    & 17.04 & 16.35 & 7808.3 & 106.23 &  SL  & -49.10  & -10.50 &  -14.60  &  -15.20 & SF   &   SF   &   9.08  &   -0.43   &   -       &   -	  \\	 
150.01395 & 64.41458 & 19.46 & 19.30 & 17.60 & 17.30 & 5648.0 &  76.84 &  SL  & -15.96  & -1.31  &  -5.83	&  -8.44  &  -   &    -   &   8.00  &   -1.00	 &   -       &   -	  \\	 
150.01643 & 4.812521 & - & 21.94 & 17.49 & 16.41 & 4137.8 &  56.30 &  SL  & -0.26   & 0.055  &  0.09	&  -0.80  &  -   &    -   &   8.92  &   0.43	 &   -       &   -	  \\	 
150.02130 & 45.52088 & 17.04 & 16.79 & 15.25 & 14.75 & 1797.2 &  24.45 &  SL  & -96.57  & -7.45  &  -23.22  &  -40.67 &  SF  &   SF  &   8.26  &   0.43	 &   -       &   -	  \\	 
150.02407 & 43.19282 & 17.92 & 17.79 & 16.63 & 16.46 & 1678.7 &  22.84 &  SL  & -97.09  & -3.53  &  -22.14  &  -38.35 & SF   &   SF   &   7.18  &   -0.15   &   -       &   -	  \\	 
150.02437 &  1.91106 &  19.94    &   19.69   & 17.67 & 17.20 & 1903.5 &  25.90 &  SL  & -7.79   & -0.62  &  -3.38	&  -1.86  &  -   &    -   &   7.29  &   2.42	 &   -       &   -	  \\	 
150.02705 & 38.17220 & 16.64 & 16.22 & 14.74 & 14.19 & 7022.8 &  95.55 &  SL  & -17.15  & -7.87  &  -5.50	&  -2.28  & SF   &   SF   &   9.72  &   -0.15   &   -       &   -	  \\	 
150.02710 & 34.17198 & - & 18.20 & 16.37 & 15.67 & 5212.3 &  70.92 &  SL  & -8.64   & -1.84  &  -3.05	&  -2.99  & SF   &   SF   &   9.01  &   -0.15   &   -       &   0.36	  \\	 
150.03106 & 13.55048 & 17.72 & 16.82 & 15.11 & 14.46 & 9705.6 & 132.05 &  SL  & -7.86   & -3.18  &  -3.47	&  -1.16  & SF   &   SF   &   9.99  &   -0.15   &   11.78   &   -	  \\	 
150.03378 & 2.765159 &  17.04    &  16.18  & 15.17 & 14.62 & 8742.9 & 118.95 &  SL  & -43.11  & -21.50 &  -11.10  &  -1.24  & SF   &   SF   &   9.74  &   -0.72   &   -       &   -	  \\	 
150.03499 &  0.98479 & 16.95 & 16.32 & 15.68 & 15.33 & 9539.5 & 129.79 &  SL  & -61.18  & -34.80 &  -15.70  &  -20.90 & Comp &   SEY  &   9.32  &   -0.15   &   -       &   -	  \\	 
150.03706 & 38.45670 & 17.95 & 17.55 & 15.60 & 14.83 & 8146.0 & 110.82 &  SL  & -13.42  & -4.76  &  -4.12	&  -1.22  & SF   &   SF   &   9.82  &   -1.00	 &   -       &   -	  \\	 
150.03863 &  2.71320 &  18.62    &  18.08   & 16.85 & 16.36 & 9837.4 & 133.84 &  SL  & -49.86  & -10.30 &  -13.10  &  -8.51  & SF   &   SF   &   9.08  &   -1.00	 &   -       &   -0.37    \\	 
150.04333 &  2.15583 & -   & 18.96 & 14.40 & 13.30 & 1793.0 &  24.39 &  ZW  & 10.0    &  -     &   -	&   -	  &  -   &    -   &   9.45  &   2.42	 &   -       &   -	  \\	 
150.05021 &  9.64873 & 19.42 & 18.02 & 13.82 & 15.61 & 5411.0 &  73.62 &  LOI & -1.78    &  -5.38    &   -	&   -	  &  -   &    LIN   &   6.51  &   -0.72   &   -       &   0.16	  \\	 
150.05336 & 13.56379 & 18.95 & 18.09 & 17.18 & 16.90 & 9649.2 & 131.28 &  SL  & -36.41  & -1.95  &  -7.70	&  -18.40 & SF   &   SF   &   8.62  &   -0.15   &   11.78   &   -	  \\	 
150.05906 & 15.64372 & 19.43 & 19.42 & 17.68 & 17.17 & 9996.0 & 134.41 &  SL  & -9.74   & -2.66  &  -2.46	&  -2.87  & SF   &   SF   &   8.78  &   -0.15   &   -       &   -	  \\	 
150.06091 &  9.61748 & 20.62 & 20.07 & 17.82 & 18.47 & 5375.1 &  73.13 &  SL  & -49.27  & -4.89  &  -12.6	&  -17.50 & SF   &   SF   &   6.53  &   -0.72   &   -       &   -	  \\	 
150.06597 &  0.51725 & 20.26 & 19.65 & 17.65 & 16.97 & 9955.6 & 135.45 &  SL  & -6.981  & -1.11  &  -2.68	&  -3.94  & SF   &   SF   &   9.04  &   0.71	 &   -       &   -	  \\	 
150.06851 & 24.81398 & 18.81 & 18.29 & 16.53 & 16.14 & 6630.8 &  90.22 &  SL  & -12.77  & -2.11  &  -5.09	&  -5.24  & SF   &   SF   &   8.72  &   -1.00	 &   -       &   -	  \\	 
150.06854 & 60.72017 & 17.67 & 17.57 & 16.47 & 16.16 & 7258.5 &  98.76 &  SL  & -29.84  & -3.96  &  -9.16	&  -12.90 & SF   &   SF   &   8.71  &   -1.00	 &   -       &   -	  \\	 
150.07132 &  4.88875 & 19.66 & 19.50 & 16.25 & 15.39 & 4032.2 &  54.86 &  SL  & -93.62  & -13.90 &  -29.10  &  -15.90 & SF   &   SF   &   9.22  &   0.42	 &   -       &   -	  \\	 
150.07878 & 54.53856 & 16.52 & 16.13 & 14.11 & 13.40 & 1645.6 &  22.39 &  SL  & -30.97  & -7.28  &  -9.63	&  -5.67  & SF   &   SF   &   8.93  &   1.28	 &   -       &   0.58	  \\	 
150.08479 &  0.50598 & 19.56 & 18.44 & 14.54 & 13.68 & 9943.2 & 132.74 &  SL  & -2.26   & -2.87  &  -0.72	&  -2.43  & LIN  &   LIN  &  10.52  &   0.71	 &   -       &   -	  \\	 
150.08620 & 15.97847 &  -     & 21.15  & 16.59 & 15.43 & 8190.3 & 111.43 &  SL  & -0.42   & -0.06  &  -0.11	&  -0.24  &  -   &    -   &   9.99  &   -0.15   &   -       &   -	  \\	 
150.08645 & 15.95577 & 17.64 & 17.47 & 16.43 & 16.15 & 8171.6 & 111.17 &  SL  & -60.72  & -10.60 &  -14.30  &  -15.40 & SF   &   SF   &   8.78  &   -0.15   &   -       &   -	  \\	 
150.08708 & 11.33583 &  -     & -      & 15.52 & 14.66  & 7160.0 &  97.41 &  SL  & -4.01   &  -1.55 &  -0.46	&  -0.62  &  -   &   SF   &   9.86  &  -0.43	 &   -       &   -0.24    \\
  \hline                                                                                          \hline                                                                                          \noalign{\smallskip}                                                                            \noalign{\smallskip}
   \end{longtable}
   \normalsize
 \end{landscape}
\end{onecolumn}
\end{document}